\documentclass[11pt]{article}

\usepackage{geometry}
\usepackage{fullpage}
\usepackage{graphicx}
\usepackage{pdflscape}
\usepackage{multirow}
\usepackage{ragged2e}

\usepackage{dcolumn}
\usepackage{booktabs,calc}
\usepackage{longtable}
\usepackage{array}
\usepackage{paralist}
\usepackage{verbatim}
\usepackage{subfig}
\usepackage{setspace}
\usepackage{amsmath}
\usepackage{amssymb}
\usepackage{amsthm}
\usepackage{natbib}
\usepackage[linktocpage=true, colorlinks=true, linkcolor=blue, citecolor=black]{hyperref}
\usepackage{enumerate}
\usepackage{authblk}
\usepackage{placeins}
\usepackage{multicol}
\usepackage{array}
\usepackage[usenames, dvipsnames]{xcolor}
\usepackage{url}
\usepackage{listings}

\lstdefinestyle{myListingStyle} 
    {
        basicstyle = \small\ttfamily,
        breaklines = true,
    }

\usepackage{threeparttable}

\usepackage{siunitx} 
    \defcitealias{ETH1}{CSA, 2014}
    \defcitealias{ETH2}{CSA, 2015}
    \defcitealias{ETH3}{CSA, 2017}
    
    \defcitealias{MWI1}{NSO, 2012}
    \defcitealias{MWI2}{NSO, 2015}
    \defcitealias{MWI3}{NSO, 2017}
    
    \defcitealias{NGR1}{NIS, 2014}
    \defcitealias{NGR2}{NIS, 2016}
    
    \defcitealias{NGA1}{NBS, 2012}
    \defcitealias{NGA2}{NBS, 2014}
    \defcitealias{NGA3}{NBS, 2019}
    
    \defcitealias{TZA1}{TNBS, 2011}
    \defcitealias{TZA2}{TNBS, 2012}
    \defcitealias{TZA3}{TNBS, 2015}
    
    \defcitealias{UGA1}{UBOS, 2014a}
    \defcitealias{UGA2}{UBOS, 2014b}
    \defcitealias{UGA3}{UBOS, 2016}

\begin{document}
	
\title{Privacy Protection, Measurement Error, and the Integration of Remote Sensing and Socioeconomic Survey Data\thanks{Correspondence to \href{mailto:jdmichler@arizona.edu}{jdmichler@arizona.edu} or \href{mailto:tkilic@worldbank.org}{tkilic@worldbank.org}. A pre-analysis plan for this research has been filed with Open Science Framework (OSF): \href{https://osf.io/8hnz5/}{https://osf.io/8hnz5/}. We gratefully acknowledge funding from the World Bank Living Standards Measurement Study (LSMS) and the Knowledge for Change Program (KCP). This paper has been shaped by conversations with Leah Bevis as well as seminar participants at the Methods and Measurement Conference 2021, the AAEA annual meetings in Chicago and Atlanta, the $31^{st}$ triennial ICAE conference, and participants in presentations at Arizona State University, the University of Minnesota, the World Bank, and Virginia Tech. We are especially grateful to Alison Conley, Emil Kee-Tui, and Brian McGreal for their diligent work as research assistants and to Oscar Barriga Cabanillas and Aleksandr Michuda for early help in developing the Stata \texttt{wxsum} package. We are solely responsible for any errors or misunderstandings.}}

	\author[1]{Jeffrey D. Michler}
	\author[1]{Anna Josephson}
	\author[2]{Talip Kilic}
	\author[2]{Siobhan Murray}
	\affil[1]{\small \emph{Department of Agricultural and Resource Economics, University of Arizona}}
	\affil[2]{\small \emph{Development Data Group, World Bank}}

\date{February 2022}
\maketitle

\thispagestyle{empty}

\begin{center}\begin{abstract}
		\noindent When publishing socioeconomic survey data, survey programs implement a variety of statistical methods designed to preserve privacy but which come at the cost of distorting the data. We explore the extent to which spatial anonymization methods to preserve privacy in the large-scale surveys supported by the World Bank Living Standards Measurement Study - Integrated Surveys on Agriculture (LSMS-ISA) introduce measurement error in econometric estimates when that survey data is integrated with remote sensing weather data. Guided by a pre-analysis plan, we produce 90 linked weather-household datasets that vary by the spatial anonymization method and the remote sensing weather product. By varying the data along with the econometric model we quantify the magnitude and significance of measurement error coming from the loss of accuracy that results from protect privacy measures. We find that spatial anonymization techniques currently in general use have, on average, limited to no impact on estimates of the relationship between weather and agricultural productivity. However, the degree to which spatial anonymization introduces mismeasurement is a function of which remote sensing weather product is used in the analysis. We conclude that care must be taken in choosing a remote sensing weather product when looking to integrate it with publicly available survey data.
	\end{abstract}\end{center}

	{\small \noindent\emph{JEL Classification}: C38, C81, D83, O13, Q12 }

\newpage
\onehalfspacing


\section{Introduction}

Public use datasets from large-scale household surveys play a central role in tracking progress towards national and international development goals and in formulating a wide array of development research. These surveys include those that are supported by the World Bank's Living Standards Measurement Study (LSMS), the USAID-funded Demographic and Health Surveys (DHS), and UNICEF's Multiple Indicator Cluster Surveys (MICS). In making these datasets public, survey programs must balance the demand for accurate data with the need for privacy protection. The more accurate the public data, the more privacy is lost \citep{DinurandNissim03}.

To preserve privacy when publishing data, survey programs implement statistical disclosure limitation (SDL). SDL methods distort data, preserving privacy but reducing data accuracy and interoperability, both key requirements for data to generate value for development \citep{Jolliffeetal21}. Interoperability relates to the ease with which different data sources can be linked through various means, including geographic coordinates or common geographic identifiers. In large-scale household surveys, the use of Global Positioning System (GPS) technology to capture sampled enumeration area (EA), household, and agricultural plot locations has dramatically increased the interoperability of the survey data by allowing the integration of survey data with remote sensing data \citep{Burkeetal21}. Although capturing precise GPS coordinates increases interoperability, and thus the relevance and cost-effectiveness, of household surveys, such data are confidential and must be ``spatially anonymized'' before public release. International survey programs have, therefore, adopted SDL coordinate masking techniques such that public use datasets that include anonymized unit-record microdata are also inclusive of spatially anonymized GPS coordinates. While a range of coordinate masking techniques exist (see Figure~\ref{fig:features}), the technique that is currently used by the DHS and LSMS randomly offsets precise EA coordinates by zero to two kilometers (km) in urban areas and two to five km in rural areas \citep{Murray21}.

This paper contributes to the nascent economics literature on privacy protection and statistical accuracy. We integrate nine remotely sensed geospatial weather datasets with the georeferenced longitudinal household survey data that have been collected across six Sub-Saharan African countries under the World Bank LSMS-Integrated Surveys on Agriculture (LSMS-ISA) initiative. Prior to the integration process, we used the confidential household GPS coordinates to generate ten different spatial representations of the precise household locations. Linking the weather data to the household survey data using each of these ten spatial representations allows us to quantify the magnitude and significance of measurement error coming from the loss of accuracy that results from different SDL methods to protect privacy. We test this by modeling the relationship between weather and smallholder agricultural productivity, as measured through the LSMS-ISA-supported household surveys.\footnote{Besides being an area of research itself, agricultural production and productivity is often used to proxy for a variety of economic outcomes, including economic growth \citep{DescheneGreenstone07}, intra-household bargaining power \citep{CornoEtAl20}, and migration \citep{Jayachandran06}.} Our goal is to provide guidance to researchers looking to integrate geospatial data with socioeconomic survey data regarding the degree to which their results may be mismeasured due to privacy protection methods.

There are three headline findings from our research. First, we find that spatial anonymization techniques currently in general use, such as those currently employed by the LSMS and the DHS, have, on average, limited to no impact on estimates of agricultural productivity. At this time, the spatial resolution of publicly available remote sensing weather products are generally too coarse for any of the spatial anonymization methods to make a substantial difference in which pixel a household ends up in. Second, and not unexpectedly, the degree to which spatial anonymization introduces mismeasurement is a function of which remote sensing weather product is used in the analysis. Remote sensing products that merge gauge and satellite data, such as ARC2, CHIRPS, and TAMSAT, are seemingly of a high enough resolution to be sensitive to some spatial anonymization techniques.\footnote{Section~\ref{sec:weatherdata} includes a full description of each of these products.} Remote sensing products that rely on assimilation models, such as ERA5 and MERRA-2, or products that primarily rely on gauge data, such as CPC, are of a low enough resolution that commonly used spatial anonymization techniques have no discernible impact on estimates of agricultural productivity. Third, estimates of weather's impact on agricultural productivity are also a function of the remote sensing data source, regardless of the degree of/approach to spatial anonymization. The extent to which weather impacts agricultural productivity varies substantially both in sign, significance, and magnitude, across remote sensing weather data products for the same spatial anonymization technique. These results suggest the need for care when choosing a remote sensing data product to integrate with socioeconomic survey data, as results can vary depending on the choice of product and the spatial anonymization technique used to protect privacy.

As noted above, there is scope for the impact of spatial anonymization to vary in accordance with the measurement error in geospatial data sources that household survey data are linked to - in our case, remote sensing weather data. The goal of a remote sensing weather product is to document an objective fact: that is, the volume of precipitation or the temperature in a given location at a given time. Inaccuracies introduced by either the sensor (e.g., infrared, microwave, optical) or the algorithm used to convert sensor data into rainfall or temperature (e.g., reanalysis, interpolation) means remote sensing products may mismeasure the objective fact. Simply with respect to the ``raw'' weather data, there can be substantial variation in what a remote sensing product reports as the actual rainfall or temperature in a given location. Figures~\ref{fig:rain_res} and \ref{fig:temp_res} show this variation across six remote sensing precipitation products and three temperature products. One precipitation product reports rainfall of zero to five millimeters (mm) in the southeast corner of the grid cell while a different product reports 47-64 mm for the same location on the same day. Temperature also varies by remote sensing product, with one product reporting a maximum temperature of 23$^{\circ}$ Celsius while another reports the maximum temperature that day as 27$^{\circ}$ Celsius.

That variation exists not only in the spatial resolution of the remote sensing data but in the precipitation and temperature reported by each product informs how we implemented our research design. First, we developed a pre-analysis plan and registered it at Open Science Framework \citep{PAP}. While pre-analysis plans have become common in experimental economics, they are still relatively uncommon for binding researchers' hands when using observational data \citep{JanzenMichler21}. The use of a pre-analysis plan allowed us to pre-define the sources of data for inclusion in the study, what metrics would be tested using what functional forms, and how we would compare results across models in the absence of formal statistical tests. Second, we adopted a blinding strategy to help ensure objectivity in the implementation of the pre-analysis plan. As such, the authors were divided into two groups: the Data Generating Group and the Data Analysis Group. Authors Kilic and Murray were in the Data Generating Group and had full responsibility for extracting the remote sensing data and matching it to the household records in the household survey data to create a number of different paired weather-survey datasets.\footnote{For example, in one dataset the remote sensing weather data product may be matched with the exact household coordinates, while in another dataset the remote sensing weather data may be matched with low-level Administrative area.} In these datasets, the source of the weather data and the spatial anonymization method was anonymized prior to sharing with the Data Analysis Group. Authors Josephson and Michler made up the Data Analysis Group and had full responsibility for cleaning the agricultural productivity data, running the regressions, and conducting and writing the analysis. The pre-specified analysis was carried out on the blinded datasets and these results were posted to \href{arXiv.org}{arXiv.org} prior to unblinding \citep{MichlerEtAl20}. The generation of datasets in this manner preserves the objectivity of any findings regarding differences in outcomes between different spatial anonymization techniques and different remote sensing products. 

Against this background, this paper provides, to our knowledge, the first empirical evidence on the extent to which spatial anonymization of public use survey datasets affects econometric analysis when those datasets are linked to remote sensing data. We also provide evidence on how the significance and magnitude of the effect of spatial anonymization varies in accordance with the remote sensing data source. In our case, the unique access of the Data Generating Group to the confidential household GPS coordinates in the LSMS-ISA's nationally-representative, panel datasets allows us to execute the comparative assessment and isolate the role of spatial anonymization in subsequent econometric analyses of smallholder agricultural productivity.

The issues surrounding privacy-preserving data analysis are well-known in computer science but have come to the widespread attention of economists only since the announcement by the US Census Bureau to implement differential privacy for the 2020 Census of Population \citep{AbowdSchmutte19}. The issue of accuracy in privacy-preserving data remains largely unexplored in the development economics literature, despite the proliferation of research on accuracy and measurement error in household  survey  data \citep{Carlettoetal17,AbayEtAl19,KosmowskiEtAl19,GollinUdry21, KilicEtAl21}. To date, there is limited evidence on how the use of spatially anonymized public use datasets may impact the findings of research efforts that are centered on the integration of georeferenced socioeconomic survey data with satellite imagery and/or processed geospatial data. This is despite the rapid expansion in publicly available high-resolution satellite imagery, which has been used in combination with household survey data for small area estimation of poverty, wealth, health, nutrition, and agricultural outcomes in low-income contexts \citep{AzzariEtAl21, BurkeandLobell17, Graetzetal18, OsgoodZimmermanetal18, DwyerLindetal19, YehEtAl20}.

Relatedly, a large body of economic research has relied on remotely-sensed weather data for identification of causal effects \citep{DellEtAl14, DonaldsonStoreygard16}. This includes important contributions that rely on the availability of georeferenced household survey data and that relate to human capital formation \citep{MacciniYang09, ShahSteinberg17, GargEtAl20}, labor markets \citep{Jayachandran06, ChenEtAl17, Kaur19, Morten19}, conflict and institutions \citep{BruckerCiccone11, Sarsons15, KonigEtAl18}, agricultural production and economic growth \citep{MiguelEtAl04, DescheneGreenstone07, BarriosEtAl10, DellEtAl12, YehEtAl20}, intra-household bargaining power \citep{CornoEtAl20}, technology adoption \citep{Suri11, Taraz18, JagnaniEtAl21, AragonEtAl21, TesfayeEtAl21}, and extreme weather impacts \citep{WinemanEtAl17, MichlerEtAl19, McCarthyEtAl21a}. Our findings suggest that economists should exercise caution when seeking to combine remote sensing data with public use socioeconomic survey data.

The paper is organized as follows: in Section 2 we discuss the issue of privacy loss, different methods for privacy protection, and their implications for economic analysis. We also discuss the current coordinate masking techniques used by the DHS and the LSMS to ensure spatial anonymity in their published datasets. Section 3 details the sources and characteristics of the weather data and the household data used in this analysis. We provide details on how data was integrated, including specifics on how the blinded data was combined. The section concludes by presenting some descriptive evidence of mismeasurement in the remotely sensed weather data. Section 4 gives details of the pre-analysis plan, specifically our estimation strategy and approach to inference. Section 5 discusses results while Section 6 concludes with a set of recommended best practices for researchers looking to integrate remote sensing data with socioeconomic survey data.


\section{Privacy Protection in Socioeconomic Data}

Socioeconomic data, including personal data and household survey data, are collected with the understanding that the identity of individual respondents will be protected when the data are disseminated or used in research. This is the case with the large, public use datasets most commonly used in development economics, including those made available by the LSMS, the DHS, and the MICS. Statistical disclosure limitation (SDL) methods such as noise infusion, aggregation, record swapping, or suppression may be employed to reduce the uniqueness of any single record in the sample and maintain confidentiality. In the spatial dimension, SDL is often achieved through coordinate masking and noise infusion on derived spatial variables. SDL inherently distorts the data, which can lead to bias in statistical analysis \citep{AbowdEtAl19}. Because data providers do not publish SDL critical parameters, so as to reduce the potential for database reconstruction, it is not possible to determine the magnitude or direction of the bias \citep{AbowdSchmutte15}.

Regardless of the SDL methods employed to protect privacy, the database reconstruction theorem demonstrates that publishing too many statistics too accurately from a confidential database exposes the entire database with near certainty \citep{DinurandNissim03}. Additionally, the expanding availability of personal data that can be linked to survey data, as well as the wide availability of software and computational resources for mining these data, means that data de-identified via traditional SDL are vulnerable to re-identification via record linkage. In recent years, companies like Apple, Facebook, and Google have used differential privacy (DP) techniques in preserving privacy of user data \citep{WoodEtAl18}. This is also the method adopted by the US Census Bureau in preparing the 2020 Census data for release \citep{AbowdEtAl19}. DP techniques allow for the precise measurement of disclosure risk, thereby avoiding excessive data manipulation, while meeting anonymization objectives \citep{DworkEtAl06}. The use of DP, or any privacy protecting statistical technique, raises important questions about social choice, privacy protection, data accuracy, and the transparency and reproducibility of research. This is a debate which economists are just now beginning to enter.\footnote{See the symposium at the 2019 AEA Annual Meeting \citep{AbowdEtAl19, Abraham19, ChettyFriedman19, RugglesEtAl19}.}

As of 2022, DP has only just begun to be adopted by the statistical agencies and the managers of the databases most commonly used by economists. This includes the US Census Bureau, which adopted DP for the 2020 census. Privacy in the Opportunity Atlas, which is published at the Census tract level, is also protected by methods that build on DP \citep{ChettyFriedman19}. However, to date, public use household survey datasets in development economics still rely on SDL to protect participant privacy. While DP may hold promise for future household survey data dissemination, in this analysis we make use of existing LSMS-ISA public datasets which rely on SDL to anonymize location data. In the remainder of this section, we detail the SDL methods currently used in the LSMS-ISA data in addition to the various methods we test in our analysis.


\subsection{Geomasking}

Spatial anonymization has dual objectives: (1) to provide a geographic reference that enables users to integrate information from spatial datasets into a household survey and, at the same time, (2) to preserve confidentiality of place, preventing re-identification of the location of survey respondents. Geomasking, or coordinate perturbation, serves to conceal the actual location and, when mask parameters are revealed, also enables users to incorporate uncertainty into spatial variables derived using the anonymized locations. The geomasking technique applied to LSMS-ISA public microdata is a type of SDL developed by the DHS Program and has been used in the dissemination of survey datasets since the early 2010s \citep{Murray21}. 

Specifically, the coordinate modification strategy relies on noise infusion through random offset of EA centerpoint coordinates (or average of sample household GPS locations by EA) within a specified range determined by an urban/rural classification. For urban areas, a range of 0-2 km is used. For rural areas, where communities are further dispersed and risk of disclosure could be greater, a range of 0-5 km offset is used. An additional 0-10 km offset for a small percentage (ranging from $1\%-10\%$) of rural clusters effectively increases the known range for all rural points to 10 km while introducing only a small amount of noise. The result is a set of coordinates, representative at the EA level, that fall within limits of accuracy known to the data user.

With the geomasking method described, there is no guarantee that specific anonymization objectives are achieved. Further, this geomasking method does not take into account location-specific characteristics, other than official rural/urban classification. Adaptive approaches, where displacement is a function of site characteristics or the offset range defined by a target population count, have been explored by both the LSMS and DHS. An adaptive approach has the potential to avoid instances of excessive displacement in densely populated urban areas, as well as inadequate protection in sparsely populated areas. However, uncertainty in gridded population data inputs at large scale remains a barrier to implementation of the adaptive approach in many settings \citep{Murray21}. As a result, the strata-based method remains the primary spatial anonymization for dissemination of the LSMS datasets at this time. In addition to the current geomasking method, we evaluate the use of spatial features generated by two other aggregation methods in this analysis.


\subsection{Spatial Feature Representation}

Most household survey datasets include location variables (e.g., region, district, or other place names), that define a base level of spatial disclosure risk. Any additional spatial information, including anonymized coordinates, allows for refinement of the anonymizing region, or area within which the survey respondent is known to reside. The trade-off for this increased exposure risk is an expected gain in the accuracy of derived spatial variables, such as precipitation or temperature. As the unit of analysis in many analyses - this one included - is the household, variables derived using exact household coordinates are assumed to contain the least amount of noise but produce the greatest risk of re-identification. 

Starting with the exact, confidential, household coordinates, we provide a comparative assessment of five additional spatial feature representations: (1) the average of sampled household locations within an EA (EA center); (2) an anonymized EA location (EA center modified); (3) the area of anonymizing region (EA zone of uncertainty); (4) the centerpoint of the administrative unit associated with lowest-level locality variable in the public microdata (Administrative area center); and (5) the Administrative area. In Table~\ref{tab:spatialanon} these spatial features are described in terms of the average displacement distance and a qualitative assessment of the impact on spatial disclosure risk associated with the dissemination of the reference location or spatial variables derived using the respective features.

The average point displacement, which could be viewed as representing potential mismeasurement in the derived variables, varies somewhat by country and strata, depending on factors such as the areal extent of EAs and administrative units. However, the direction and magnitude of difference between feature types is common across all surveys in the analysis. While the effect of displacement distance may be generally progressive for landscape-level phenomena like weather and medium resolution datasets, this impact is scale-dependent. One could expect that hyperlocal characteristics, like field-level vegetation indices, from high resolution imagery would be rendered unusable by insertion of almost any noise.


\subsection{Extraction Method}

The spatial features discussed above are a mix of point and polygon, or area, representations (see Figure~\ref{fig:features}). In this analysis we make use of multiple gridded, or raster, weather data sources produced at different spatial resolutions (see Figures~\ref{fig:rain_res} and~\ref{fig:temp_res}). The method by which raster values are linked to different spatial features can compensate to some degree for differences in feature size and grid resolution. For example, the EA zone of uncertainty or Administrative area may be smaller than a single grid cell or cover multiple cells. A point feature may lie on the boundary of two grid cells or be located near a cell center. Extraction method refers to the way underlying grid cell values are processed.

We evaluate three commonly employed techniques for merging values from raster data to household roster records using the six spatial feature representations. For the four point features we extract weather time series data using both simple and bilinear methods, resulting in eight outputs. The simple method extracts raster cell values by spatial intersection alone, not accounting for the point location within cell boundaries. The bilinear method computes the distance weighted average of values at four nearest cell centers. It is important to note that the bilinear method is generally preferred for integration of continuous data like precipitation and temperature. However, as we are aiming to assess the added value of the more complex calculations in this context, both are considered in our analysis. For the two polygon feature types we extract values using a zonal mean, or average of all cells overlapped by the polygon. The use of polygon features can account for uncertainty in location, as with the zone of uncertainty or modified EA location. Zonal means will also smooth the results, reducing the effect of extreme cell values.

All together, the combination of spatial feature representations and extraction methods gives us ten spatial anonymization methods that we test against each other in the following analysis. For reference, the LSMS-ISA public datasets include EA center modified coordinates. Geovariables disseminated with the microdata are currently generated using the EA center modified location and bilinear extraction, unless the underlying spatial dataset is categorical, in which case the simple extraction method is used.


\section{Data} \label{sec:data}

To understand the privacy/accuracy trade-off in anonymizing spatial data, we combine publicly available satellite-based weather data products with publicly available unit-record survey data that have been generated as part of the World Bank LSMS-ISA initiative and that are made available through the World Bank Microdata Library. In this section, we first describe the weather data and household data. We then discuss the blinding of the research team and the data integration process. We conclude with a discussion of some descriptive statistics for the combined weather-household datasets.


\subsection{Remote Sensing Weather Data} \label{sec:weatherdata}

We use a number of public domain sources of weather datasets representing different modeling types, input sources, and spatial resolutions. Although there are many possible weather products to consider, we sought to include the remote sensing data products most commonly used by economists. To ensure consistency and enable the production of common metrics across the analysis, we imposed two inclusion criteria. The source had to have (1) high temporal resolution, i.e., daily, and (2) a minimum 30-year length of record, from 1987 to, at least, 2017. Unfortunately, this criteria meant that some data sources frequently used by economists, such as the various versions of the monthly \emph{Terrestrial Air Temperature and Precipitation} from the Center for Climatic Research at the University of Delaware was excluded. Table~\ref{tab:weather} describes each data sources, including the length of record, spatial and temporal resolution, and the type of data recorded. See online Appendix~\ref{sec:appRS} for more details on each remote sensing product and guidance for economists on merging these data with survey data.

The remote sensing weather data that we use can be categorized by its method of generating precipitation and temperature values. The first type of product we use merges gauge data, which provide site-level observations, with data from meteorological satellites, which provide valuable indirect information at full coverage. Remote sensing products of this type include the African Rainfall Climatology version 2 (ARC2), the Tropical Applications of Meteorology using SATellite data and ground-based observations (TAMSAT), and the Climate Hazards group InfraRed Precipitation with Station Data (CHIRPS) \citep{ARC2, TAMSAT, CHIRPS}.

The second type of product uses assimilation models to combine a large number of observations from different sources (e.g., satellites, weather stations, ships, aircraft) to produce a model of the global climate system or a particular atmospheric phenomenon. Outputs are inferred or predicted based on the system state and understanding of interactions between model variables. We use two reanalysis datasets for both rainfall and temperature in this analysis: the European Centre for Medium-Range Weather Forecasts ERA5 and the NASA Modern-Era Retrospective analysis for Research and Applications (MERRA-2) \citep{ERA5, MERRA2}.

Last, we consider a data product produced primarily from gauge data, using only spatial interpolation techniques to produce a continuous surface from observed measurements. The NOAA Climate Prediction Center (CPC) Unified Gauge-Based Analysis of Daily Precipitation and Temperature datasets were created using all information sources available at CPC and undergoes extensive pre-processing and cleaning, including comparison with contemporaneous data from satellite and other sources \citep{CPC}.


\subsection{Household Survey Data} \label{sec:householddata}

The World Bank Living Standards Measurement Study - Integrated Surveys on Agriculture (LSMS-ISA) is a household survey program that provides financial and technical assistance to national statistical offices in Sub-Saharan Africa for the design and implementation of national, multi-topic longitudinal household surveys with a focus on agriculture. As detailed below, our analysis leverages data from several rounds of panel household surveys conducted over the last decade in Ethiopia, Malawi, Niger, Nigeria, Uganda and Tanzania. Table~\ref{tab:lsms} provides a summary of the countries, years, and observations used in the analysis. Online Appendix~\ref{sec:appHH} provides greater details on each country's sampling frame and data collection process.

In Ethiopia, we use the data from the 2011/12, 2013/14 and 2015/16 rounds of the Ethiopia Socioeconomic Survey (ESS), which has been conducted by the Central Statistical Agency of Ethiopia \citepalias{ETH1, ETH2, ETH3}. The Wave 1 data is representative at the regional level for the most populous regions in the country while Wave 2 and 3 expanded to include 1,500 households in urban areas. After data cleaning to remove urban and non-agricultural rural households, we are left with 7,272 household observations across three survey waves.

In Malawi, the LSMS-ISA data includes two separate surveys: the cross-sectional Integrated Household Survey (IHS), and the longitudinal Integrated Household Panel Survey (IHPS) \citepalias{MWI1, MWI2, MWI3}. This analysis relies on the data from the IHPS, which is representative at the national-, urban/rural-, and regional-level. Data comes from 2010/11, 2013, and 2016/17. After data cleaning to remove tracked and non-agricultural households, we are left with 3,250 household observations across three survey waves.

In Niger, we use two waves, the first from 2011 and the second from 2014 \citepalias{NGR1, NGR2}. The sample is representative at the national and urban/rural-level. Data cleaning and removal of non-agricultural households gives us 3,913 household observations across two survey waves.

In Nigeria, we use the data from the 2010/11, 2012/13, and 2015/16 rounds of the General Household Survey - Panel, which is representative at the national and urban/rural-level \citepalias{NGA1, NGA2, NGA3}. Data cleaning and removal of non-agricultural households yields 8,384 household observations across three survey waves.

In Tanzania, the data come from the 2008/09, 2010/11, and 2012/13 rounds of the Tanzania National Panel Survey (TZNPS) \citepalias{TZA1, TZA2, TZA3}. The sample is representative for the nation, and provides estimates of key socioeconomic variables for mainland rural areas, Dar es Salaam, other mainland urban areas, and Zanzibar. Focusing on rural, crop producing households that do not move, we have 5,669 household observations across three survey waves.

In Uganda, we use the data from the 2009/10, 2010/11, and 2011/12 rounds of the Uganda National Panel Survey (UNPS) \citepalias{UGA1, UGA2, UGA3}. As with the other LSMS-ISA data, the Uganda sample was designed to be representative at the national-, urban/rural- and regional-level. We include 5,250 household observations after cleaning and removing non-agricultural households.

For the analysis, we combine data from the six countries and all waves to generate a single cross-country panel dataset which includes 33,738 household observations. For estimation, we include two measures of agricultural productivity: yield (kg/ha) of the primary cereal crop and the value (2010 USD/ha) of all seasonal crop productivity on the farm.


\subsection{Data Integration} \label{sec:integration}

Methods of data integration are often overlooked in the process of merging spatial data, in particular weather data, with household surveys. Publicly available datasets obfuscate the exact GPS coordinates of unit-records to ensure privacy. If underlying datasets are fairly smooth and areas of interest are small relative to the resolution of spatial data, then the effect of integration method could be negligible. However, this is not known and so our analysis sheds light on this privacy/accuracy trade-off.

As defined in our pre-analysis plan, the authors divided themselves into two groups to blind the Data Analysis Group from the identity of the spatial anonymization technique as well as the source of the remote sensing data \citep{PAP}. The entire team participated in the development and registration of the pre-analysis plan, which included defining the remote sensing products to be used and the anonymization methods to be employed. At that point, the Data Generating Group accessed the publicly available remote sensing data for use in the study. They also used the privately available household coordinate data to generate the ten different sets of anonymization methods to be assessed. The actual GPS household location is not part of the publicly available LSMS-ISA data and is known only to a limited number of individuals at the World Bank. 

After pre-processing, the Data Generating Group extracted the relevant remote sensing data for the LSMS-ISA households based on the ten spatial anonymization methods for all remote sensing sources. This generated time series datasets of daily precipitation or temperature from January 1, 1983 until December 31, 2017. For each country in each of these years, a growing season was defined based on FAO recommendations.\footnote{For more details on the definitions of growing seasons in each country, see Appendix~\ref{sec:appRS_gs} and Table~\ref{tab:growseason}.} And so, for each of the 17 LSMS-ISA country-wave household datasets, this generated 90 remote sensing weather datasets (six precipitation sources $+$ three temperature sources $\times$ ten anonymization methods). The time series weather datasets include daily observations and the unique household identifiers made part of the publicly available LSMS-ISA data. datasets were named and labeled \texttt{x0, ..., x9} for each anonymization method, \texttt{rf1, ..., rf6} for each precipitation data source, and \texttt{tp1, ..., tp3} for each temperature data source. These 1,530 blinded datasets were then shared, via a secure server, with the Data Analysis Group. 

The Data Analysis Group then processed each of the time series weather datasets using a user-written Stata package \texttt{wxsum} which is available through \href{https://github.com/AIDELabAZ}{Github}. This package processes daily precipitation or temperature data and outputs up to 22 different weather metrics. See Table~\ref{tab:Wvar} in the Online Appendix for a complete list of weather metrics used in the analysis. These weather metrics from each of the 1,530 weather datasets were then merged to the relevant country-wave LSMS-ISA dataset using the unique household identifier (90 weather datasets per country-wave dataset). All country-wave datasets containing the productivity data and the weather metrics from each remote sensing source and extraction method were then appended to create a single panel dataset covering all countries, waves, remote sensing sources, and anonymization methods. Table~\ref{tab:sourcesetc} summarizes the scope of the resulting data.

Following \cite{Dufloetal20}, we have produced a ``populated pre-analysis plan'' that completely reproduces the results of all pre-specified analysis. After the Data Analysis Group conducted all of the analysis on the blinded dataset, they posted the populated pre-analysis plan to arXiv.org on 19 August 2021. That version of the populated pre-analysis plan (\href{https://arxiv.org/abs/2012.11768v2}{arXiv:2012.11768v2}) refers to all results based on their randomly assigned identifier (\texttt{x0, ..., x9}; \texttt{rf1, ..., rf6}; and \texttt{tp1, ..., tp3}). On 23 August 2021, the Data Generating Group shared the key so that the Data Analysis Group could de-anonymize the data. The populated pre-analysis plan was then updated to replace the randomly assigned identifiers with the actual anonymization methods and names of remote sensing sources (\href{https://arxiv.org/abs/2012.11768v3}{arXiv:2012.11768v3}).\footnote{The populated pre-analysis plan is also available as a World Bank Policy Research Working Paper \citep{MichlerEtAl21WB}.} The current research paper presents the subset of the pre-specified results that focused on the issue of spatial anonymization.


\subsection{Descriptive Statistics} \label{sec:summarystats}

Our pre-analysis plan specifies that we will examine 22 different ways to measure precipitation and temperature in order to evaluate certain weather metrics are more or less accurate to spatial anonymization methods used to ensure participant privacy. A complete list of these variables with their exact definitions are in Table~\ref{tab:Wvar} in the Online Appendix. For parsimony, we focus on only four of these 22 variables in this analysis: (1) mean daily rainfall, (2) number of days without rain, (3) mean seasonal temperature, and (4) growing degree days (GDD). These four variables are indicative of a number of different ways to measure precipitation (volume v. count) and temperature (measured temperature v. bounded count).

Figure~\ref{fig:density_aez_rf} presents the distribution of mean daily rainfall (measured in mm) during the growing season, by anonymization method and remote sensing product. Looking across panels there are substantial differences in the distribution of rainfall as reported by each remote sensing product. CHIRPS, CPC, ARC2, and TAMSAT each report maximums in the eight to 12mm range. By comparison, MERRA-2 reports a maximum average of 15mm a day and ERA5 reports maximum average rainfall of nearly 42mm. Recall, this is the mean of daily rainfall for a single growing season in a single year. While there is substantial disagreement between remote sensing products regarding the volume of precipitation in a given location, there is much less variation between anonymization methods. In general, different anonymization methods implemented to protect privacy have a small effect on the accuracy of measuring the volume of precipitation. Where differences occur, they tend to be deviations due to mismeasurement introduced by using Administrative boundaries (either bilinear, simple, or zonal mean methods) instead of Household or EA. These deviations appear to be focused in the lower and center part of the distribution. However, deviations are not limited to only one remote sensing product: mismeasurment occurs in all six remote sensing products.

Figure~\ref{fig:norain_aez_rf} further explores these differences by estimating the mean number of days without rain reported by each remote sensing product for each anonymization method in each season. Mean estimates are generated using a fractional-polynomial and graphs include $95\%$ confidence intervals on the mean estimates. First, considering the consistencies across panels, CHIRPS, CPC, and ARC2 frequently report a similar number of days without rain (100-150). Similarly, MERRA-2 and ERA5 are often in agreement (40-80). TAMSAT is similar to CHIRPS, CPC, and ARC2 in the early years ($\approx100$), though deviates from these products in later years ($110 < 140$). Measurements from CHIRPS, CPC, ARC2, and TAMSAT suggest that there are substantially more days without rain, relative to the measurements from MERRA-2 and ERA5. Considering the variation by anonymization method, Administrative bilinear and Administrative zonal mean clearly under count the days without rain while modified EA simple and Administrative simple tend to over count days without rain. These differences are less pronounced in products based on assimilation models.

In Figure~\ref{fig:density_aez_tp} we present the distribution of mean seasonal temperature (measured in $^{\circ}$Celsius), by anonymization method and remote sensing product. Compared to the distribution of mean daily rainfall, the figures show much tighter distributions around mean temperature, though MERRA-2 and CPC report temperatures of zero degrees, giving them long left tails. Again, the use of Administrative linear, Administrative simple, and Administrative zonal mean frequently result in mismeasurement, though in these cases, the deviations are almost exclusively at the lower end of the distribution. All ten anonymization methods produce essentially the same results for temperatures above 25$^{\circ}$ Celsius.

Figure~\ref{fig:gdd_aez_tp} estimates the mean GDDs in a year using a fractional-polynomial and includes $95\%$ confidence intervals on the mean estimates. As with number of days without rain, GDD represents a relative coarsening of the data by converting measured temperature into a count variable for the number of days in which temperature fell within a given range. Unlike the number of days without rain, we see no statistical differences in GDD across the ten anonymization methods or across the three remote sensing products. Confidence intervals overlap for all methods, for all remote sensing products, and in all years.

Summarizing the descriptive evidence: the use of some anonymization methods to protect privacy induces a loss of accuracy. This loss of accuracy, however, is primarily limited to the use of administrative area for spatial feature representation. Not surprisingly, administrative area provides the greatest degree of privacy protection but is also the least accurate in representing the precipitation and temperature experienced by the household. Reducing privacy protection by using anonymization methods that are closer to the true household location produce more accurate measurements of the weather. Mismeasurement also varies by remote sensing product, which makes intuitive sense since the products differ in their spatial resolution. Finally, there is also evidence of mismeasurement in the remote sensing products themselves, with large disagreements between some products regarding daily precipitation and smaller disagreements regarding the daily temperature.


\section{Analysis Plan}  \label{sec:analysisplan}

The following analysis, and the associated results, was pre-specified in our pre-analysis plan \citep{PAP} and was registered with Open Science Framework (OSF). If methods, approaches, or inference criteria differ from our plan, we highlight these differences. Results arising from these deviations in our plan should be interpreted as exploratory.


\subsection{Estimation}\label{sec:eststart}

Our basic model specification follows \cite{DescheneGreenstone07}:

\begin{equation}
Y_{ht} = \alpha_{h} + \gamma_{t} + \sum_{j}^{J} \beta_{j} f_{j} \left( W_{jht} \right) + u_{ht}
\end{equation}

\noindent where $Y_{ht}$ is our outcome variables from the LSMS-ISA-supported household surveys, described above, for household $h$ in year $t$, log transformed using the inverse hyperbolic sine. We control for year fixed-effects $(\gamma_{t})$ and include household fixed-effects $(\alpha_{h})$ in some specifications. The function $f_{j} \left( W_{jht} \right)$ represents our weather variables of interest where $j$ represents a particular measurement of weather. Finally, $u_{ht}$ is an idiosyncratic error term clustered at the household-level.

From this general set-up, we estimate four versions of the model: two linear and two quadratic.\footnote{In our pre-analysis plan we defined two additional models that include measured inputs (fertilizer, labor, pesticide, herbicide, and irrigation). However, we find that controlling for inputs has no discernible effect on results, relative to the household fixed effects model and so we exclude these results from this paper. The populated pre-analysis plan on \href{arXiv.org}{arXiv.org} and through the World Bank contain all of these results \citep{MichlerEtAl20, MichlerEtAl21WB}.} For each model, a single weather variable is considered. For the linear specification:

\begin{subequations}
\begin{align}
Y_{ht} &= \alpha + \beta_{1} W_{ht} + u_{ht} \label{eq:linear} \\
Y_{ht} &= \alpha_{h} + \gamma_{t} +  \beta_{1} W_{ht} + u_{ht} \label{eq:linearFE} \\
\end{align}
\end{subequations}

\noindent For the quadratic specification:

\begin{subequations}
\begin{align}
Y_{ht} &= \alpha + \beta_{1} W_{ht} + \beta_{2} W^{2}_{ht} + u_{ht} \label{eq:quad} \\
Y_{ht} &= \alpha_{h} + \gamma_{t} +  \beta_{1} W_{ht} + \beta_{2} W^{2}_{ht} + u_{ht} \label{eq:quadFE} \\
\end{align}
\end{subequations}

\noindent All of the regression models are estimated for each permutation of the data (see Table~\ref{tab:sourcesetc}). This is a substantial number of regressions, given the number of variables defined (14 rainfall, eight temperature variables), the number of countries (six), the number of remote sensing products (six rainfall, three temperature), the number of extraction methods (ten), and the number of outcomes (two). This gives us a total of 51,840 different regressions: each of our four models and two outcomes on the 540 different versions of the data. By varying both specifications and data, we seek to define a robust set of outcomes by combining the multiple analysis approach of \cite{SimonsohnEtAl20} with the multiverse approach of \cite{SteegenEtAl16}.


\subsection{Inference}

In a ``typical'' economics paper, empirical results would be presented in a table, which would include coefficient estimates and some statistic for inference, such as standard errors, $p$-values, $t$-statistics, or confidence intervals. In our case, because of the large number of regressions that we estimate, standard modes of inference and traditional presentations of results are not appropriate. Instead, per our pre-analysis plan, we rely on a series of methods and criteria to make inference, evaluate the results, and present our findings.\footnote{As specified in our pre-analysis plan, we intended to examine the CDFs of coefficient estimates, following \cite{SaliMartin971, SaliMartin972}. However, using this approach in our context did not yield informative results. As such, we instead graph coefficients and confidence intervals ordered by the size of the coefficient estimate in specification charts. While not the same as the CDFs of coefficients in \cite{SaliMartin972, SaliMartin971}, the graphs communicate roughly the same information and are more appropriate for the variation in metrics, data products, anonymization methods, and so on, which are relevant for this analysis.}

As no formal statistical test exists to compare results across model, we develop three heuristics that allow us to describe similarities and differences in our results. Before describing these heuristics, it is useful to reflect on what sort of characteristics a heuristic would need to be useful for our purposes (i.e., comparing across tens of thousands of model-data combinations). First, some weather metrics that we test are likely to be positively correlated with outcomes (mean rainfall) while others are likely to be negatively correlated (days without rain). So, a heuristic should be agnostic about the sign of the coefficient. Second, our prior is that weather is significantly correlated with outcomes, regardless of direction. This maintained assumption is based on the frequency with which weather is used in the economics literature to predict all sorts of outcomes, from crop production to migration to economic growth. So, one would want a heuristic that is able to determine when a weather metric is significantly correlated with outcomes and when it is not. Finally, and in line with our prior, we expect weather to reduce the amount of unexplained variance in a model, all else being equal. So, one would want a heuristic that can measure the amount of unexplained variance in the model after controlling for weather.

With these three characteristics in mind, we adopt three general metrics to evaluate our results and two methods to test differences between these metrics. The three metrics are (1) mean log likelihood values, (2) share of coefficient $p$-values significant at standard levels ($0.01$, $0.05$, and $0.10$), and (3) coefficient size with $95\%$ confidence intervals. To compare our metrics across regressions, we apply two tests:

\begin{enumerate}
    \item Weak difference test: the value of a result (either mean log likelihood, share of significant $p$-values, or coefficients) from one regression lies outside the 95\% confidence interval on the value of a result from a competing regression. The confidence intervals \emph{can} overlap.
    \item Strong difference test: the $95\%$ confidence interval on the value of a result (either mean log likelihood, share of significant $p$-values, or coefficients) from one regression lies outside the $95\%$ confidence interval on the value of a result from a competing regression. The confidence intervals \emph{cannot} overlap.
\end{enumerate}

\noindent Our approach builds on the extreme bounds approach to assessing difference in estimates from \cite{LevineRenelt92} and the graphical methods to visualize these differences in \cite{SaliMartin972, SaliMartin971}.

While the three metrics are formal statistics, our weak and strong tests are not and we do not treat them that way. Rather, we use the combination of metrics and informal tests as heuristics in evaluating the loss of accuracy (mismeasurement) induced by anonymization methods used to protect participant privacy. All comparisons of one obfuscation/metric/source combination are made relative to the Household bilinear/metric/source combination. Our heuristics do not allow us to make claims regarding a formal definition of statistical accuracy, such as the expected squared-error loss in \cite{AbowdSchmutte19}. Rather, we quantify the significance and magnitude of measurement error by comparing results from one anonymization method with results from Household bilinear always bearing in mind that, for a given metric and country, if there was no measurement error induced by anonymization method, then the results from our tens of thousands of regressions would be exactly the same regardless of the obfuscation/source combination.

An important caveat to bear in mind with respect to our results, in particular all of the results focused on $p$-values, is that the significance of a point estimate does not imply that the model is correctly specified, that the point estimate is agronomically meaningful, or that the point estimate has the correct sign. These results and the associated figures simply allow us to visualize the variability in the number of significant coefficients across these specifications of interest. And any variability in results is a sign that obfuscation/source combinations provide different measures of weather and measurement error thus exists.


\section{Results} \label{sec:ext}

Due to the large number of regressions and estimated values produced in our analyses, we present results in a series of figures, which allow us to evaluate the significance, magnitude, and general trends in the effects of methods undertaken to preserve privacy on accuracy.

To examine the impact that different obfuscation procedures have on agricultural productivity, we pool the results from the 51,840 regressions and then divide the pool into ten bins, one for each anonymization method. In order to evaluate these outcomes, following the heuristics for inference discussed above, we then calculate descriptive statistics for each bin of results. These include the mean log likelihood value and the share of coefficients $(\beta_{1})$ with $p$-values of $p>0.90, p>0.95$ or $p>0.99$. For each of these values, we calculate the $95\%$ confidence interval on the mean. We then compare mean log likelihood values or the share of $p>0.95$s across all ten anonymization methods and use the $95\%$ confidence interval on the mean to evaluate differences using our weak and strong test criteria. Finally, we use specification charts to examine the actual regression coefficients and estimated confidence intervals for a subset of regressions.


\subsection{Log Likelihood}\label{sec:extR2}

We use specification charts to examine log likelihood values across the ten types of anonymization methods. Figure~\ref{fig:r2_ext} shows the mean log likelihood and the $95\%$ confidence interval on the mean by anonymization method. We further disaggregate results by model specification, as a model with fixed effects will have a different log likelihood value than a model without fixed effects. The top panels of Figure~\ref{fig:r2_ext} displays results from model specifications~\eqref{eq:linear} and~\eqref{eq:quad}, which are the linear and quadratic models without household or year fixed effects. The bottom panel displays results from model specifications~\eqref{eq:linearFE} and~\eqref{eq:quadFE}, which include household and year fixed effects. Within each specification chart, at the top of each ``column'' is the mean log likelihood and the $95\%$ confidence interval on the mean for the set of 1,296 regressions run. Below, markers on the chart indicate the anonymization method associated with the statistics.

Considering first the specification charts in the top panel which include only weather as an explanatory variable. Mean log likelihood values are not different across anonymization method within model specifications~\eqref{eq:linear}. The mean log likelihood value for any one anonymization method fails to pass even our weak difference test when compared to any of the other anonymization methods. Similarly, when comparing across anonymization methods within model specification~\eqref{eq:quad}, no mean log likelihood is weakly different from any other.

We conduct the same exercise for results presented in the bottom panels from model specification that include fixed effects. As with the top panel, the mean log likelihood value for any one anonymization method is not even weakly different from any other method. Our heuristic fails to identify significant differences within any model specification. Based on this, we conclude that remote sensing weather data from any one anonymization method does not explain a substantially larger amount of the variance in our outcome variables relative to any other anonymization method.

Despite the failure to identify differences in anonymization method, based on either the strong or weak criteria, the pattern of which anonymization methods result in the largest log likelihood values is remarkably consistent. Bilinear extraction methods for Household coordinates, EA centerpoint, and modified EA centerpoint always make up three of the top four models. Recall that the bilinear method computes the distance weighted average of values at the four nearest cell centers. Thus, unlike the simple extraction method, the bilinear method accounts for the point location within the arbitrary cell boundaries of the gridded data product. This approach seems to produce slightly better results than the simple extraction method or the zonal means for small areas, that is households or EAs. Administrative area appears to be too large of an area to produce strong results, as using Administrative area, regardless of extraction method (simple, bilinear, or zonal mean), tends to produce the smallest log likelihood values. While the pattern is consistent, it is important to recall that differences between each spatial anonymization method is not substantial enough to pass even our weak test, and we fail to identify significant differences across methods.


\subsection{\texorpdfstring{$p$}{p}-values}\label{sec:extpval}

We next consider if different anonymization methods produce substantially different counts of significant coefficients. Although while examining log likelihood values we disaggregated each bin of regression results by model specification, when examining $p$-values we disaggregate by whether the remote sensing data is rainfall or temperature. Figure~\ref{fig:pval_ext} presents the share of significant coefficient estimates for three standard $p$-values: for $p>0.90, p>0.95$ or $p>0.99$. To these bars we add the $95\%$ confidence interval on the mean number of significant coefficients. The top panel presents results from precipitation products while the bottom panel presents results from temperature products. Each bar and confidence interval in the rainfall panel is based on 4,032 regressions while each bar and confidence interval in the temperature panel is based on 1,152 regressions. To facilitate comparison, we draw red lines to designate the top and bottom of the confidence interval on the mean for the Household bilinear method, which are the actual Household coordinates.

A quick visual inspection of the results in the top panel of Figure~\ref{fig:pval_ext} does not reveal many, if any, differences across anonymization method. Comparing numerical values for the share of significant coefficients from Household bilinear to the $95\%$ confidence interval on the mean of any other extraction reveals that there are no comparisons that are strongly different from each other. There is only one weak difference, that of Administrative zonal mean, which produces slightly more significant $p$-values than those produced by data matched to the true Household coordinates. Similarly, the results in the lower panel on temperature look fairly uniform across anonymization methods. No pairwise comparisons are strongly different or weakly different. As with our examination of log likelihood values, the preponderance of evidence here implies that different anonymization methods used to protect privacy do not introduce substantial mismeasurement into the analysis. 

However, there is a possibility of heterogeneity across or within countries. As such, we next consider this same metric, disaggregated by country. Figures~\ref{fig:pval_ext_rf} and~\ref{fig:pval_ext_tp} present different anonymization methods across all rainfall and temperature metrics, for each of the six countries. Now that we have divided the results by anonymization method, rainfall/temperature, and country, each bar represents the share of significant coefficients from 672 regressions for rainfall and 192 for temperature. We simplify the graph by only presenting the share of coefficients with $p > 0.95$.

We see some variation within countries based on anonymization method. While no anonymization method is strongly different from Household bilinear, in Ethiopia, Niger, Nigeria, and Uganda, there are some methods that are weakly different. In all cases, these differences are for Administrative anonymization methods. In Ethiopia, Administrative simple and Administrative bilinear are weakly different from Household bilinear. In Niger, both Administrative bilinear and Administrative zonal mean are weakly different from Household bilinear while in Nigeria, Administrative zonal mean is weakly different from Household bilinear. In Uganda, Administrative simple is weakly different from Household bilinear. There are no significant differences in Malawi or Tanzania. That all significant differences are associated with Administrative area suggests that this approach to privacy protect does come at the cost of some data accuracy, though again the differences are only weak and are not present in all countries. 

Considering temperature, the evidence for differences in anonymization method is noisy (larger confidence intervals) relative to rainfall. As a result, there is no apparent pattern of one anonymization method differing from Household bilinear. One exception to this is the case of Ethiopia, in which there are weak differences between Household bilinear and Household simple, EA simple, modified EA simple, EA zonal mean, and Administrative zonal mean. But, no other countries show any differences, weak or strong, between any of the pairwise comparisons. 

Besides some small variation in anonymization method across countries, there are patterns to the variation across countries with respect to the share of significant $p$-values. Tanzania and Uganda produce substantially fewer significant estimates on rainfall relative to the other four countries. These differences pass our strong difference heuristic. For temperature, Ethiopia and Tanzania produce strongly different results compared to Malawi, Niger, and Nigeria, while Uganda produces weakly different from Malawi and Niger and strong different results from Nigeria. But, this does not speak to mismeasurement due to differences in anonymization method, the pattern is interesting to note for the discussion of cross-country differences in weather's relationship to agricultural productivity.\footnote{\cite{MichlerEtAl21WB} explores in more detail these relationships and their implication for integrating remote sensing weather data with household survey data.}

Taken together, the preponderance of evidence from all of our 51,840 regressions regarding our heuristics lead us to conclude that, generally, there is no clear evidence that different SDL methods implemented to preserve privacy of farms or households have substantially different impacts on estimates of agricultural productivity. One exception to this is that Administrative measurements produce some differences, though relatively small discrepancies, in the share of significant $p$-values. As in the descriptive statistics, we find evidence that while anonymization methods that rely on Administrative area provide the greatest degree of privacy protection they result in losses in accuracy for measurement of precipitation experienced by the household and correspondingly mismeasure the relationship between weather and agricultural productivity. Outside of the use of Administrative area, however, our findings suggest that any measurement error which may arise from the use of different anonymization methods does not substantially affect estimates. While household, EA, and modified EA bilinear appear to provide slightly better results than the other anonymization methods, when researchers use publicly available data with obfuscated GPS information, they should feel confident that matching those coordinates with remote sensing data will not introduce substantial measurement error into the analysis.


\subsection{Coefficients} \label{sec:extcoef}

In order to be able to examine individual regression coefficients, we first must narrow our focus to a subset of the 51,840 results. To do this, we consider four weather metrics: mean daily rainfall, number of days without rain, mean seasonal temperature, and growing degree days.\footnote{Results and conclusions do not change in a meaningful way if we use any of the other 18 weather metrics instead of these four. These four were chosen to provide evidence from different ways to measure precipitation (volume v. count) and temperature (actual temperature v. bounded count). Complete results for all 22 weather metrics are available in our populated pre-analysis plan \citep{MichlerEtAl21WB}.} We also focus in the body of the paper on two models: weather only and weather with year and household fixed effects.\footnote{Results and conclusions do not change in a meaningful way if we instead use the quadratic specifications. Results for the quadratic specifications are in Online Appendix~\ref{sec:appextraction}.} Similar to the specification charts for log likelihood, labels identify characteristics of the results are presented at the bottom of the specification chart. Unlike the log likelihood charts, we now present coefficients and confidence intervals for single regressions - 120 results per rainfall metric per country and 60 results per temperature metric per country - and not means of aggregated results and confidence intervals on the mean. Thus we present specific coefficient estimates from 4,320 regressions. In the following discussion, the term significance defines a point estimate with $p>0.95$.

Figures~\ref{fig:line_cty1_rf} through~\ref{fig:line_cty7_rf} present specification charts for coefficients and confidence intervals on mean daily rainfall and the number of no rain days by country. A number of patterns are immediately obvious. Results vary systematically by country, model, remote sensing product, and dependant variable. What is not clear is how results vary by anonymization method. In many countries and in both models, markers indicating remote sensing product or dependent variable tend to cluster within a specification chart, suggesting a pattern to results. Consider, as an example, in Ethiopia rainfall tends to be more strongly correlated (measured by a large absolute value of coefficient size) with yield than with value of harvest. No pattern of clustering exists for anonymization method, regardless of country, model, remote sensing product, or weather metric. The markers for anonymization method appear as random noise in each specification chart, suggesting that relative to other sources of variation, anonymization method does not have a meaningful impact on coefficient size or significance.

While we fail to observe patterns in coefficients as a function of anonymization method, there are strong patterns based on country, remote sensing product, and dependent variable. Focusing on models with only the weather metric on the right hand side, results in Ethiopia and Malawi are quite consistent. Mean daily rainfall is either positively correlated with outcomes or it is not significant. Conversely, the number of days without rain is a either negatively correlated with outcomes or it is not significant. This pattern persists in Niger and Nigeria, though precipitation measured by MERRA-2 in Niger and ERA5 in Nigeria produces coefficients with opposite signs (negative for mean rain and positive for no rain days). In Tanzania and Uganda, there is little consistency across regressions, with about an equal number of regressions reporting positive and negative coefficients. In Tanzania, this appears to be driven by the choice of dependant variable (more rain reduces the value of harvest but increases yield) while in Uganda it appears to be driven by the choice of remote sensing product (for ARC2 and TAMSAT more rain is negatively correlated with outcomes).

The primary impact of including fixed effects in the regressions is to weaken the correlation between rainfall and outcomes. In Ethiopia, without fixed effects rainfall is always significantly correlated with outcomes but by including fixed effects rainfall is no longer significantly related to outcomes in a majority of regressions. Results are similar in Malawi, Niger, and Nigeria, suggesting that once time-invariant household unobservables are controlled for, rainfall matters little in agricultural productivity. Tanzania and Uganda again prove to be outliers. Where without fixed effects, rainfall could be both positively and negatively correlated with outcomes, by including fixed effects results in these countries become much more consistent. In Tanzania rainfall tends to be uncorrelated with value of harvest but is consistently significantly correlated with yield. In Uganda, the results are the opposite, with rainfall significantly correlated with value of harvest but not with yield.

Turning to temperature, results regarding the impact of anonymization method are qualitatively similar to rainfall. In Figures~\ref{fig:line_cty1_tp} through~\ref{fig:line_cty7_tp}, markers for anonymization method appear to be nearly random while markers for remote sensing weather product and dependent variable cluster depending on the country, model, and temperature metric. As with rainfall, variation from country, model, remote sensing product, or weather metric appears to be more of a factor in determining coefficient sign, size, and significance than anonymization method.

Digging further into the specification charts reveals intriguing patterns in terms of these other sources of variation. Focusing on models with only the weather metric on the right hand side, mean seasonal temperature is either negatively correlated with outcomes or not significant in Ethiopia, Malawi, Niger, Nigeria, and Uganda. Only in Tanzania do results vary, with higher temperatures reducing yields but increasing the total value of harvest. For GDD, the metric is either positively correlated with outcomes or not significant in Malawi, Niger, and Uganda. In Ethiopia, Nigeria, and Tanzania, an increase in GDD can be either positively or negatively correlated with outcomes, depending on the remote sensing weather product that the data comes from and the dependent variable used in the regression.

When household and year fixed effects are added to the regressions, most temperature variables are no longer correlated with outcomes. The impact of including fixed effects varies by country and by temperature metric. As an example, in Ethiopia, without fixed effect mean seasonal temperature is always negative or not significant but with fixed effects the correlation can be both positive (MERRA-2), negative (ERA5), or not significant (CPC). Conversely, GDD was both positively and negatively correlated with outcomes in Ethiopia without fixed effects. Including fixed effects changes the results so that coefficients are always positively correlated or not significant. Similarly confounding patterns exist in Niger, Nigeria, Tanzania, and Uganda. Variables that were always of the same sign without fixed effects (mean and GDD in Niger and Uganda, mean in Nigeria) can have opposite signs when fixed effects are included. Or, variables that had opposite signs without fixed effects (mean and GDD in Tanzania) have consistent signs or are not significant when fixed effects are included. Which coefficients change signs with the inclusion of fixed effects is a function of both the source of the weather data and the choice of dependent variable. Only in Malawi do coefficients on temperature variables maintain consistent signs with and without fixed effects.


\section{Towards a Set of Best Practices} \label{sec:best}

Having examined the results from 51,840 regressions on a panel survey database with 33,738 total household observations that span a decade and six countries in Eastern, Western, and Southern Africa with significant heterogeneity in agro-ecological conditions and rainfall patterns, it is useful to recapitulate the key takeaways towards the formulation of of best practices and the identification of areas for future research.

Based on descriptive evidence and our heuristics, we find only minor evidence that SDL methods undertaken to protect privacy in the LSMS-ISA has an impact on the accuracy of results. The vast majority of spatial anonymization methods have no meaningful impact on estimates of the relationship between weather and agricultural productivity when compared to estimates from data that integrates weather and survey data using the exact household coordinates. To the extent that weak differences exist, they are in estimates from data that uses Administrative area center or Administrative area to match household locations to the gridded weather data products. Locations derived from administrative area provides the most privacy protection by introducing the most uncertainty regarding the exact location of a sampled household. And this privacy protect comes at a small cost in terms of data accuracy, resulting in some mismeasurement of the relationship between weather and agricultural productivity.

Though the results are generally robust to SDL methods to protect privacy, they are not robust to the choice of remote sensing weather product or the choice of weather metric. The correlation between rainfall or temperature and agricultural productivity varies by country depending on if the weather data comes from ARC2, CPC, CHIRPS, ERA5, MERRA-2, or TAMSAT. The relationship also varies depending on how one chooses to measure rainfall (e.g., mean daily or number of days without rain) and temperature (e.g., mean seasonal or GDD). Finally, the relationship can vary depending on the choice of how to measure agricultural productivity (harvest value or yield). In extreme cases, the relationship between rainfall or temperature and agricultural productivity can have opposite signs depending on the source of the weather data, the metric to measure weather, and the metric to measure agricultural productivity. We were only able to touch briefly on these issues here, but our populated pre-analysis plan explores these questions extensively \citep{MichlerEtAl21WB}.

Remotely sensed weather data has become a common component of economic analysis \citep{DellEtAl14, DonaldsonStoreygard16}. Yet, there has been little recognition in the economics literature that the need for privacy protection in public use survey data can introduce mismeasurement when integrating this data with remote sensing data. The need to protect privacy while producing accurate analysis has long been discussed in the computer science literature but has only recently been taken up in the economics literature \citep{AbowdEtAl19, Abraham19, ChettyFriedman19, RugglesEtAl19}. Neither has there been a convergence on a set of best practices for dealing with measurement error in the remote sensing data itself. Few empirical papers today would fail to verify the robustness of the results to different specifications \citep{SimonsohnEtAl20} or different iterations of the data \cite{SteegenEtAl16}. Yet economics papers rarely, if ever, verify the robustness of results to the choice of remote sensing data source or weather metric.

In trying to formulate a set of best practices for researchers interested in the integration of public use survey data with publicly available remote sensing weather datasets we recommend the following:

\begin{enumerate}
    \item At this time, researchers need not be concerned about potential inaccuracies that may be introduced into their analysis by integrating spatially anonymized survey datasets with publicly available remote sensing weather products. The current spatial resolution of the latter geospatial data is not fine enough for common SDL methods, such as $k$-anonymity, to result in mismeasurement of weather events that are experienced by sampled households.
    \item Researchers must carefully choose which remote sensing source to use in their analysis. Despite the volume of precipitation and the temperature in a given location on a given day being objective facts, remote sensing products can differ substantially in how they measure these objective facts. Because of this, remote sensing products can and do disagree on what the weather was.
    \item Researchers may want to demonstrate the robustness of their results to the choice of weather data drawn from different remote sensing products, or different weather metrics. When weather is critical to the identification strategy, results should not be sensitive to the choice of remote sensing product or the weather metric.
\end{enumerate}

Despite the thematic focus of our paper on weather and agricultural productivity, future research should work towards building a robust body of knowledge regarding the impacts of using spatially anonymized survey data in a wide range of analytical and mapping applications. In specific cases, such as high-resolution crop area or crop yield mapping, it is clear that spatially anonymized public use datasets will not be useful since researchers need access to survey data with precise agricultural plot locations for integration with higher-resolution satellite imagery, such as Sentinel-2 \citep{AzzariEtAl21}. However, there is a high degree of thematic heterogeneity in research applications that rest on the integration of georeferenced socioeconomic survey datasets with geospatial data sources, and it is not always clear, ex-ante, to what extent, if any, spatial anonymization may lead to biased insights. A comprehensive body of evidence on the potential impacts of using spatially anonymized survey data will ultimately have implications for both survey data users and producers. While it can enable data users to better identify research questions whose answers may or may not be mediated by spatial anonymization of survey data, it can also provide further impetus for data producers to invest in physical and technological infrastructure to provide secure access to scientific use datasets that include confidential geolocation data that are not included in public use datasets but that may be needed to answer specific research questions.



\newpage
\singlespacing
\bibliographystyle{chicago}
\bibliography{LSMSref}

\begin{thebibliography}{}

\bibitem[\protect\citeauthoryear{Abay, Abate, Barrett, and Bernard}{Abay
  et~al.}{2019}]{AbayEtAl19}
Abay, K.~A., G.~T. Abate, C.~B. Barrett, and T.~Bernard (2019).
\newblock Correlated non-classical measurement errors, ‘second best’ policy
  inference, and the inverse size-productivity relationship in agriculture.
\newblock {\em Journal of Development Economics\/}~{\em 139}, 171--84.

\bibitem[\protect\citeauthoryear{Abowd and Schmutte}{Abowd and
  Schmutte}{2015}]{AbowdSchmutte15}
Abowd, J.~M. and I.~M. Schmutte (2015).
\newblock Economic analysis and statistical disclosure limitation.
\newblock {\em Brookings Papers on Economic Activity\/}~{\em 46}, 221--93.

\bibitem[\protect\citeauthoryear{Abowd and Schmutte}{Abowd and
  Schmutte}{2019}]{AbowdSchmutte19}
Abowd, J.~M. and I.~M. Schmutte (2019).
\newblock An economic analysis of privacy protection and statistical accuracy
  as social choice.
\newblock {\em American Economic Review\/}~{\em 109\/}(1), 171--202.

\bibitem[\protect\citeauthoryear{Abowd, Schmutte, Sexton, and Vilhuber}{Abowd
  et~al.}{2019}]{AbowdEtAl19}
Abowd, J.~M., I.~M. Schmutte, W.~N. Sexton, and L.~Vilhuber (2019).
\newblock Why the economics profession must actively participate in the privacy
  protection debate.
\newblock {\em AEA Paers and Proceedings\/}~{\em 109}, 397--402.

\bibitem[\protect\citeauthoryear{Abraham}{Abraham}{2019}]{Abraham19}
Abraham, K.~G. (2019).
\newblock Reconciling data access and privacy: Building a sustainable model for
  the future.
\newblock {\em AEA Paers and Proceedings\/}~{\em 109}, 409--13.

\bibitem[\protect\citeauthoryear{Arag\'{o}n, Oteiza, and Pablo~Rud}{Arag\'{o}n
  et~al.}{2021}]{AragonEtAl21}
Arag\'{o}n, F.~M., F.~Oteiza, and J.~Pablo~Rud (2021).
\newblock Climate change and agriculture: Subsistence farmers’ response to
  extreme heat.
\newblock {\em American Economic Journal: Economic Policy\/}~{\em 13\/}(1),
  1--35.

\bibitem[\protect\citeauthoryear{Azzari, Jain, Jeffries, Kilic, and
  Murray}{Azzari et~al.}{2021}]{AzzariEtAl21}
Azzari, G., S.~Jain, G.~Jeffries, T.~Kilic, and S.~Murray (2021).
\newblock Understanding the requirements for surveys to support satellite-based
  crop type mapping: Evidence from sub-saharan africa.
\newblock {\em Remote Sensing\/}~{\em 13\/}(23), 4749.

\bibitem[\protect\citeauthoryear{Barrios, Bertinelli, and Strobl}{Barrios
  et~al.}{2010}]{BarriosEtAl10}
Barrios, S., L.~Bertinelli, and E.~Strobl (2010).
\newblock Trends in rainfall and economic growth in {A}frica: A neglected cause
  of the {A}frican growth tragedy.
\newblock {\em Review of Economics and Statistics\/}~{\em 92\/}(2), 350--66.

\bibitem[\protect\citeauthoryear{Blankespoor, Croft, Dontamsetti, Mayala, and
  Murray}{Blankespoor et~al.}{2021}]{Murray21}
Blankespoor, B., T.~Croft, T.~Dontamsetti, B.~Mayala, and S.~Murray (2021).
\newblock Spatial anonymization: Guidance note prepared for the
  {I}nter-{S}ecretariat working group on household surveys.
\newblock UN Inter-secretariat Working Group on Household Surveys Task Force on
  Spatial Anonymization in Public-Use Household Survey Datasets.
  \href{https://unstats.un.org/iswghs/task-forces/documents/Spatial_Anonymization_Report_submit01272021_ISWGHS.pdf}{https://unstats.un.org/iswghs/task-forces/documents/Spatial\_Anonymization\_Report\_submit01272021\_ISWGHS.pdf}.

\bibitem[\protect\citeauthoryear{Bosilovich, Lucchesi, and Suarez}{Bosilovich
  et~al.}{2016}]{MERRA2}
Bosilovich, M., R.~Lucchesi, and M.~Suarez (2016).
\newblock {MERRA}-2: File specification.
\newblock GMAO Office Note No. 9 (Version 1.1).
  \url{http://gmao.gsfc.nasa.gov/pubs/office_notes}.

\bibitem[\protect\citeauthoryear{Br\"{u}ckner and Ciccone}{Br\"{u}ckner and
  Ciccone}{2011}]{BruckerCiccone11}
Br\"{u}ckner, M. and A.~Ciccone (2011).
\newblock Rain and the democratic window of opportunity.
\newblock {\em Econometrica\/}~{\em 79\/}(3), 923--47.

\bibitem[\protect\citeauthoryear{Burke, Driscoll, Lobell, and Ermon}{Burke
  et~al.}{2021}]{Burkeetal21}
Burke, M., A.~Driscoll, D.~B. Lobell, and S.~Ermon (2021).
\newblock Using satellite imagery to understand and promote sustainable
  development.
\newblock {\em Science\/}~{\em 371\/}(6536), eabe8628.

\bibitem[\protect\citeauthoryear{Burke and Lobell}{Burke and
  Lobell}{2017}]{BurkeandLobell17}
Burke, M. and D.~B. Lobell (2017).
\newblock Satellite-based assessment of yield variation and its determinants in
  smallholder {A}frican systems.
\newblock {\em Proceedings of the National Academy of Sciences\/}~{\em
  114\/}(9), 2189--94.

\bibitem[\protect\citeauthoryear{Carletto, Gourlay, Murray, and Zezza}{Carletto
  et~al.}{2017}]{Carlettoetal17}
Carletto, C., S.~Gourlay, S.~Murray, and A.~Zezza (2017).
\newblock Cheaper, faster, and more than good enough: Is {GPS} the new gold
  standard in land area measurement.
\newblock {\em Survey Research Methods\/}~{\em 11\/}(3), 235--65.

\bibitem[\protect\citeauthoryear{{Central Statistics Agency of Ethiopia
  (CSA)}}{{Central Statistics Agency of Ethiopia (CSA)}}{2014}]{ETH1}
{Central Statistics Agency of Ethiopia (CSA)} (2014).
\newblock {R}ural {S}ocioeconomic {S}urvey 2011-2012.
\newblock Public Use Dataset. Ref: ETH\_2011\_ERSS\_v01\_M. Downloaded from
  \url{https://microdata.worldbank.org/index.php/catalog/2053} on 6 September
  2019.

\bibitem[\protect\citeauthoryear{{Central Statistics Agency of Ethiopia
  (CSA)}}{{Central Statistics Agency of Ethiopia (CSA)}}{2015}]{ETH2}
{Central Statistics Agency of Ethiopia (CSA)} (2015).
\newblock {E}thiopia {S}ocioeconomic {S}urvey 2013-2014.
\newblock Public Use Dataset. Ref: ETH\_2013\_ESS\_v02\_M. Downloaded from
  \url{https://microdata.worldbank.org/index.php/catalog/2053} on 6 September
  2019.

\bibitem[\protect\citeauthoryear{{Central Statistics Agency of Ethiopia
  (CSA)}}{{Central Statistics Agency of Ethiopia (CSA)}}{2017}]{ETH3}
{Central Statistics Agency of Ethiopia (CSA)} (2017).
\newblock {E}thiopia {S}ocioeconomic {S}urvey, wave 3 ({ESS}3) 2015-2016.
\newblock Public Use Dataset. Ref: ETH\_2015\_ESS\_v02\_M. Downloaded from
  \url{https://microdata.worldbank.org/index.php/catalog/2783} on 6 September
  2019.

\bibitem[\protect\citeauthoryear{Chen, Mueller, Jia, and Tseng}{Chen
  et~al.}{2017}]{ChenEtAl17}
Chen, J.~J., V.~Mueller, Y.~Jia, and S.~K.-H. Tseng (2017).
\newblock Validating migration responses to flooding using satellite and vital
  registration data.
\newblock {\em American Economic Review\/}~{\em 107\/}(5), 441--45.

\bibitem[\protect\citeauthoryear{Chen, Shi, Xie, Silva, Kousky, Higgins, and
  Janowiak}{Chen et~al.}{2008}]{CPC}
Chen, M., W.~Shi, P.~Xie, V.~B. Silva, V.~E. Kousky, R.~W. Higgins, and J.~E.
  Janowiak (2008).
\newblock Assessing objective techniques for gauge‐based analyses of global
  daily precipitation.
\newblock {\em Journal of Geophysical Research\/}~{\em 113}, D04110.

\bibitem[\protect\citeauthoryear{Chetty and Friedman}{Chetty and
  Friedman}{2019}]{ChettyFriedman19}
Chetty, R. and J.~N. Friedman (2019).
\newblock A practical method to reduce privacy loss when disclosing statistics
  based on small samples.
\newblock {\em AEA Paers and Proceedings\/}~{\em 109}, 414--20.

\bibitem[\protect\citeauthoryear{Corno, Hildebrandt, and Voena}{Corno
  et~al.}{2020}]{CornoEtAl20}
Corno, L., N.~Hildebrandt, and A.~Voena (2020).
\newblock Age of marriage, weather shocks, and the direction of marriage
  payments.
\newblock {\em Econometrica\/}~{\em 88\/}(3), 879--915.

\bibitem[\protect\citeauthoryear{Dell, Jones, and Olken}{Dell
  et~al.}{2012}]{DellEtAl12}
Dell, M., B.~F. Jones, and B.~A. Olken (2012).
\newblock Temperature shocks and economic growth: Evidence from the last half
  century.
\newblock {\em American Economic Journal: Macroeconomics\/}~{\em 4\/}(3),
  66--95.

\bibitem[\protect\citeauthoryear{Dell, Jones, and Olken}{Dell
  et~al.}{2014}]{DellEtAl14}
Dell, M., B.~F. Jones, and B.~A. Olken (2014).
\newblock What do we learn from the weather? the new climate-economy
  literature.
\newblock {\em Journal of Economic Literature\/}~{\em 52\/}(3), 740--98.

\bibitem[\protect\citeauthoryear{Desch{\^e}ne and Greenstone}{Desch{\^e}ne and
  Greenstone}{2007}]{DescheneGreenstone07}
Desch{\^e}ne, O. and M.~Greenstone (2007).
\newblock The economic impacts of climate change: Evidence from agricultural
  output and random fluctuations in weather.
\newblock {\em American Economic Review\/}~{\em 97\/}(1), 354--85.

\bibitem[\protect\citeauthoryear{Dinur and Nissim}{Dinur and
  Nissim}{2003}]{DinurandNissim03}
Dinur, I. and K.~Nissim (2003).
\newblock Revealing information while preserving privacy.
\newblock {\em Proceedings of {ACM SIGMOD-SIGACT-SIGART} {S}myposium on
  {P}rinciples of {D}atabase {S}ystems\/}~{\em 22}, 202--10.

\bibitem[\protect\citeauthoryear{Donaldson and Storeygard}{Donaldson and
  Storeygard}{2016}]{DonaldsonStoreygard16}
Donaldson, D. and A.~Storeygard (2016).
\newblock The view from above: Applications of satellite data in economics.
\newblock {\em Journal of Economic Perspectives\/}~{\em 30\/}(4), 171--98.

\bibitem[\protect\citeauthoryear{Duflo, Banerjee, Finkelstein, Katz, Olken, and
  Sautman}{Duflo et~al.}{2020}]{Dufloetal20}
Duflo, E., A.~Banerjee, A.~Finkelstein, L.~F. Katz, B.~A. Olken, and A.~Sautman
  (2020).
\newblock In praise of moderation: Suggestions for the scope and use of
  pre-analysis plans for rcts in economics.
\newblock {NBER} Working Paper 26993.

\bibitem[\protect\citeauthoryear{Dwork, McSherry, Nissim, and Smith}{Dwork
  et~al.}{2006}]{DworkEtAl06}
Dwork, C., F.~McSherry, K.~Nissim, and A.~Smith (2006).
\newblock Calibrating noise to sensitivity in private data analysis.
\newblock In S.~Halevi and T.~Rabin (Eds.), {\em Theory of Cryptography}, pp.\
  265--84. Springer Berlin Heidelberg.

\bibitem[\protect\citeauthoryear{Dwyer-Lindgren, Cork, Sligar, Steuben, Wilson,
  Provost, Mayala, VanderHeide, Collison, Hall, Biehl, Carter, Frank,
  Douwes-Schultz, Burstien, Casey, Deshpande, Earl, Bcheraoui, .Farag, Henry,
  Kinyoki, Marczak, Nixon, Osgood-Zimmerman, Pigott, Jr., Ross, Schaeffer,
  Smith, Weaver, Wiens, Easton, Justman, Opio, Sartorius, Tanser, Wabiri, Piot,
  Murray, and Hay}{Dwyer-Lindgren et~al.}{2019}]{DwyerLindetal19}
Dwyer-Lindgren, L., M.~A. Cork, A.~Sligar, K.~M. Steuben, K.~F. Wilson, N.~R.
  Provost, B.~K. Mayala, J.~D. VanderHeide, M.~L. Collison, J.~B. Hall, M.~H.
  Biehl, A.~Carter, T.~Frank, D.~Douwes-Schultz, R.~Burstien, D.~C. Casey,
  A.~Deshpande, L.~Earl, C.~E. Bcheraoui, T.~H. .Farag, N.~J. Henry,
  D.~Kinyoki, L.~B. Marczak, M.~R. Nixon, A.~Osgood-Zimmerman, D.~Pigott,
  R.~C.~R. Jr., J.~M. Ross, L.~E. Schaeffer, D.~L. Smith, N.~D. Weaver, K.~E.
  Wiens, J.~W. Easton, J.~E. Justman, A.~Opio, B.~Sartorius, F.~Tanser,
  N.~Wabiri, P.~Piot, C.~J. Murray, and S.~I. Hay (2019).
\newblock Mapping {HIV} prevalence in sub-{S}aharan {A}frica between 2000 and
  2017.
\newblock {\em Nature\/}~{\em 570}, 189--93.

\bibitem[\protect\citeauthoryear{Funk, Peterson, Landsfeld, Pedreros, Verdin,
  Shukla, Husak, Rowland, Harrison, Hoell, and Michaelsen}{Funk
  et~al.}{2015}]{CHIRPS}
Funk, C., P.~Peterson, M.~Landsfeld, D.~Pedreros, J.~Verdin, S.~Shukla,
  G.~Husak, J.~Rowland, L.~Harrison, A.~Hoell, and J.~Michaelsen (2015).
\newblock The climate hazards infrared precipitation with stations—a new
  environmental record for monitoring extremes.
\newblock {\em Scientific Data\/}~{\em 2}, 150066.

\bibitem[\protect\citeauthoryear{Garg, Jagnani, and Taraz}{Garg
  et~al.}{2020}]{GargEtAl20}
Garg, T., M.~Jagnani, and V.~Taraz (2020).
\newblock Temperature and human capital in {I}ndia.
\newblock {\em Journal of the Association of Environmental and Resource
  Economists\/}~{\em 7\/}(6), 1113--50.

\bibitem[\protect\citeauthoryear{Gollin and Udry}{Gollin and
  Udry}{2021}]{GollinUdry21}
Gollin, D. and C.~Udry (2021).
\newblock Heterogeneity, measurement error, and misallocation: Evidence from
  {A}frican agriculture.
\newblock {\em Journal of Political Economy\/}~{\em 129\/}(1), 1--80.

\bibitem[\protect\citeauthoryear{Graetz, Friedman, Osgood-Zimmerman, Burstien,
  Biehl, Shields, Mosser, Casey, Deshpande, Earl, Reiner, Ray, Fullerman,
  Levine, Stubbs, Mayala, Longbottom, Browne, Bhatt, Weiss, Gething, Mokdad,
  Lim, Murray, Gakidou, and Hay}{Graetz et~al.}{2018}]{Graetzetal18}
Graetz, N., J.~Friedman, A.~Osgood-Zimmerman, R.~Burstien, M.~H. Biehl,
  C.~Shields, J.~F. Mosser, D.~C. Casey, A.~Deshpande, L.~Earl, R.~C. Reiner,
  S.~E. Ray, N.~Fullerman, A.~J. Levine, R.~W. Stubbs, B.~K. Mayala,
  J.~Longbottom, A.~J. Browne, S.~Bhatt, D.~J. Weiss, P.~W. Gething, A.~H.
  Mokdad, S.~S. Lim, C.~J. Murray, E.~Gakidou, and S.~I. Hay (2018).
\newblock Mapping local variation in education attainment across {A}frica.
\newblock {\em Nature\/}~{\em 555}, 48--53.

\bibitem[\protect\citeauthoryear{Hennermann and Berrisford}{Hennermann and
  Berrisford}{2020}]{ERA5}
Hennermann, K. and P.~Berrisford (2020).
\newblock {ERA}5: data documentation.
\newblock Last modified Nov 18, 2020.
  \url{https://confluence.ecmwf.int/display/CKB/ERA5%3A+data+documentation}.

\bibitem[\protect\citeauthoryear{Jagnani, Barrett, Liu, and You}{Jagnani
  et~al.}{2021}]{JagnaniEtAl21}
Jagnani, M., C.~B. Barrett, Y.~Liu, and L.~You (2021).
\newblock Within-season producer response to warmer temperatures: Defensive
  investments by kenyan farmers.
\newblock {\em Economics Journal\/}~{\em 131\/}(633), 392--419.

\bibitem[\protect\citeauthoryear{Janzen and Michler}{Janzen and
  Michler}{2021}]{JanzenMichler21}
Janzen, S.~A. and J.~D. Michler (2021).
\newblock Ulysses' pact or {U}lysses' raft: Using pre-analysis plans in
  experimental and nonexperimental research.
\newblock \emph{Applied Economic Perspectives and Policy}, forthcoming.

\bibitem[\protect\citeauthoryear{Jayachandran}{Jayachandran}{2006}]{Jayachandran06}
Jayachandran, S. (2006).
\newblock Selling labor low: Wage responses to productivity shocks in
  developing countries.
\newblock {\em Journal of Political Economy\/}~{\em 114\/}(3), 538--75.

\bibitem[\protect\citeauthoryear{Jolliffe, Mahler, Veerappan, Kilic, and
  Wollburg}{Jolliffe et~al.}{2021}]{Jolliffeetal21}
Jolliffe, D., D.~G. Mahler, M.~Veerappan, T.~Kilic, and P.~Wollburg (2021).
\newblock Under what conditions are data valuable for development?
\newblock World Bank Policy Research Working Paper, No. 9811.

\bibitem[\protect\citeauthoryear{Kaur}{Kaur}{2019}]{Kaur19}
Kaur, S. (2019).
\newblock Nominal wage rigidity in village labor markets.
\newblock {\em American Economic Review\/}~{\em 109\/}(10), 3585--616.

\bibitem[\protect\citeauthoryear{Kilic, Moylan, Ilukor, Mtengula, and
  Pangapanga-Phiri}{Kilic et~al.}{2021}]{KilicEtAl21}
Kilic, T., H.~Moylan, J.~Ilukor, C.~Mtengula, and I.~Pangapanga-Phiri (2021).
\newblock Root for the tubers: Extended-harvest crop production and
  productivity measurement in surveys.
\newblock {\em Food Policy\/}~{\em 102}, 102033.

\bibitem[\protect\citeauthoryear{K\"{o}nig, Rohner, Thoenig, and
  Zilibotti}{K\"{o}nig et~al.}{2017}]{KonigEtAl18}
K\"{o}nig, M.~D., D.~Rohner, M.~Thoenig, and F.~Zilibotti (2017).
\newblock Networks in conflict: Theory and evidence from the {G}reat {W}ar of
  {A}frica.
\newblock {\em Econometrica\/}~{\em 85\/}(4), 1093--1132.

\bibitem[\protect\citeauthoryear{Kosmowski, Aragaw, Kilian, Ambel, Ilukor,
  Yigezu, and Stevenson}{Kosmowski et~al.}{2019}]{KosmowskiEtAl19}
Kosmowski, F., A.~Aragaw, A.~Kilian, A.~Ambel, J.~Ilukor, B.~Yigezu, and
  J.~Stevenson (2019).
\newblock Varietal identification in household surveys: Results from three
  household-based methods against the benchmark of {DNA} fingerprinting in
  southern {E}thiopia.
\newblock {\em Experimental Agriculture\/}~{\em 55\/}(3), 371--85.

\bibitem[\protect\citeauthoryear{Levine and Renelt}{Levine and
  Renelt}{1992}]{LevineRenelt92}
Levine, R. and D.~Renelt (1992).
\newblock A sensitivity analysis of cross-country growth regressions.
\newblock {\em American Economic Review\/}~{\em 82\/}(4), 942--63.

\bibitem[\protect\citeauthoryear{Maccini and Yang}{Maccini and
  Yang}{2009}]{MacciniYang09}
Maccini, S. and D.~Yang (2009).
\newblock Under the weather: Health, schooling, and economic consequences of
  early-life rainfall.
\newblock {\em American Economic Review\/}~{\em 99\/}(3), 1006--26.

\bibitem[\protect\citeauthoryear{McCarthy, Kilic, Brubaker, and
  Murray}{McCarthy et~al.}{2021}]{McCarthyEtAl21a}
McCarthy, N., T.~Kilic, J.~Brubaker, and S.~Murray (2021).
\newblock Droughts and floods in {M}alawi: Impacts on crop production and the
  performance of sustainable land management practices under climate extremes.
\newblock {\em Environment and Development Economics\/}~{\em 26}, 432–49.

\bibitem[\protect\citeauthoryear{Michler, Baylis, Arends-Kuenning, and
  Mazvimavi}{Michler et~al.}{2019}]{MichlerEtAl19}
Michler, J.~D., K.~Baylis, M.~Arends-Kuenning, and K.~Mazvimavi (2019).
\newblock Conservation agriculture and climate resilience.
\newblock {\em Journal of Environmental Economics and Management\/}~{\em 93},
  148--69.

\bibitem[\protect\citeauthoryear{Michler, Josephson, Kilic, and Murray}{Michler
  et~al.}{2019}]{PAP}
Michler, J.~D., A.~Josephson, T.~Kilic, and S.~Murray (2019).
\newblock Empirically estimating the impact of weather on agriculture.
\newblock OSF Registries. July 1.
  \href{https://doi.org/10.17605/OSF.IO/Z3SNH}{https://doi.org/10.17605/osf.io/z3snh}.

\bibitem[\protect\citeauthoryear{Michler, Josephson, Kilic, and Murray}{Michler
  et~al.}{2021a}]{MichlerEtAl20}
Michler, J.~D., A.~Josephson, T.~Kilic, and S.~Murray (2021a).
\newblock Estimating the impact of weather on agriculture.
\newblock arXiv.org. August 19.
  \href{https://arxiv.org/abs/2012.11768v2}{https://arxiv.org/abs/2012.11768v2}.

\bibitem[\protect\citeauthoryear{Michler, Josephson, Kilic, and Murray}{Michler
  et~al.}{2021b}]{MichlerEtAl21WB}
Michler, J.~D., A.~Josephson, T.~Kilic, and S.~Murray (2021b).
\newblock Estimating the impact of weather on agriculture.
\newblock World Bank Policy Research Working Paper, No. 9867.

\bibitem[\protect\citeauthoryear{Miguel, Satyanath, and Sergenti}{Miguel
  et~al.}{2004}]{MiguelEtAl04}
Miguel, E., S.~Satyanath, and E.~Sergenti (2004).
\newblock Economic shocks and civil conflict: An instrumental variables
  approach.
\newblock {\em Journal of Political Economy\/}~{\em 112\/}(4), 725--53.

\bibitem[\protect\citeauthoryear{Morten}{Morten}{2019}]{Morten19}
Morten, M. (2019).
\newblock Temporary migration and endogenous risk sharing in village {I}ndia.
\newblock {\em Journal of Political Economy\/}~{\em 127\/}(1), 1--46.

\bibitem[\protect\citeauthoryear{{National Bureau of Statistics
  (NBS)}}{{National Bureau of Statistics (NBS)}}{2012}]{NGA1}
{National Bureau of Statistics (NBS)} (2012).
\newblock {N}igeria {G}eneral {H}ousehold {S}urvey ({GHS}), panel 2010, wave 1.
\newblock Public Use Dataset. Ref: NGA\_2010\_GHSP-W1\_v04\_M. Downloaded from
  \url{https://microdata.worldbank.org/index.php/catalog/1002} on 6 September
  2019.

\bibitem[\protect\citeauthoryear{{National Bureau of Statistics
  (NBS)}}{{National Bureau of Statistics (NBS)}}{2014}]{NGA2}
{National Bureau of Statistics (NBS)} (2014).
\newblock {N}igeria {G}eneral {H}ousehold {S}urvey, panel 2012-2013, wave 2.
\newblock Public Use Dataset. Ref: NGA\_2012\_GHSP-W2\_v02\_M. Downloaded from
  \url{https://microdata.worldbank.org/index.php/catalog/1952} on 6 September
  2019.

\bibitem[\protect\citeauthoryear{{National Bureau of Statistics
  (NBS)}}{{National Bureau of Statistics (NBS)}}{2019}]{NGA3}
{National Bureau of Statistics (NBS)} (2019).
\newblock {N}igeria {G}eneral {H}ousehold {S}urvey, panel (ghs-panel)
  2015-2016.
\newblock Public Use Dataset. Ref: NGA\_2015\_GHSP-W3\_v02\_M. Downloaded from
  \url{https://microdata.worldbank.org/index.php/catalog/2734} on 6 September
  2019.

\bibitem[\protect\citeauthoryear{{National Statistical Office (NSO)}}{{National
  Statistical Office (NSO)}}{2012}]{MWI1}
{National Statistical Office (NSO)} (2012).
\newblock {T}hird {M}alawi {I}ntegrated {H}ousehold {S}urvey 2010-2011.
\newblock Public Use Dataset. Ref: MWI\_2010\_IHS-III\_v01\_M. Downloaded from
  \url{https://microdata.worldbank.org/index.php/catalog/1003} on 6 September
  2019.

\bibitem[\protect\citeauthoryear{{National Statistical Office (NSO)}}{{National
  Statistical Office (NSO)}}{2015}]{MWI2}
{National Statistical Office (NSO)} (2015).
\newblock {M}alawi {I}ntegrated {H}ousehold {P}anel {S}urvey 2010-2013
  (short-term panel, 204 {EA}s).
\newblock Public Use Dataset. Ref: MWI\_2010-2013\_IHPS\_v01\_M. Downloaded
  from \url{https://microdata.worldbank.org/index.php/catalog/2248} on 6
  September 2019.

\bibitem[\protect\citeauthoryear{{National Statistical Office (NSO)}}{{National
  Statistical Office (NSO)}}{2017}]{MWI3}
{National Statistical Office (NSO)} (2017).
\newblock {M}alawi {I}ntegrated {H}ousehold {P}anel {S}urvey 2010-2013-2016
  (long-term panel, 102 eas).
\newblock Public Use Dataset. Ref: MWI\_2010-2016\_IHPS\_v02\_M. Downloaded
  from \url{https://microdata.worldbank.org/index.php/catalog/2939} on 6
  September 2019.

\bibitem[\protect\citeauthoryear{Novella and Thiaw}{Novella and
  Thiaw}{2013}]{ARC2}
Novella, N.~S. and W.~M. Thiaw (2013).
\newblock African rainfall climatology version 2 for famine early warning
  systems.
\newblock {\em Journal of Applied Meteorology and Climatology\/}~{\em 52\/}(3),
  588--606.

\bibitem[\protect\citeauthoryear{Osgood-Zimmerman, Millear, Stubbs, Shields,
  Pickering, Earl, Garaetz, Kinyoki, Ray, Bhatt, Browne, Burnstien, Cameron,
  .Casey, Despande, Fullman, Gething, Gibson, Henry, Herrero, Krause,
  Letourneau, Levine, Lui, Longbottom, Mayala, Mosser, Noor, Pigott, Piwoz,
  Rao, Rawat, Reiner, Smith, Weiss, Wiens, Mokdad, Lim, Murray, Kassebaum, and
  Hay}{Osgood-Zimmerman et~al.}{2018}]{OsgoodZimmermanetal18}
Osgood-Zimmerman, A., A.~I. Millear, R.~W. Stubbs, C.~Shields, B.~V. Pickering,
  L.~Earl, N.~Garaetz, D.~K. Kinyoki, S.~E. Ray, S.~Bhatt, A.~J. Browne,
  R.~Burnstien, E.~Cameron, D.~C. .Casey, A.~Despande, N.~Fullman, P.~W.
  Gething, H.~S. Gibson, N.~J. Henry, M.~Herrero, L.~K. Krause, I.~D.
  Letourneau, A.~J. Levine, P.~Y. Lui, J.~Longbottom, B.~K. Mayala, J.~F.
  Mosser, A.~M. Noor, D.~M. Pigott, E.~G. Piwoz, P.~Rao, R.~Rawat, R.~C.
  Reiner, D.~L. Smith, D.~J. Weiss, K.~E. Wiens, A.~H. Mokdad, S.~S. Lim, C.~J.
  Murray, N.~J. Kassebaum, and S.~I. Hay (2018).
\newblock Mapping child growth failure in {A}frica between 2000 and 2015.
\newblock {\em Nature\/}~{\em 555}, 41--47.

\bibitem[\protect\citeauthoryear{Parker}{Parker}{2016}]{Parker16}
Parker, W.~S. (2016).
\newblock Reanalyses and observations: What’s the difference?
\newblock {\em Bulletin of the American Meteorological Society\/}~{\em
  97\/}(9), 1565--72.

\bibitem[\protect\citeauthoryear{Ritchie and NeSmith}{Ritchie and
  NeSmith}{1991}]{RS1991}
Ritchie, J. and D.~NeSmith (1991).
\newblock Temperature and crop development.
\newblock In R.~J. Hanks and J.~T. Ritchie (Eds.), {\em Modeling Plant and Soil
  Systems}, pp.\  5--29. American Society of Agronomy, Crop Science Society of
  America, Soil Science Society of America.

\bibitem[\protect\citeauthoryear{Ruggles, Fitch, Magnuson, and
  Schroeder}{Ruggles et~al.}{2019}]{RugglesEtAl19}
Ruggles, S., C.~Fitch, D.~Magnuson, and J.~Schroeder (2019).
\newblock Differential privacy and census data: Implications for social and
  economic research.
\newblock {\em AEA Paers and Proceedings\/}~{\em 109}, 403--8.

\bibitem[\protect\citeauthoryear{{Sala-i-Martin}}{{Sala-i-Martin}}{1997a}]{SaliMartin972}
{Sala-i-Martin}, X.~X. (1997a).
\newblock I just ran four million regressions.
\newblock {NBER} Working Paper 6252.

\bibitem[\protect\citeauthoryear{{Sala-i-Martin}}{{Sala-i-Martin}}{1997b}]{SaliMartin971}
{Sala-i-Martin}, X.~X. (1997b).
\newblock I just ran two million regressions.
\newblock {\em American Economic Review\/}~{\em 87\/}(2), 174--83.

\bibitem[\protect\citeauthoryear{Sarsons}{Sarsons}{2015}]{Sarsons15}
Sarsons, H. (2015).
\newblock Rainfall and conflict: A cautionary tale.
\newblock {\em Journal of Development Economics\/}~{\em 115}, 62--72.

\bibitem[\protect\citeauthoryear{Shah and Steinberg}{Shah and
  Steinberg}{2017}]{ShahSteinberg17}
Shah, M. and B.~M. Steinberg (2017).
\newblock Drought of opportunities: Contemporaneous and long-term impacts of
  rainfall shocks on human capital.
\newblock {\em Journal of Political Economy\/}~{\em 125\/}(2), 527--61.

\bibitem[\protect\citeauthoryear{Simonsohn, Simmons, and Nelson}{Simonsohn
  et~al.}{2020}]{SimonsohnEtAl20}
Simonsohn, U., J.~P. Simmons, and L.~D. Nelson (2020).
\newblock Specification curve analysis descriptive and inferential statistics
  for all plausible specifications.
\newblock {\em Nature Human Behaviour\/}~{\em 4}, 1208--14.

\bibitem[\protect\citeauthoryear{Steegen, Tuerlinckx, Gelman, and
  Vanpaemel}{Steegen et~al.}{2016}]{SteegenEtAl16}
Steegen, S., F.~Tuerlinckx, A.~Gelman, and W.~Vanpaemel (2016).
\newblock Increasing transparency through a multiverse analysis.
\newblock {\em Perspectives on Psychological Science\/}~{\em 11\/}(5), 702--12.

\bibitem[\protect\citeauthoryear{Suri}{Suri}{2011}]{Suri11}
Suri, T. (2011).
\newblock Selection and comparative advantage in technology adoption.
\newblock {\em Econometrica\/}~{\em 79\/}(1), 159--209.

\bibitem[\protect\citeauthoryear{{Survey and Census Division, National
  Institute of Statistics, Niger (NIS)}}{{Survey and Census Division, National
  Institute of Statistics, Niger (NIS)}}{2014}]{NGR1}
{Survey and Census Division, National Institute of Statistics, Niger (NIS)}
  (2014).
\newblock {N}ational {S}urvey on {H}ousehold {L}iving {C}onditions and
  {A}griculture 2011.
\newblock Public Use Dataset. Ref: NER\_2011\_ECVMA\_v01\_M. Downloaded from
  \url{https://microdata.worldbank.org/index.php/catalog/2050} on 6 September
  2019.

\bibitem[\protect\citeauthoryear{{Survey and Census Division, National
  Institute of Statistics, Niger (NIS)}}{{Survey and Census Division, National
  Institute of Statistics, Niger (NIS)}}{2016}]{NGR2}
{Survey and Census Division, National Institute of Statistics, Niger (NIS)}
  (2016).
\newblock {N}ational {S}urvey on {H}ousehold {L}iving {C}onditions and
  {A}griculture (ecvma) 2014.
\newblock Public Use Dataset. Ref: NER\_2014\_ECVMA\_v02\_M. Downloaded from
  \url{https://microdata.worldbank.org/index.php/catalog/2676} on 6 September
  2019.

\bibitem[\protect\citeauthoryear{{Tanzania National Bureau of Statistics
  (TNBS)}}{{Tanzania National Bureau of Statistics (TNBS)}}{2011}]{TZA1}
{Tanzania National Bureau of Statistics (TNBS)} (2011).
\newblock {N}ational {P}anel {S}urvey 2008-2009, wave 1.
\newblock Public Use Dataset. Ref: TZA\_2008\_NPS-R1\_v03\_M. Downloaded from
  \url{https://microdata.worldbank.org/index.php/catalog/76} on 6 September
  2019.

\bibitem[\protect\citeauthoryear{{Tanzania National Bureau of Statistics
  (TNBS)}}{{Tanzania National Bureau of Statistics (TNBS)}}{2012}]{TZA2}
{Tanzania National Bureau of Statistics (TNBS)} (2012).
\newblock {N}ational {P}anel {S}urvey 2010-2011, wave 2.
\newblock Public Use Dataset. Ref: TZA\_2010\_NPS-R2\_v03\_M. Downloaded from
  \url{https://microdata.worldbank.org/index.php/catalog/1050} on 6 September
  2019.

\bibitem[\protect\citeauthoryear{{Tanzania National Bureau of Statistics
  (TNBS)}}{{Tanzania National Bureau of Statistics (TNBS)}}{2015}]{TZA3}
{Tanzania National Bureau of Statistics (TNBS)} (2015).
\newblock {N}ational {P}anel {S}urvey 2012-2013, wave 3.
\newblock Public Use Dataset. Ref: TZA\_2012\_NPS-R3\_v01\_M. Downloaded from
  \url{https://microdata.worldbank.org/index.php/catalog/2252} on 6 September
  2019.

\bibitem[\protect\citeauthoryear{Taraz}{Taraz}{2018}]{Taraz18}
Taraz, V. (2018).
\newblock Can farmers adapt to higher temperatures? evidence from {I}ndia.
\newblock {\em World Development\/}~{\em 112}, 205--19.

\bibitem[\protect\citeauthoryear{Tarnavsky, Grimes, Maidment, Black, Allan,
  Stringer, Chadwick, and Kayitakire}{Tarnavsky et~al.}{2014}]{TAMSAT}
Tarnavsky, E., D.~Grimes, R.~Maidment, E.~Black, R.~P. Allan, M.~Stringer,
  R.~Chadwick, and F.~Kayitakire (2014).
\newblock Extension of the {TAMSAT} satellite-based rainfall monitoring over
  {A}frica and from 1983 to present.
\newblock {\em Journal of Applied Meteorology and Climatology\/}~{\em
  53\/}(12), 2805--22.

\bibitem[\protect\citeauthoryear{Tesfaye, Blalock, and Tirivayi}{Tesfaye
  et~al.}{2021}]{TesfayeEtAl21}
Tesfaye, W., G.~Blalock, and N.~Tirivayi (2021).
\newblock Climate-smart innovations and rural poverty in {E}thiopia: Exploring
  impacts and pathways.
\newblock {\em American Journal of Agricultural Economics\/}~{\em 103\/}(3),
  878--99.

\bibitem[\protect\citeauthoryear{{Uganda Bureau of Statistics (UBOS)}}{{Uganda
  Bureau of Statistics (UBOS)}}{2014a}]{UGA2}
{Uganda Bureau of Statistics (UBOS)} (2014a).
\newblock {U}ganda {N}ational {P}anel {S}urvey ({UNPS}) 2010-2011.
\newblock Public Use Dataset. Ref: UGA\_2010\_UNPS\_v01\_M. Downloaded from
  \url{https://microdata.worldbank.org/index.php/catalog/2166} on 6 September
  2019.

\bibitem[\protect\citeauthoryear{{Uganda Bureau of Statistics (UBOS)}}{{Uganda
  Bureau of Statistics (UBOS)}}{2014b}]{UGA3}
{Uganda Bureau of Statistics (UBOS)} (2014b).
\newblock {U}ganda {N}ational {P}anel {S}urvey ({UNPS}) 2010-2011.
\newblock Public Use Dataset. Ref: UGA\_2011\_UNPS\_v01\_M. Downloaded from
  \url{https://microdata.worldbank.org/index.php/catalog/2059} on 6 September
  2019.

\bibitem[\protect\citeauthoryear{{Uganda Bureau of Statistics (UBOS)}}{{Uganda
  Bureau of Statistics (UBOS)}}{2019}]{UGA1}
{Uganda Bureau of Statistics (UBOS)} (2019).
\newblock {N}ational {P}anel {S}urvey ({UNPS}) 2005-2009.
\newblock Public Use Dataset. Ref: UGA\_2005-2009\_UNPS\_v01\_M. Downloaded
  from \url{https://microdata.worldbank.org/index.php/catalog/1001} on 6
  September 2019.

\bibitem[\protect\citeauthoryear{Wineman, Mason, Ochieng, and Kirimi}{Wineman
  et~al.}{2017}]{WinemanEtAl17}
Wineman, A., N.~M. Mason, J.~Ochieng, and L.~Kirimi (2017).
\newblock Weather extremes and household welfare in rural {K}enya.
\newblock {\em Food Security\/}~{\em 9}, 281–300.

\bibitem[\protect\citeauthoryear{Wood, Altman, Bembenek, Bun, and
  Gaboardi}{Wood et~al.}{2018}]{WoodEtAl18}
Wood, A., M.~Altman, A.~Bembenek, M.~Bun, and M.~Gaboardi (2018).
\newblock Differential privacy: A primer for a non-technical audience.
\newblock {\em Vanderbilt Journal of Entertainment \& Technology Law\/}~{\em
  21\/}(1), 209--76.

\bibitem[\protect\citeauthoryear{Yeh, Perez, Driscoll, Azzari, Tang, Lobell,
  Ermon, and Burke}{Yeh et~al.}{2020}]{YehEtAl20}
Yeh, C., A.~Perez, A.~Driscoll, G.~Azzari, Z.~Tang, D.~Lobell, S.~Ermon, and
  M.~Burke (2020).
\newblock Using publicly available satellite imagery and deep learning to
  understand economic well-being in {A}frica.
\newblock {\em Nature Communications\/}~{\em 11}, 2583.

\bibitem[\protect\citeauthoryear{Zandler, Senftl, and Vanselow}{Zandler
  et~al.}{2020}]{ZandlerEtAl20}
Zandler, H., T.~Senftl, and K.~A. Vanselow (2020).
\newblock Reanalysis datasets outperform other gridded climate products in
  vegetation change analysis in peripheral conservation areas of {C}entral
  {A}sia.
\newblock {\em Scientific Reports\/}~{\em 10}, 22446.

\end{thebibliography}


\newpage 
\FloatBarrier


\begin{landscape}
\begin{table}[htbp]	\centering
    \caption{Spatial Feature Representation}  \label{tab:spatialanon}
	\scalebox{0.9}
	{ \setlength{\linewidth}{.1cm}\newcommand{\contents}
		{\begin{tabular}{lll p{0.4\linewidth}}
            \\[-1.8ex]\hline 
			\hline \\[-1.8ex]
            & Anonymization & Displacement & Spatial Disclosure \\
            & Method & (km) & Risk \\
            \midrule
            Household & none & 0.0 & Enables household location identification \\
            EA center & aggregation & 0.5 & High risk of community identification \\
            EA center modified & aggregation + perturbation & 2.0 & Moderate risk of community identification \\
            EA zone of uncertainty & aggregation + perturbation & N/A & Moderate risk of community identification \\
            Administrative area center & large area aggregation & 16.8 & No increase in risk if administrative unit is identified in microdata \\
            Administrative area & large area aggregation & N/A & No increase in risk if administrative unit is identified in microdata \\
			\\[-1.8ex]\hline 
			\hline \\[-1.8ex]
    		\multicolumn{4}{l}{\footnotesize \textit{Note}: Displacement calculated as mean displacement distance from household location for all households with GPS in baseline wave.}
    	\end{tabular}}
	\setbox0=\hbox{\contents}
    \setlength{\linewidth}{\wd0-2\tabcolsep-.25em}
    \contents}
\end{table}
\end{landscape}


\begin{landscape}
\begin{table}[htbp]	\centering
    \caption{Sources of Weather Data}  \label{tab:weather}
	\scalebox{0.9}
	{ \setlength{\linewidth}{.1cm}\newcommand{\contents}
		{\begin{tabular}{p{0.65\linewidth} lllll}
            \\[-1.8ex]\hline 
			\hline \\[-1.8ex]
            dataset & Length of record & Resolution & Time step & Data & Units \\
            \midrule
            \multicolumn{6}{l}{\textbf{Precipitation}} \\
            -Africa Rainfall Climatology version 2 (ARC2) & 1983-current & 0.1 deg & daily & total precip & mm \\
            -Climate Hazards group InfraRed Precipitation with Station data (CHIRPS) & 1981-current & 0.05 deg & daily & total precip & mm \\
            -CPC Global Unified Gauge-Based Analysis of Daily Precipitation & 1979-current & 0.5 deg & daily & total precip & mm \\
            -European Centre for Medium-Range Weather Forecasts (ECMWF) ERA5 & 1979-current & 0.28 deg & hourly & total precip & m \\
            -Modern-Era Retrospective analysis for Research and Applications, version 2 (MERRA-2) Surface Flux Diagnostics & 1980-current & 0.625x0.5 deg & hourly & rain rate & kg m$^2$ s$^1$ \\
            -Tropical Applications of Meteorology using SATellite data  and ground-based observations (TAMSAT) & 1983-current & 0.0375 deg & daily & total precip & mm \\
            \midrule
            \multicolumn{6}{l}{\textbf{Temperature}} \\
            -CPC Global Unified Gauge-Based Analysis of Daily Temperature & 1979-current & 0.5 deg & daily & min, max temp & C \\
            -European Centre for Medium-Range Weather Forecasts (ECMWF) ERA5 & 1979-current & 0.28 deg & hourly & mean temp & K \\
            -Modern-Era Retrospective analysis for Research and Applications, version 2 (MERRA-2) statD & 1980-current & 0.625x0.5 deg & daily & mean temp & K \\
			\\[-1.8ex]\hline 
			\hline \\[-1.8ex]
    		\multicolumn{6}{l}{\footnotesize \textit{Note}: The table summarizes the remote sensing sources and related details for precipitation and temperature data.}
    	\end{tabular}}
	\setbox0=\hbox{\contents}
    \setlength{\linewidth}{\wd0-2\tabcolsep-.25em}
    \contents}
\end{table}
\end{landscape}


\begin{table}[htbp]	\centering
    \caption{Sources of Household Data}  \label{tab:lsms}
	\scalebox{0.9}
	{ \setlength{\linewidth}{.1cm}\newcommand{\contents}
		{\begin{tabular}{llrrr}
            \\[-1.8ex]\hline 
			\hline \\[-1.8ex]
            Country & Survey Name & Years & Original $n$ & Final $n$  \\
            \midrule
            Ethiopia & Ethiopia Socioeconomic Survey (ERSS) & 2011/2012 & 3,969 & 1,689 \\
                        & & 2013/2014 & 5,262 & 2,865 \\
                        & & 2015/2016 & 4,954 & 2,718 \\
            Malawi & Integrated Household Panel Survey (IHPS)  & 2010/2011 & 3,246 & 1,241 \\
                        & & 2013 & 4,000 & 968 \\
                        & & 2016/2017 & 2,508 & 1,041 \\
            Niger & Enqu\^{e}te Nationale sur les Conditions de Vie des & 2011 & 3,968 & 2,223 \\
                        & $\:$ M\'{e}nages et l'Agriculture (ECVMA) & 2014 & 3,617 & 1,690 \\
            Nigeria & General Household Survey (GHS) & 2010/2011 & 5,000 & 2,833 \\
                        & & 2012/2013 & 4,802 & 2,768 \\
                        & & 2015/2016 & 4,613 & 2,783 \\
            Tanzania & Tanzania National Panel Survey (TZNPS) & 2008/2009 & 3,280 & 1,907 \\
                        & & 2010/2011 & 3,924 & 1,914 \\
                        & & 2012/2013 & 3,924 & 1,848 \\
            Uganda & Uganda National Panel Survey (UNPS) & 2009/2010 & 2,975 & 1,704 \\
                        & & 2010/2011 & 2,716 & 1,741 \\
                        & & 2011/2012 & 2,850 & 1,805 \\
            \midrule
            Total & 6 countries & 17 waves & 65,608 & 33,738 \\
			\\[-1.8ex]\hline 
			\hline \\[-1.8ex]
    		\multicolumn{5}{l}{\footnotesize \textit{Note}: The table summarizes the household data details for each country, per LSMS Basic Information Documents.}
    	\end{tabular}}
	\setbox0=\hbox{\contents}
    \setlength{\linewidth}{\wd0-2\tabcolsep-.25em}
    \contents}
\end{table}


\begin{table}[htbp]	
	\begin{center}\caption{Data Scope \label{tab:sourcesetc}}
		\resizebox{6in}{!}
		{\setlength{\linewidth}{.1cm}\newcommand{\contents}
			{\begin{tabular}{cc}
		        	\\[-1.8ex]\hline 
    		    	\hline \\[-1.8ex]
					\multicolumn{1}{l}{Countries (6)}& \multicolumn{1}{l}{Ethiopia, Malawi, Niger, Nigeria, Tanzania, Uganda}\\
					\midrule
					\multicolumn{1}{l}{Weather Products (9)}& \multicolumn{1}{l}{Precipitation}\\
					\multicolumn{1}{l}{}& \multicolumn{1}{l}{ \ \ \ \ \  \ ARC2, CHIRPS, CPC, ERA5, MERRA-2, TAMSAT}\\
					\multicolumn{1}{l}{}& \multicolumn{1}{l}{Temperature}\\
					\multicolumn{1}{l}{}& \multicolumn{1}{l}{ \ \ \ \ \  \ CPC, ERA5, MERRA-2}\\
					\midrule
					\multicolumn{1}{l}{Anonymization methods (10)}& \multicolumn{1}{l}{Points (simple)}\\
					\multicolumn{1}{l}{}& \multicolumn{1}{l}{ \ \ \ \ \  \ Household, EA center, EA center modified, Administrative area center}\\
					\multicolumn{1}{l}{}& \multicolumn{1}{l}{Points (bilinear)}\\
					\multicolumn{1}{l}{}& \multicolumn{1}{l}{ \ \ \ \ \  \ Household, EA center, EA center modified, Administrative area center}\\
					\multicolumn{1}{l}{}& \multicolumn{1}{l}{Area (zonal mean)}\\
					\multicolumn{1}{l}{}& \multicolumn{1}{l}{ \ \ \ \ \  \ EA zone of uncertainty, Administrative area}\\
					\midrule
					\multicolumn{1}{l}{Weather metrics (22)}& \multicolumn{1}{l}{14 rainfall}\\
					\multicolumn{1}{l}{}& \multicolumn{1}{l}{8 temperature}\\
					\midrule
					\multicolumn{1}{l}{Dependent variables (2)}& \multicolumn{1}{l}{value, quantity}\\
					\midrule
					\multicolumn{1}{l}{Specifications (4)}& \multicolumn{1}{l}{Linear}\\
					\multicolumn{1}{l}{}& \multicolumn{1}{l}{ \ \ \ \ \  \ without household \& year FEs, with household \& year FEs}\\
					\multicolumn{1}{l}{}& \multicolumn{1}{l}{Quadratic}\\
					\multicolumn{1}{l}{}& \multicolumn{1}{l}{ \ \ \ \ \  \ without household \& year FEs, with household \& year FEs}\\
		        	\\[-1.8ex]\hline 
    		    	\hline \\[-1.8ex]
    		    	\multicolumn{2}{p{\linewidth}}{\footnotesize  \textit{Note}: The table summarizes the scope of the data across country, weather product, anonymization method, weather metric, dependent variable, and econometric specification.} \\
			\end{tabular}}
			\setbox0=\hbox{\contents}
			\setlength{\linewidth}{\wd0-2\tabcolsep-.25em}
			\contents}
	\end{center}
\end{table}


\begin{figure}[!htbp]
	\begin{minipage}{\linewidth}
		\caption{Visualization of Anonymization Methods}
    	\label{fig:features}
		\begin{center}
			\includegraphics[width=.5\linewidth,keepaspectratio]{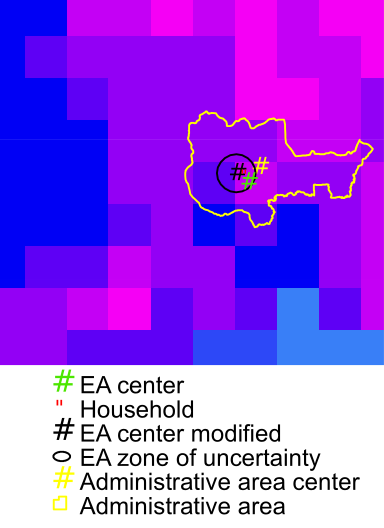}
		\end{center}
		\footnotesize  \textit{Note}: The figure presents the different anonymization methods (see Table~\ref{tab:sourcesetc}) and how the measurement of anonymization method would vary across a particular precipitation product (from Figure~\ref{fig:rain_res}).
	\end{minipage}
\end{figure}


\newpage 
\FloatBarrier 

\begin{figure}[!htbp]
	\begin{minipage}{\linewidth}
		
		\caption{Varying Resolution of Rainfall Measurement}
    	\label{fig:rain_res}
		\begin{center}
			\includegraphics[width=.9\linewidth,keepaspectratio]{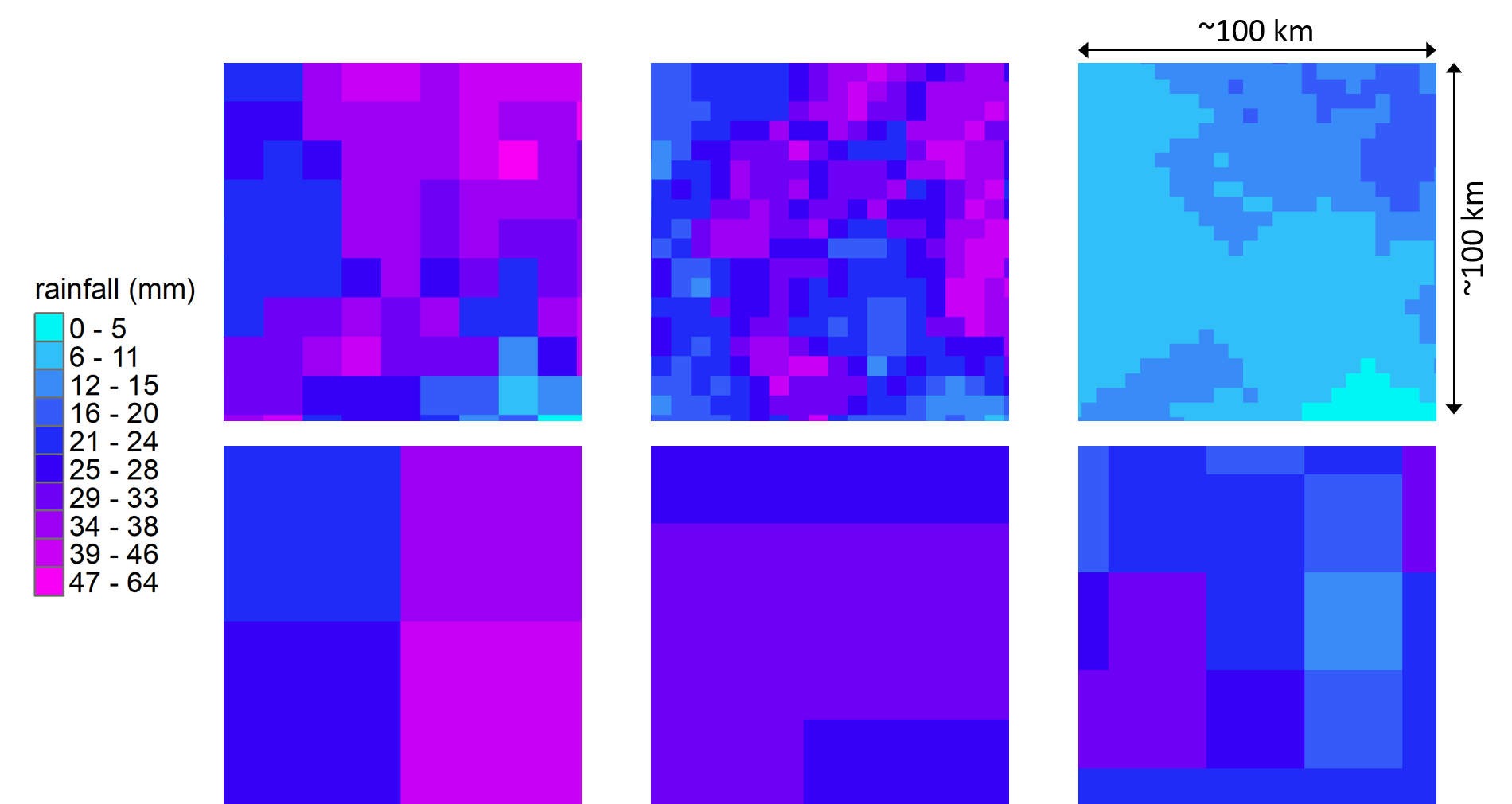}
		\end{center}
		\footnotesize  \textit{Note}: The figure captures rainfall as measured by all six precipitation products for the same 100km x 100km area on a single day (7 January 2010).

	    \vspace{2cm}
	
		\caption{Varying Resolution of Temperature Measurement}
    	\label{fig:temp_res}
		\begin{center}
			\includegraphics[width=.9\linewidth,keepaspectratio]{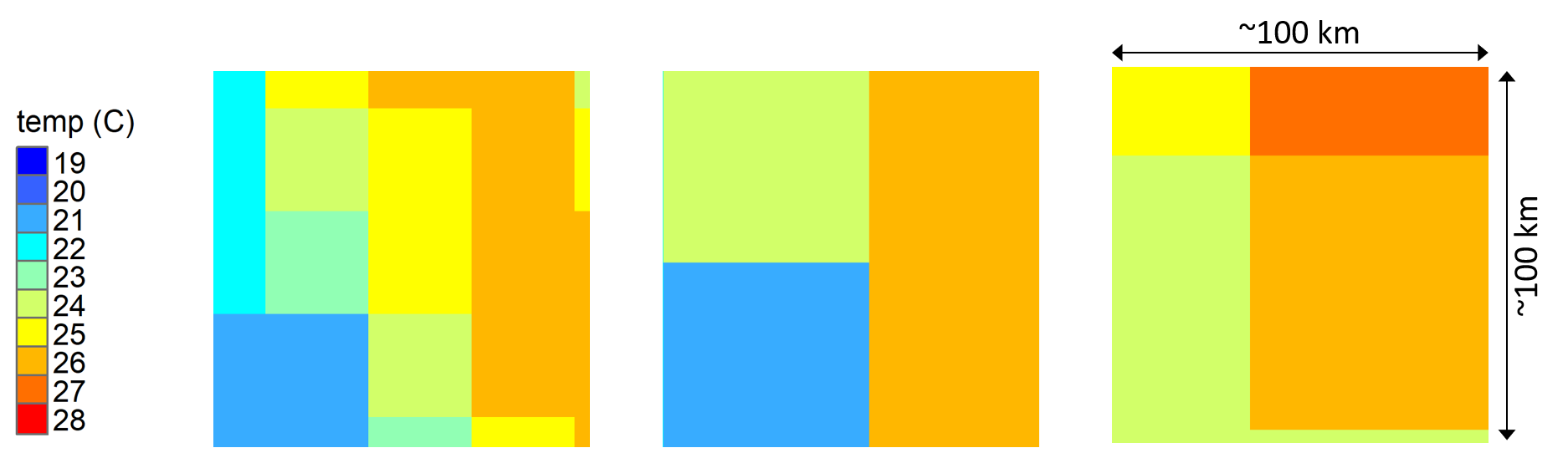}
		\end{center}
		\footnotesize  \textit{Note}: The figure captures temperature as measured by all three temperature products for the same 100km x 100km area on a single day (7 January 2010).
	\end{minipage}
\end{figure}


\newpage 

\begin{landscape}
\begin{figure}[!htbp]
	\begin{minipage}{\linewidth}	
		\caption{Distribution of Mean Daily Rainfall, by Anonymization Method and Remote Sensing Source}
    	\label{fig:density_aez_rf}
		\begin{center}
			\includegraphics[width=.88\linewidth,keepaspectratio]{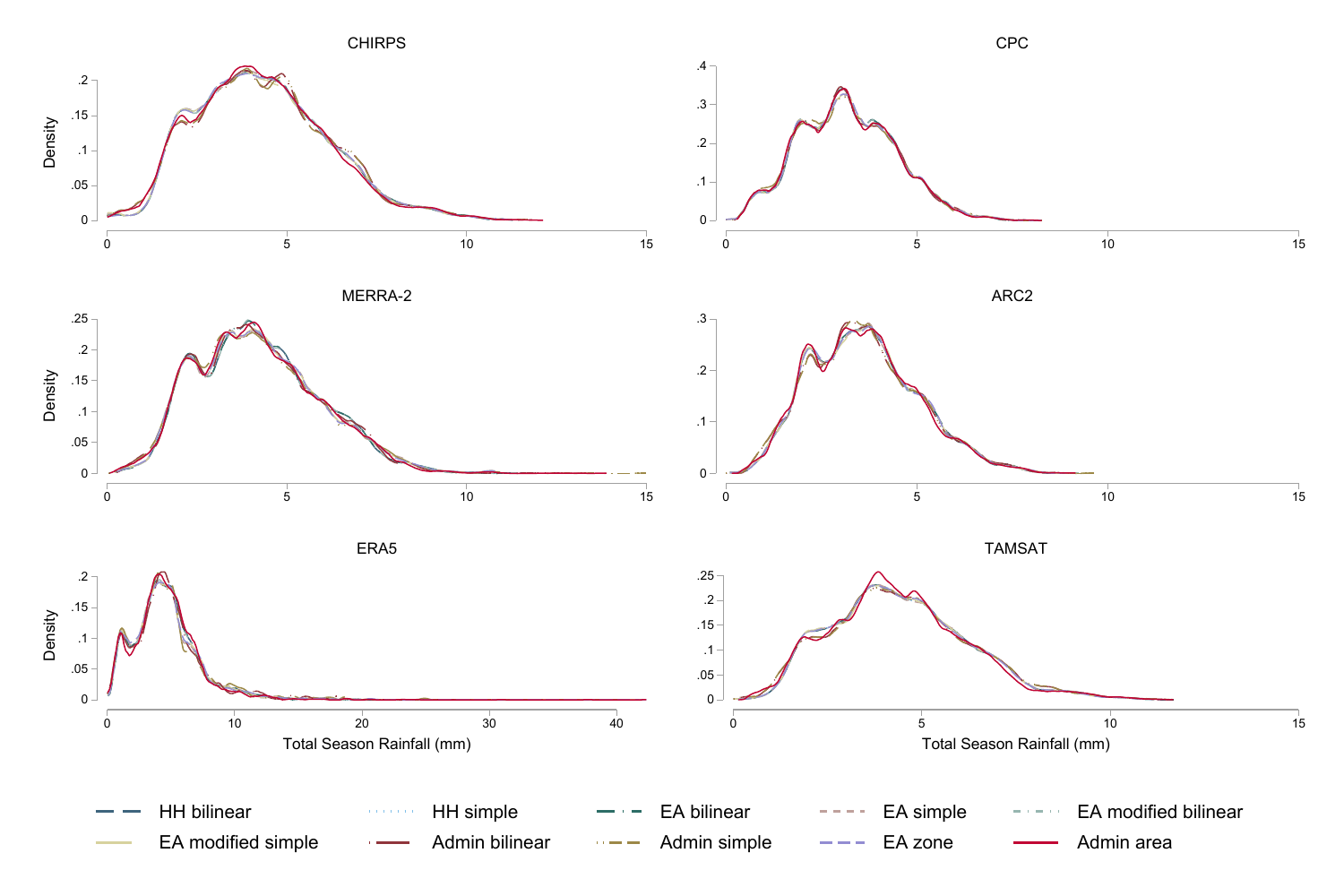}
		\end{center}
		\footnotesize  \textit{Note}: The figure presents rainfall distributions pooled across all countries and years, disaggregated by remote sensing source. Each line (anonymization method) in each panel is constructed using all $33,738$ household-year observations. Variation in lines do not come variation in the household data that is paired with the remote sensing data. Rather, variation in lines within a panel is solely due to differences in the grid cell in which the anonymization method locates the household. Variation in lines across panels is solely due to differences in the value of precipitation reported by the remote sensing source.
	\end{minipage}
\end{figure}
\end{landscape}	

\begin{landscape}
\begin{figure}[!htbp]
	\begin{minipage}{\linewidth}		
		\caption{Prediction of Mean Number of No Rain Days, by Anonymization Method and Remote Sensing Source}
		\label{fig:norain_aez_rf}
		\begin{center}
			\includegraphics[width=.88\linewidth,keepaspectratio]{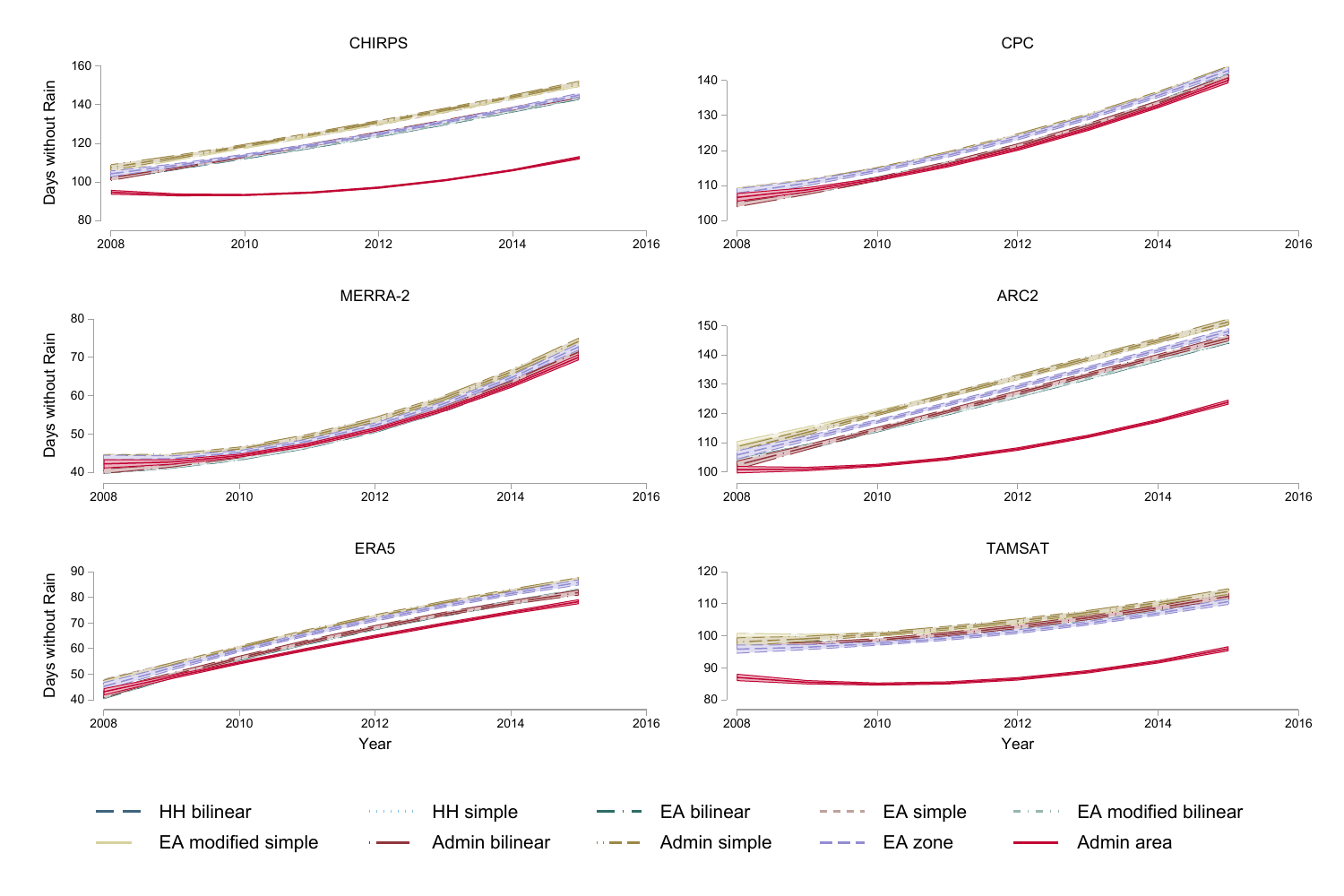}
		\end{center}
		\footnotesize  \textit{Note}: The figure presents the mean number of days without rain $(< 1mm)$ in a year, pooled across all countries, disaggregated by remote sensing source. Prediction made via Fractional-Polynomial, with $95\%$ confidence interval represented by shaded area. Each line (anonymization method) in each panel is constructed using all $33,738$ household-year observations. Variation in lines do not come variation in the household data that is paired with the remote sensing data. Rather, variation in lines within a panel is solely due to differences in the grid cell in which the anonymization method locates the household. Variation in lines across panels is solely due to differences in the number of days without rain reported by the remote sensing source.
	\end{minipage}
\end{figure}
\end{landscape}	

\begin{landscape}
\begin{figure}[!htbp]
	\begin{minipage}{\linewidth}
		\caption{Distribution of Mean Seasonal Temperature, by Anonymization Method and Remote Sensing Source}
    	\label{fig:density_aez_tp}
		\begin{center}
			\includegraphics[width=.88\linewidth,keepaspectratio]{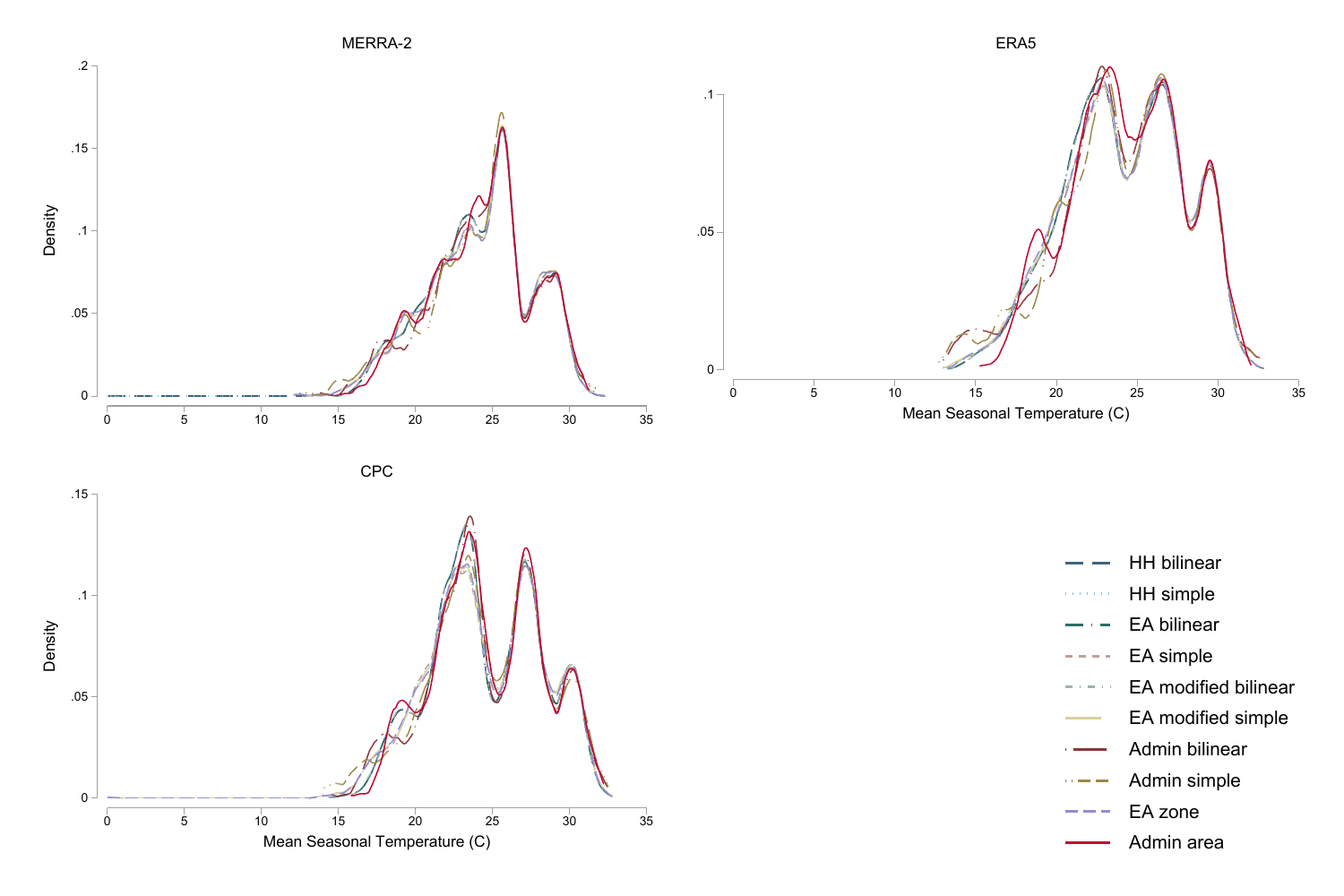}
		\end{center}
		\footnotesize  \textit{Note}: The figure presents temperature distributions pooled across all countries and years, disaggregated by remote sensing source. Each line (anonymization method) in each panel is constructed using all $33,738$ household-year observations. Variation in lines do not come variation in the household data that is paired with the remote sensing data. Rather, variation in lines within a panel is solely due to differences in the grid cell in which the anonymization method locates the household. Variation in lines across panels is solely due to differences in the value of temperature reported by the remote sensing source.
	\end{minipage}
\end{figure}
\end{landscape}	

\begin{landscape}
\begin{figure}[!htbp]
	\begin{minipage}{\linewidth}
		\caption{Prediction of Mean Number of Mean Growing Degree Days, by Anonymization Method and Remote Sensing Source}
		\label{fig:gdd_aez_tp}
		\begin{center}
			\includegraphics[width=.88\linewidth,keepaspectratio]{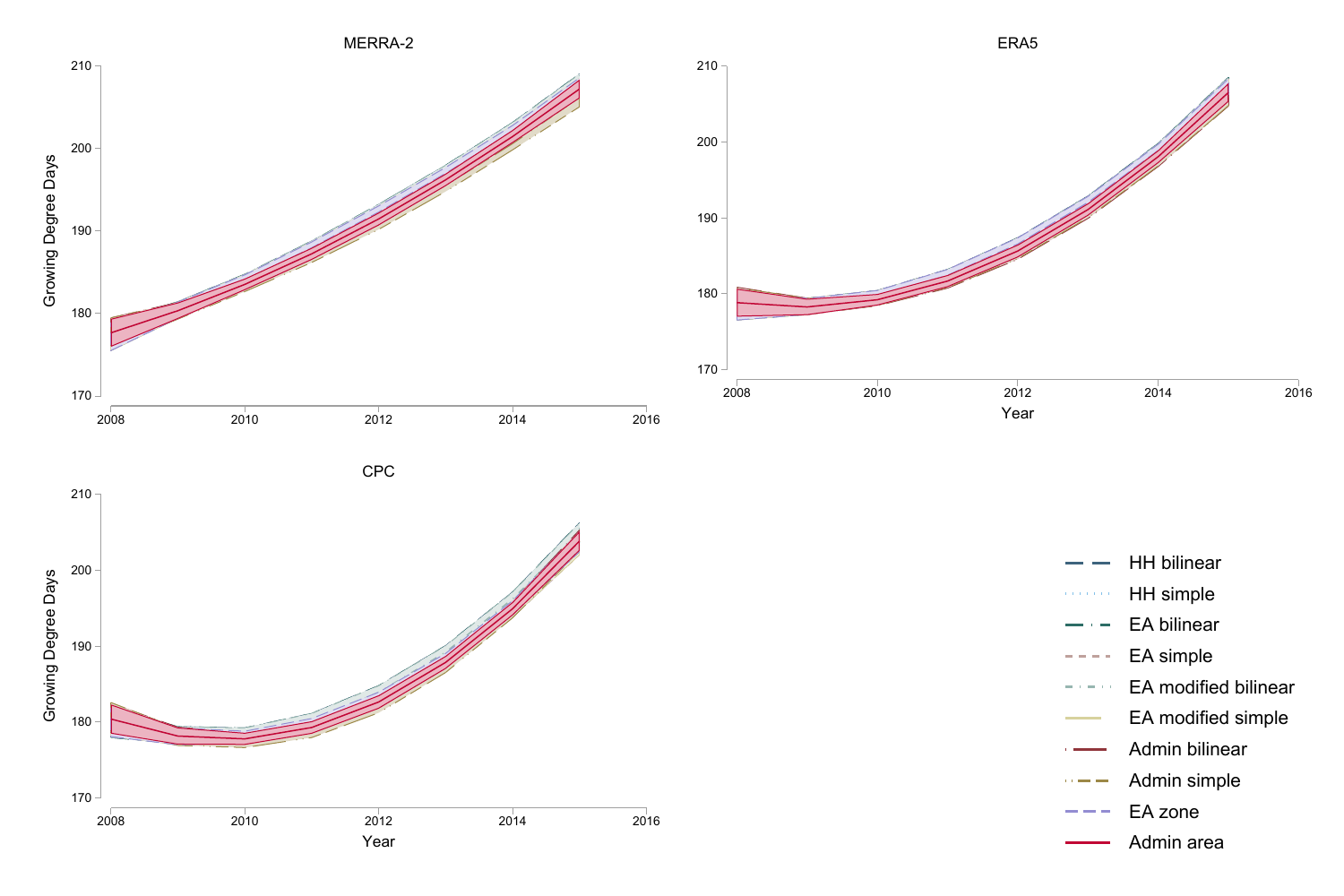}
		\end{center}
		\footnotesize  \textit{Note}: The figure presents the mean number of growing degree days (GDD) in a year, pooled across all countries, disaggregated by remote sensing source. Prediction made via Fractional-Polynomial, with $95\%$ confidence interval represented by shaded area. Each line (anonymization method) in each panel is constructed using all $33,738$ household-year observations. Variation in lines do not come variation in the household data that is paired with the remote sensing data. Rather, variation in lines within a panel is solely due to differences in the grid cell in which the anonymization method locates the household. Variation in lines across panels is solely due to differences in the value of temperature reported by the remote sensing source.
	\end{minipage}
\end{figure}
\end{landscape}	


\newpage 


\begin{landscape}
\begin{figure}[!htbp]
	\begin{minipage}{\linewidth}		
		\caption{Mean Log Likelihood, by Extraction and Model}
		\label{fig:r2_ext}
		\begin{center}
			\includegraphics[width=.9\linewidth,keepaspectratio]{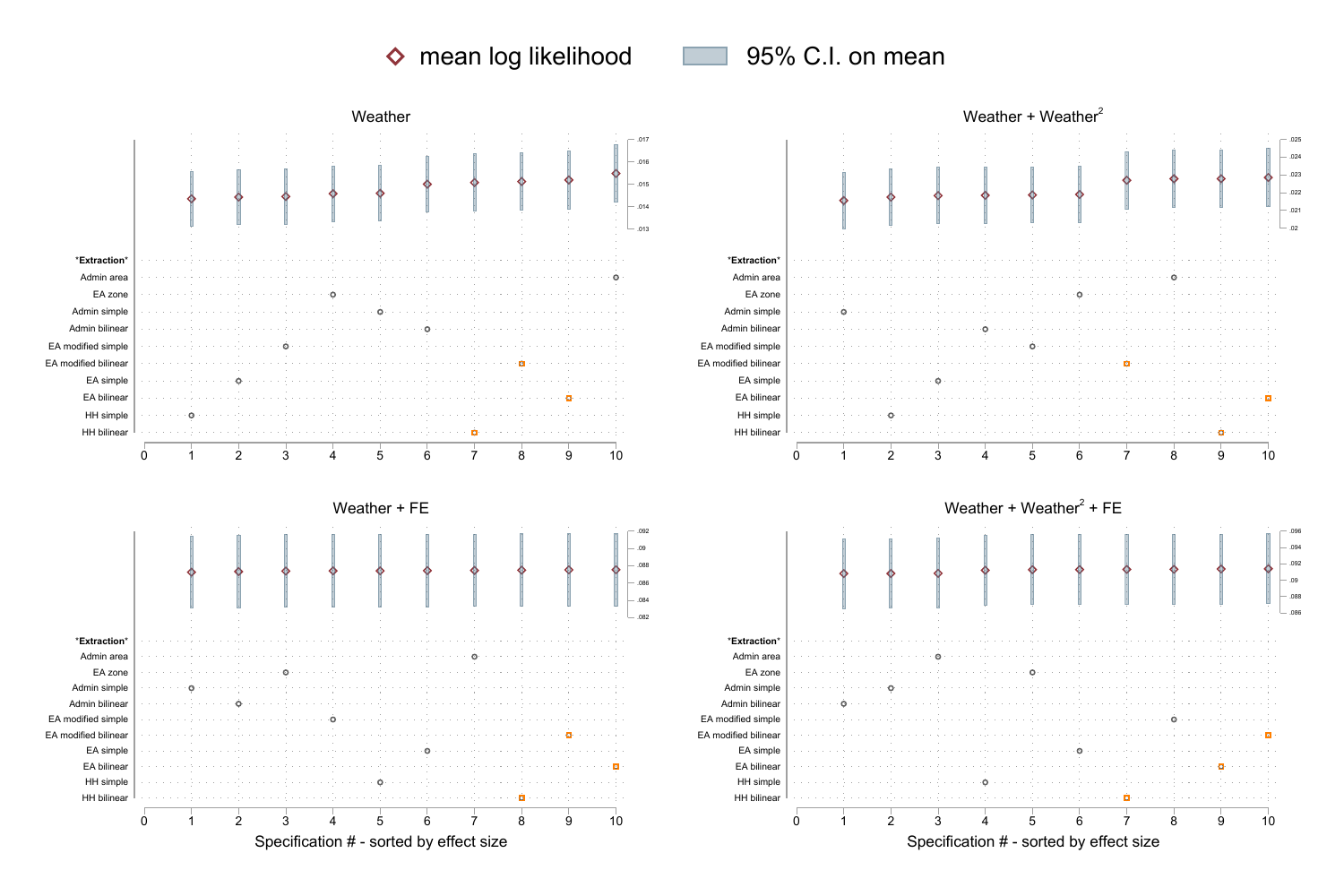}
		\end{center}
		\footnotesize  \textit{Note}: The figure presents the mean log likelihood, by anonymization method and model specification, aggregated over country, weather metric, remote sensing source, and outcome variable. The figure is derived from the results of all 51,840 regressions, with each panel summarizing the results of 12,960 regressions. Each column in each panel summarizes the results of 1,296 regressions, which are for each specification model and each anonymization method. Orange diamonds identify bilinear extract methods that tend to perform particularly well (household, EA, and modified EA).
	\end{minipage}	
\end{figure}
\end{landscape}


\begin{landscape}
\begin{figure}[!htbp]
	\begin{minipage}{\linewidth}		
		\caption{$p$-values of Rainfall and Temperature, by Anonymization Method}
		\label{fig:pval_ext}
		\begin{center}
			\includegraphics[width=.9\linewidth,keepaspectratio]{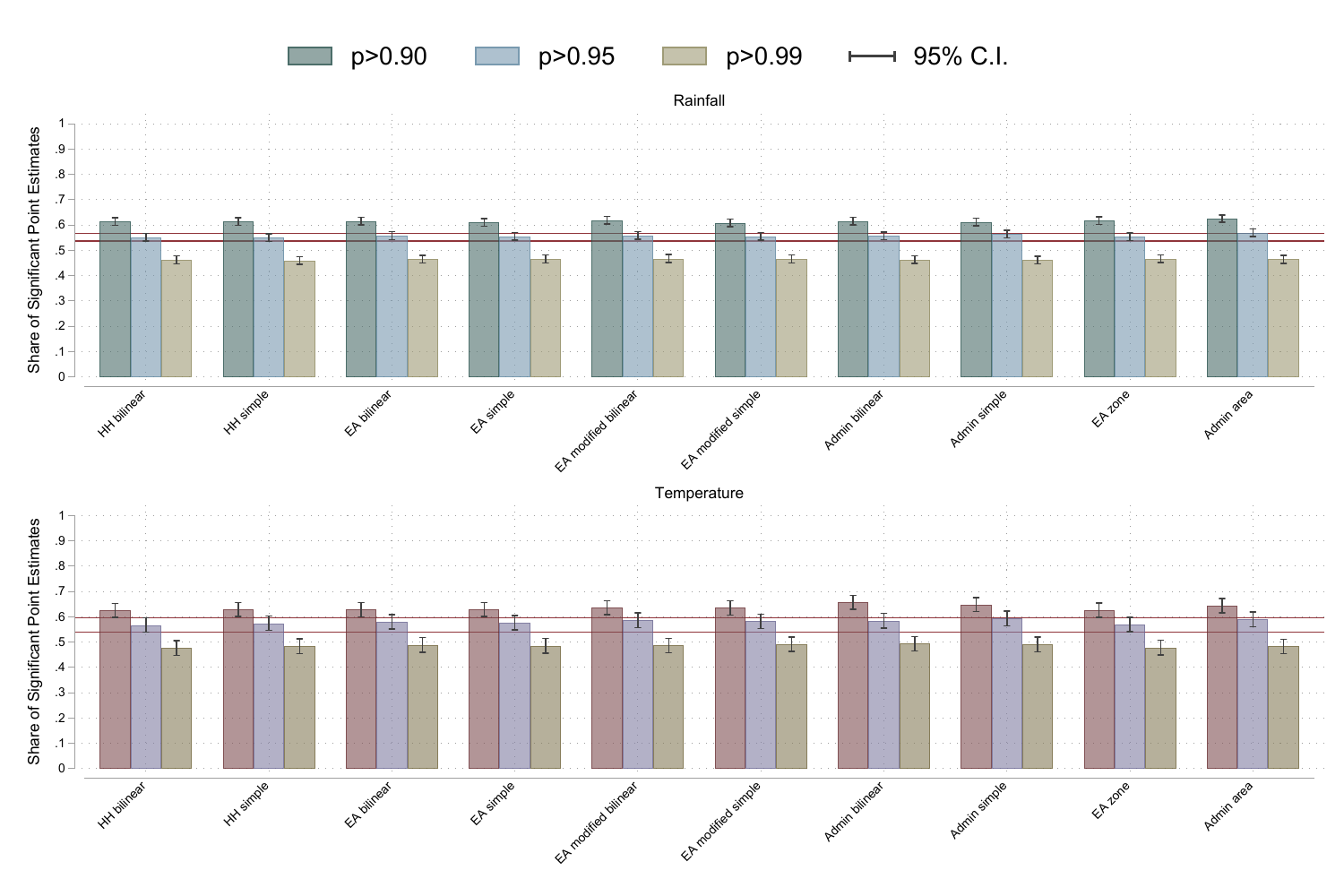}
		\end{center}
		\footnotesize  \textit{Note}: The figure displays the share of coefficients on the rainfall and temperature variables that are statistically significant from each anonymization method, aggregated over country, weather metric, remote sensing source, outcome variable, and specification. The northern panel presents rainfall while the southern panel presents temperature. The data summarized in the northern panel includes 40,320 regressions, with each column including 4,032 regressions. The data summarized in the southern panel includes 11,520 regressions, with each column including 1,152 regressions.  
	\end{minipage}	
\end{figure}
\end{landscape}

\begin{landscape}
\begin{center}
\begin{figure}[!htbp]
	\begin{minipage}{\linewidth}
		\caption{$p$-values of Rainfall, by Country and Anonymization Method}
		\label{fig:pval_ext_rf}
		\begin{center}
			\includegraphics[width=.95\linewidth,keepaspectratio]{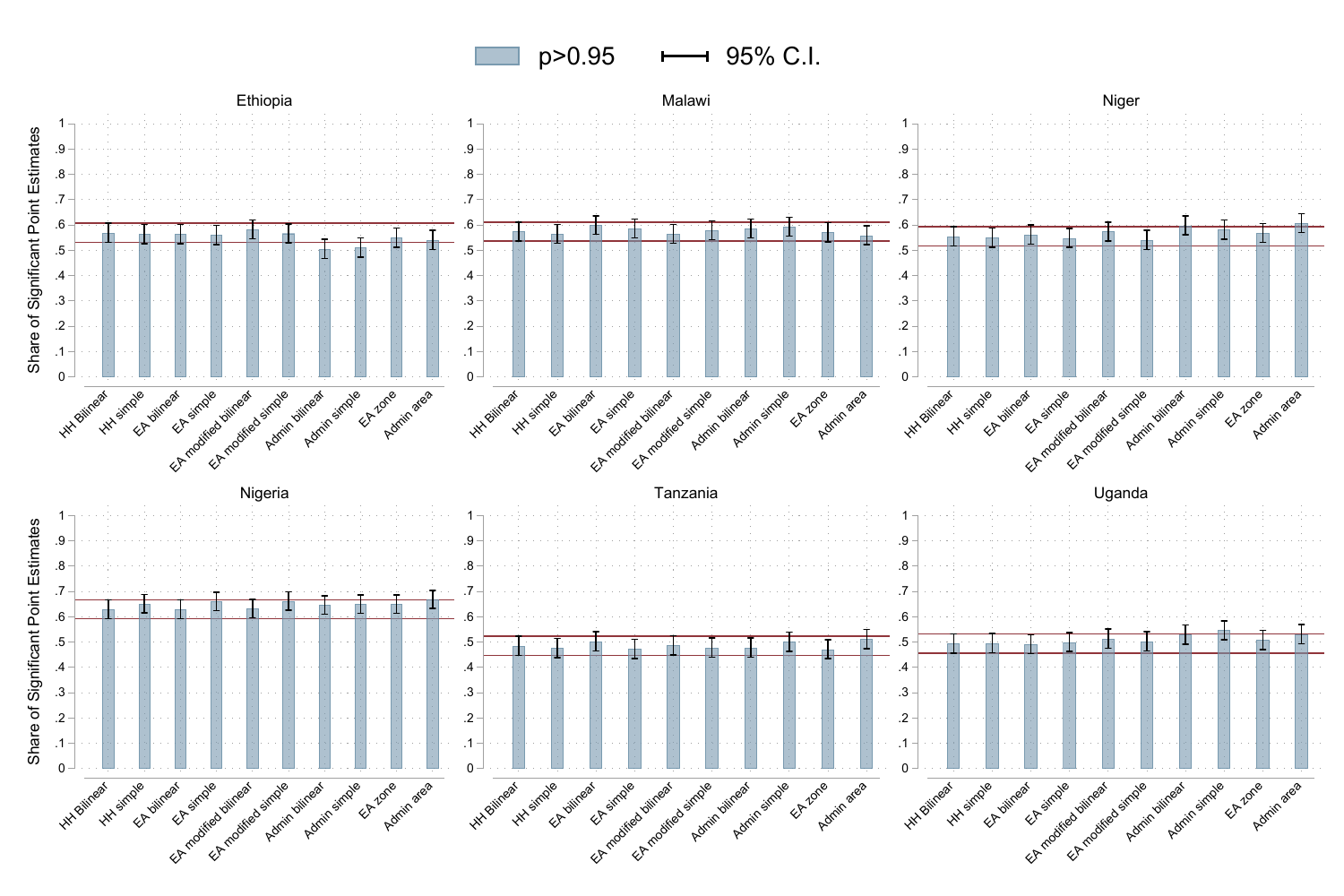}
		\end{center}
		\footnotesize  \textit{Note}: The figure displays the share of coefficients on the rainfall variables that are statistically significant from each anonymization method for each country, aggregated over weather metric, remote sensing source, outcome variable, and specification. The figure presents results from a total of 40,320 regressions. Each country includes results from 6,720 regressions and thus each column is based on 672 regressions. 
	\end{minipage}	
\end{figure}
\end{center}
\end{landscape}

\begin{landscape}
\begin{center}
\begin{figure}[!htbp]
	\begin{minipage}{\linewidth}		
		\caption{$p$-values of Temperature, by Country and Anonymization Method}
		\label{fig:pval_ext_tp}
		\begin{center}
			\includegraphics[width=.95\linewidth,keepaspectratio]{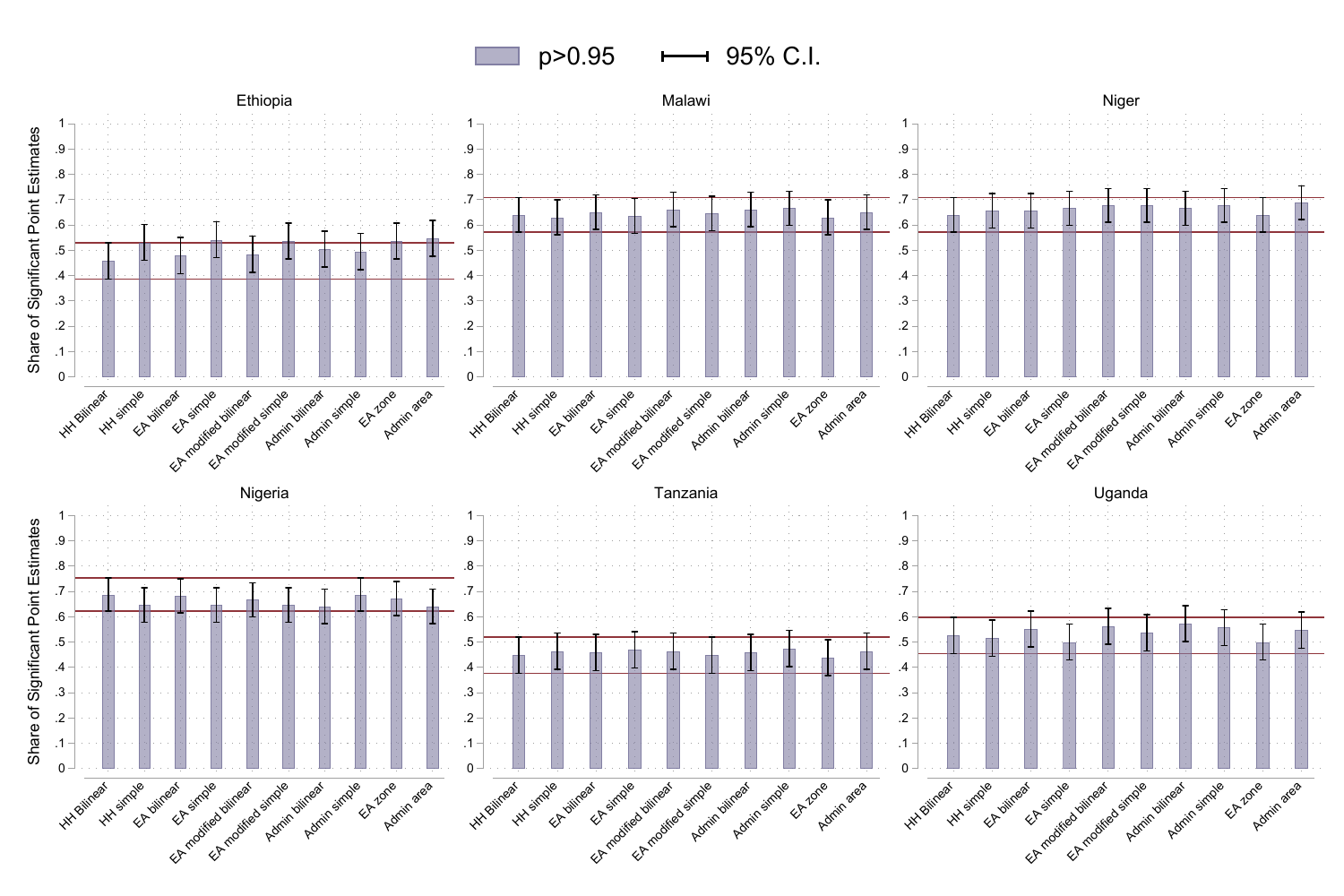}
		\end{center}
		\footnotesize  \textit{Note}: The figure displays the share of coefficients on the temperature variables that are statistically significant from each anonymization method for each country, aggregated over weather metric, remote sensing source, outcome variable, and specification. The figure presents results from a total of 11,520 regressions. Each country includes results from 1,920 regressions and thus each column is based on 192 regressions. 
	\end{minipage}	
\end{figure}
\end{center}
\end{landscape}



\begin{landscape}
\begin{figure}[!htbp]
	\begin{minipage}{\linewidth}		
		\caption{Specification Curve for Rainfall Variables in Ethiopia}
		\label{fig:line_cty1_rf}
		\begin{center}
			\includegraphics[width=.9\linewidth,keepaspectratio]{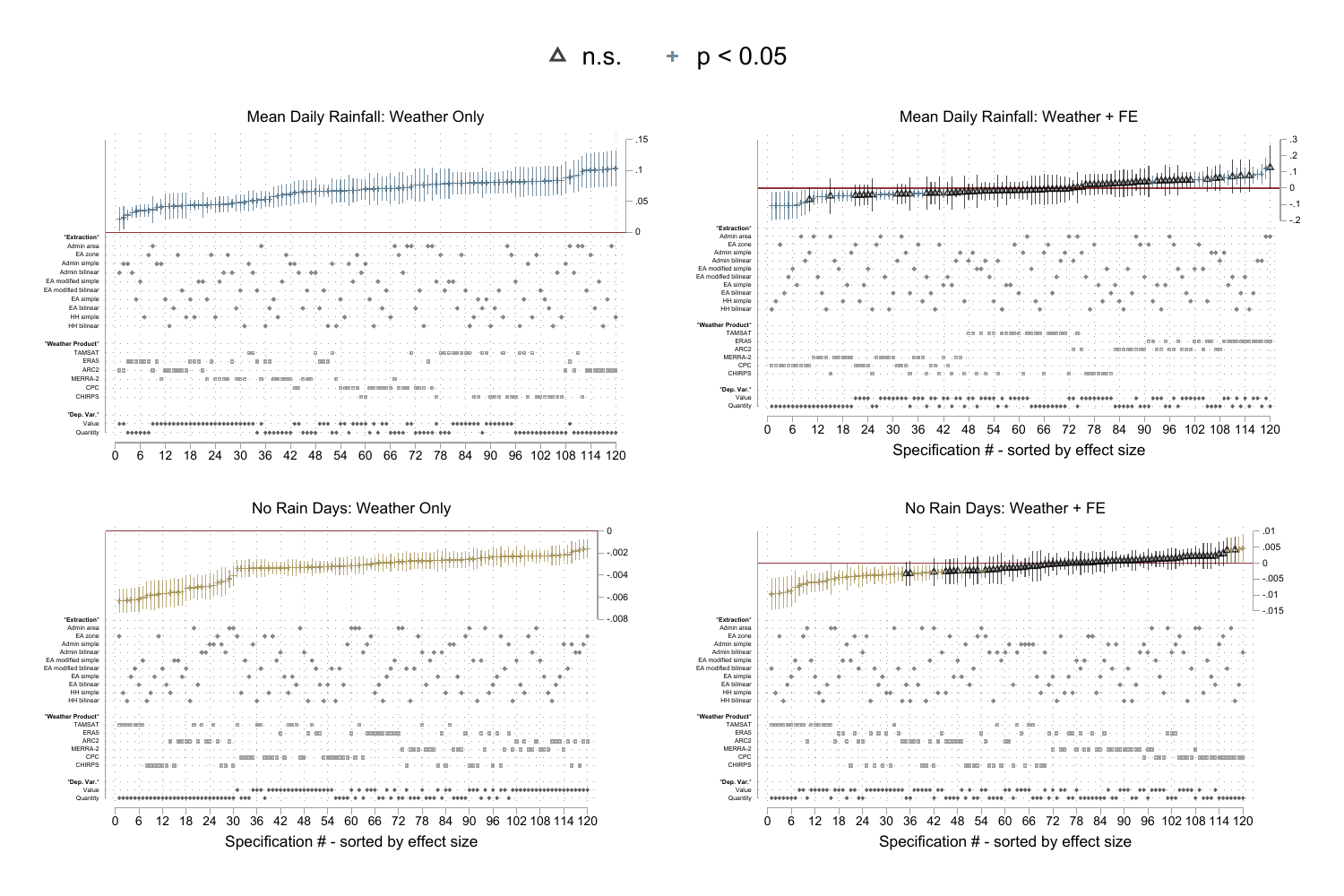}
		\end{center}
		\footnotesize  \textit{Note}: The figure presents specification curves where each panel presents results from a different model. Each panel includes 120 regressions, where each column represents a single regression. Significant and non-significant coefficients are designated at the top of the figure.
	\end{minipage}	
\end{figure}
\end{landscape}

\begin{landscape}
\begin{figure}[!htbp]
	\begin{minipage}{\linewidth}		
		\caption{Specification Curve for Rainfall Variables in Malawi}
		\label{fig:line_cty2_rf}
		\begin{center}
			\includegraphics[width=.9\linewidth,keepaspectratio]{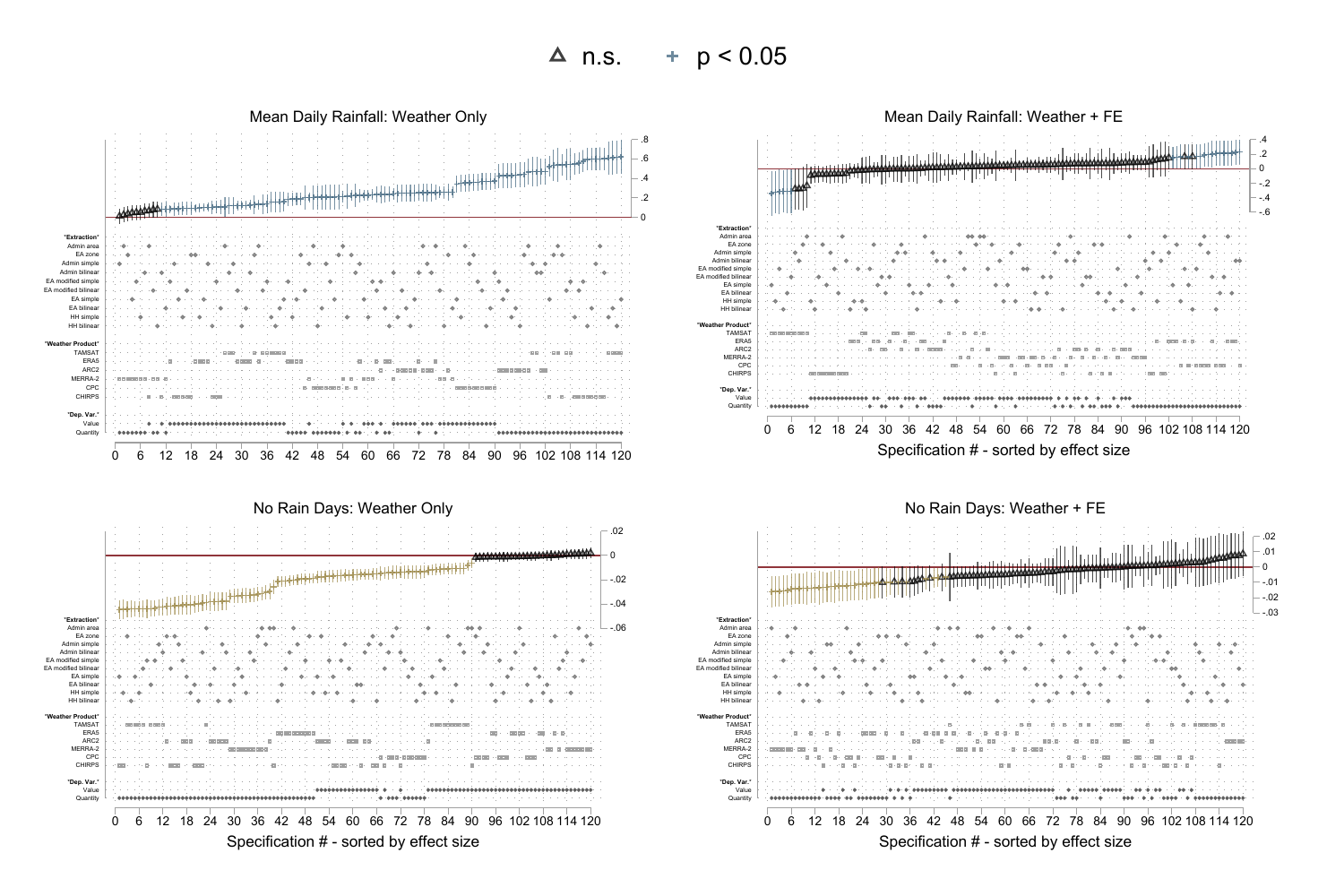}
		\end{center}
		\footnotesize  \textit{Note}: The figure presents specification curves where each panel presents results from a different model. Each panel includes 120 regressions, where each column represents a single regression. Significant and non-significant coefficients are designated at the top of the figure.
	\end{minipage}	
\end{figure}
\end{landscape}

\begin{landscape}
\begin{figure}[!htbp]
	\begin{minipage}{\linewidth}		
		\caption{Specification Curve for Rainfall Variables in Niger}
		\label{fig:line_cty4_rf}
		\begin{center}
			\includegraphics[width=.9\linewidth,keepaspectratio]{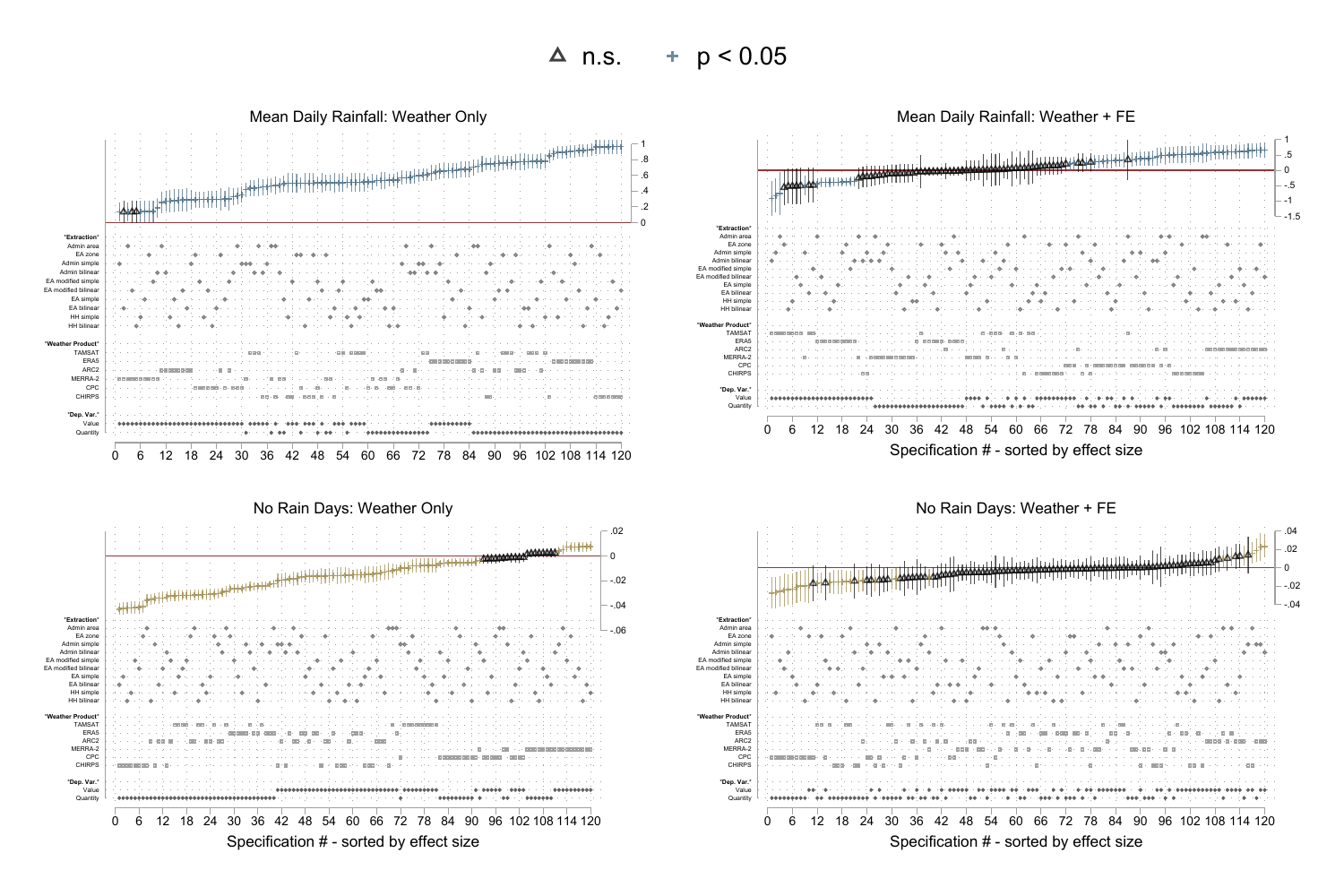}
		\end{center}
		\footnotesize  \textit{Note}: The figure presents specification curves where each panel presents results from a different model. Each panel includes 120 regressions, where each column represents a single regression. Significant and non-significant coefficients are designated at the top of the figure.
	\end{minipage}	
\end{figure}
\end{landscape}

\begin{landscape}
\begin{figure}[!htbp]
	\begin{minipage}{\linewidth}		
		\caption{Specification Curve for Rainfall Variables in Nigeria}
		\label{fig:line_cty5_rf}
		\begin{center}
			\includegraphics[width=.9\linewidth,keepaspectratio]{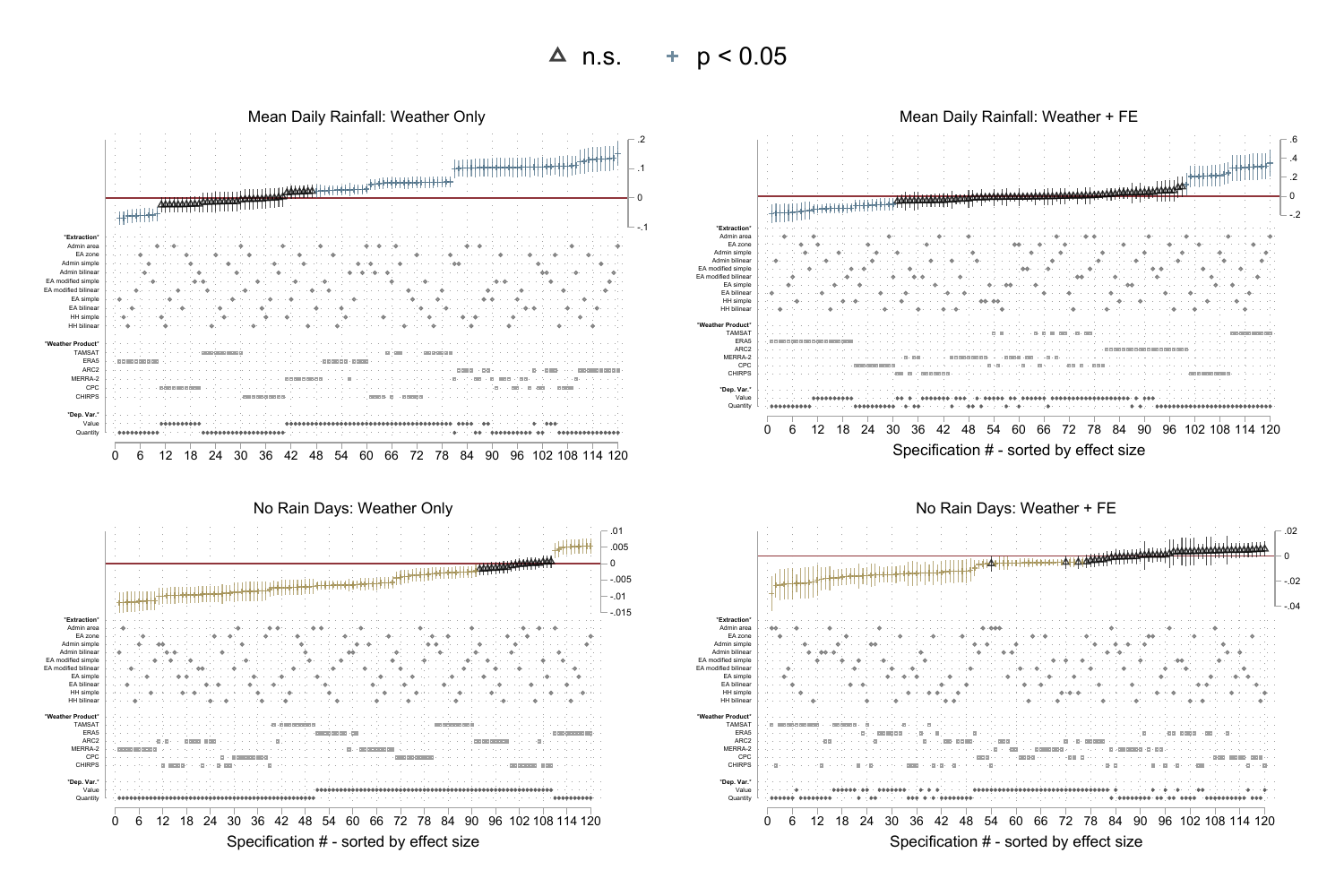}
		\end{center}
		\footnotesize  \textit{Note}: The figure presents specification curves where each panel presents results from a different model. Each panel includes 120 regressions, where each column represents a single regression. Significant and non-significant coefficients are designated at the top of the figure.
	\end{minipage}	
\end{figure}
\end{landscape}

\begin{landscape}
\begin{figure}[!htbp]
	\begin{minipage}{\linewidth}		
		\caption{Specification Curve for Rainfall Variables in Tanzania}
		\label{fig:line_cty6_rf}
		\begin{center}
			\includegraphics[width=.9\linewidth,keepaspectratio]{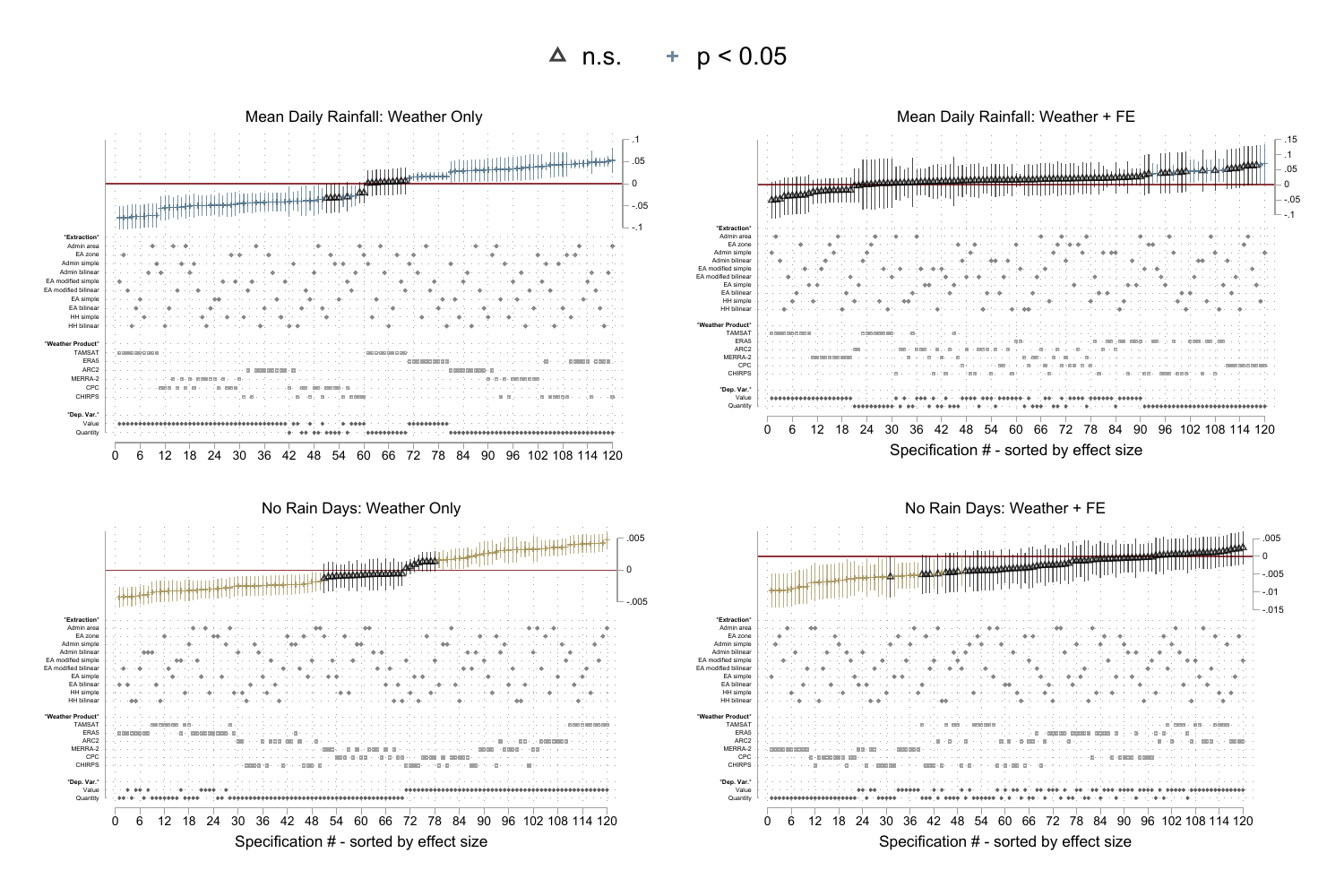}
		\end{center}
		\footnotesize  \textit{Note}: The figure presents specification curves where each panel presents results from a different model. Each panel includes 120 regressions, where each column represents a single regression. Significant and non-significant coefficients are designated at the top of the figure.
	\end{minipage}	
\end{figure}
\end{landscape}

\begin{landscape}
\begin{figure}[!htbp]
	\begin{minipage}{\linewidth}		
		\caption{Specification Curve for Rainfall Variables in Uganda}
		\label{fig:line_cty7_rf}
		\begin{center}
			\includegraphics[width=.9\linewidth,keepaspectratio]{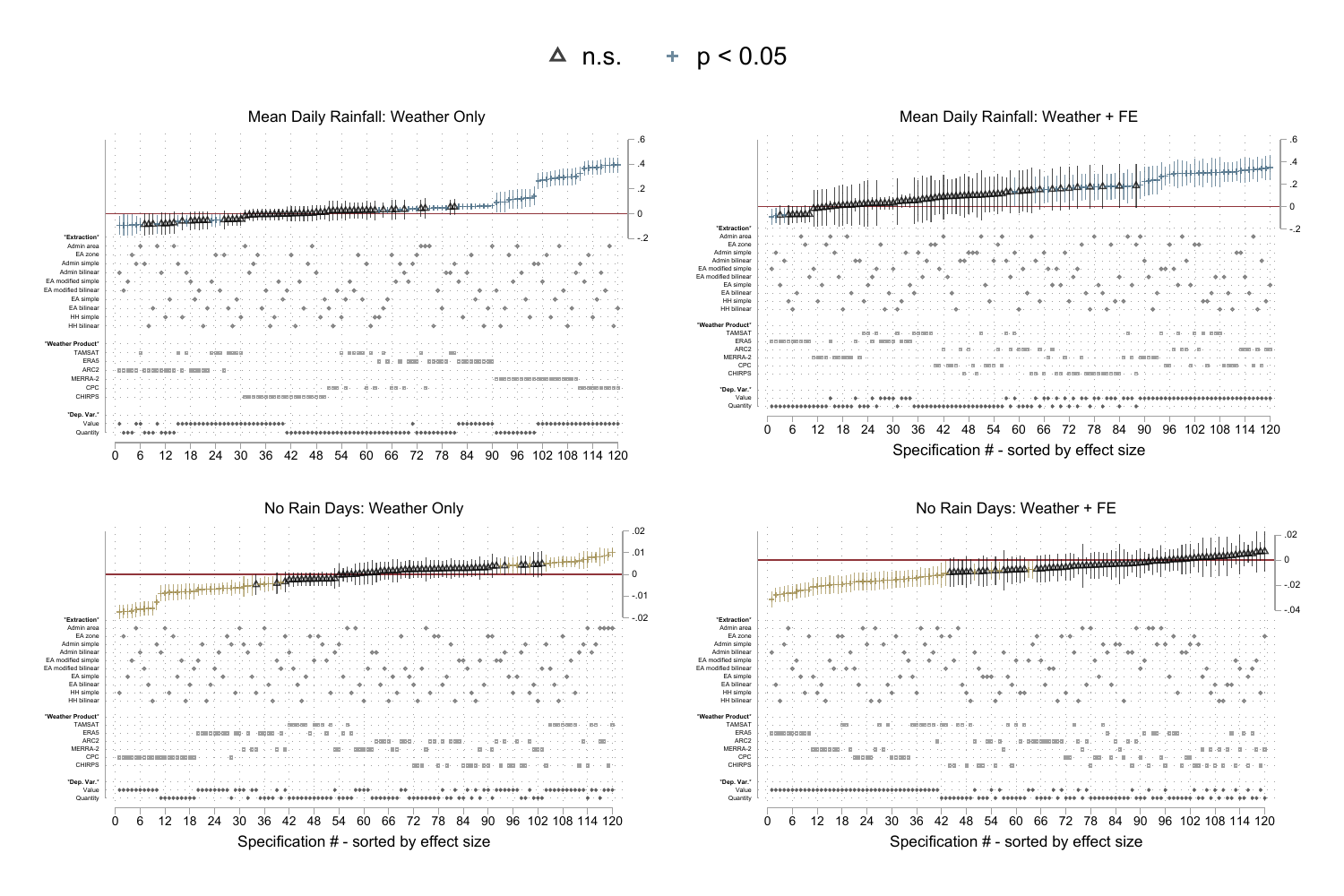}
		\end{center}
		\footnotesize  \textit{Note}: The figure presents specification curves where each panel presents results from a different model. Each panel includes 120 regressions, where each column represents a single regression. Significant and non-significant coefficients are designated at the top of the figure.
	\end{minipage}	
\end{figure}
\end{landscape}


\begin{landscape}
\begin{figure}[!htbp]
	\begin{minipage}{\linewidth}		
		\caption{Specification Curve for Temperature Variables in Ethiopia}
		\label{fig:line_cty1_tp}
		\begin{center}
			\includegraphics[width=.9\linewidth,keepaspectratio]{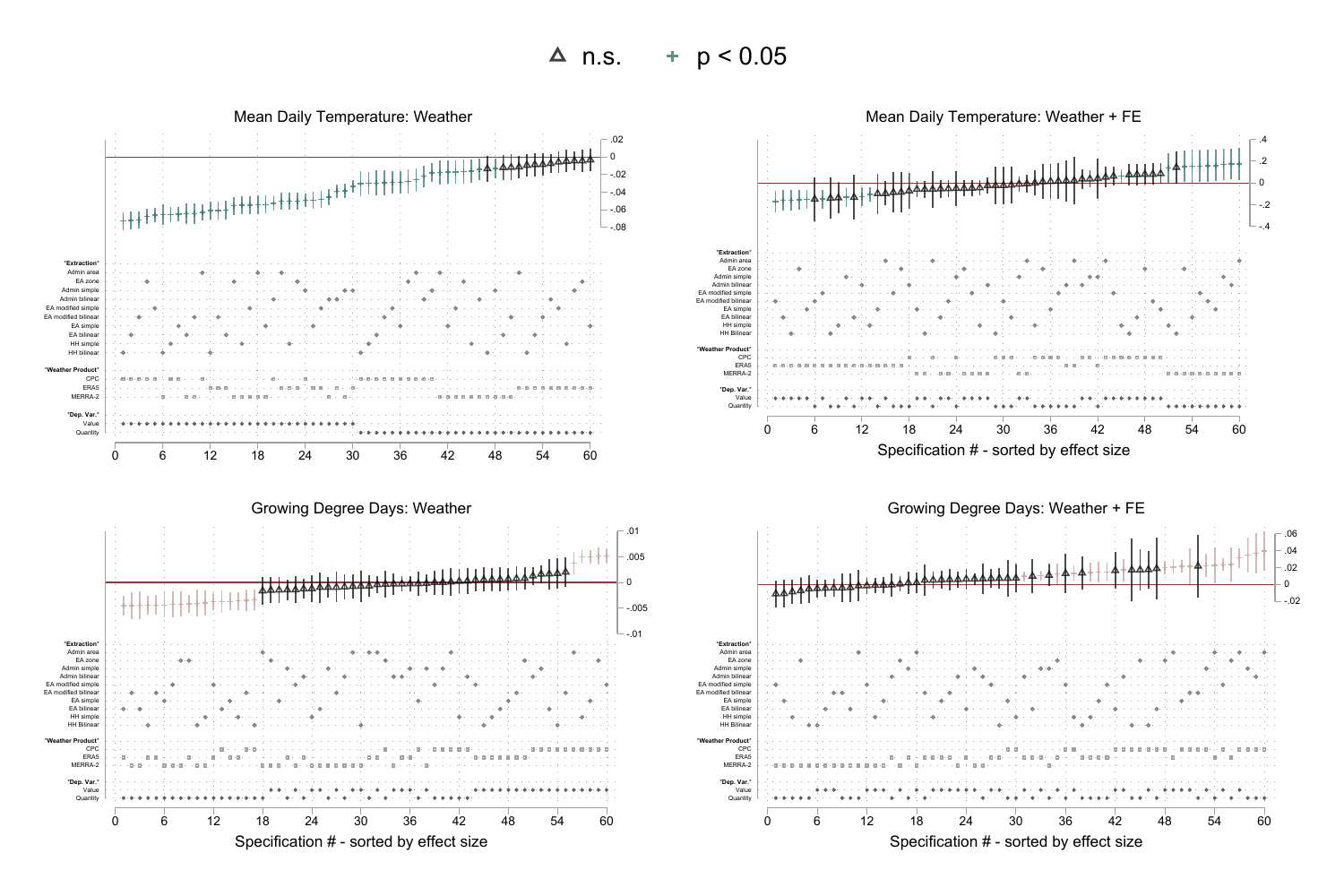}
		\end{center}
		\footnotesize  \textit{Note}: The figure presents specification curves where each panel presents results from a different model. Each panel includes 60 regressions, where each column represents a single regression. Significant and non-significant coefficients are designated at the top of the figure.
	\end{minipage}	
\end{figure}
\end{landscape}

\begin{landscape}
\begin{figure}[!htbp]
	\begin{minipage}{\linewidth}		
		\caption{Specification Curve for Temperature Variables in Malawi}
		\label{fig:vline_cty2_tp}
		\begin{center}
			\includegraphics[width=.9\linewidth,keepaspectratio]{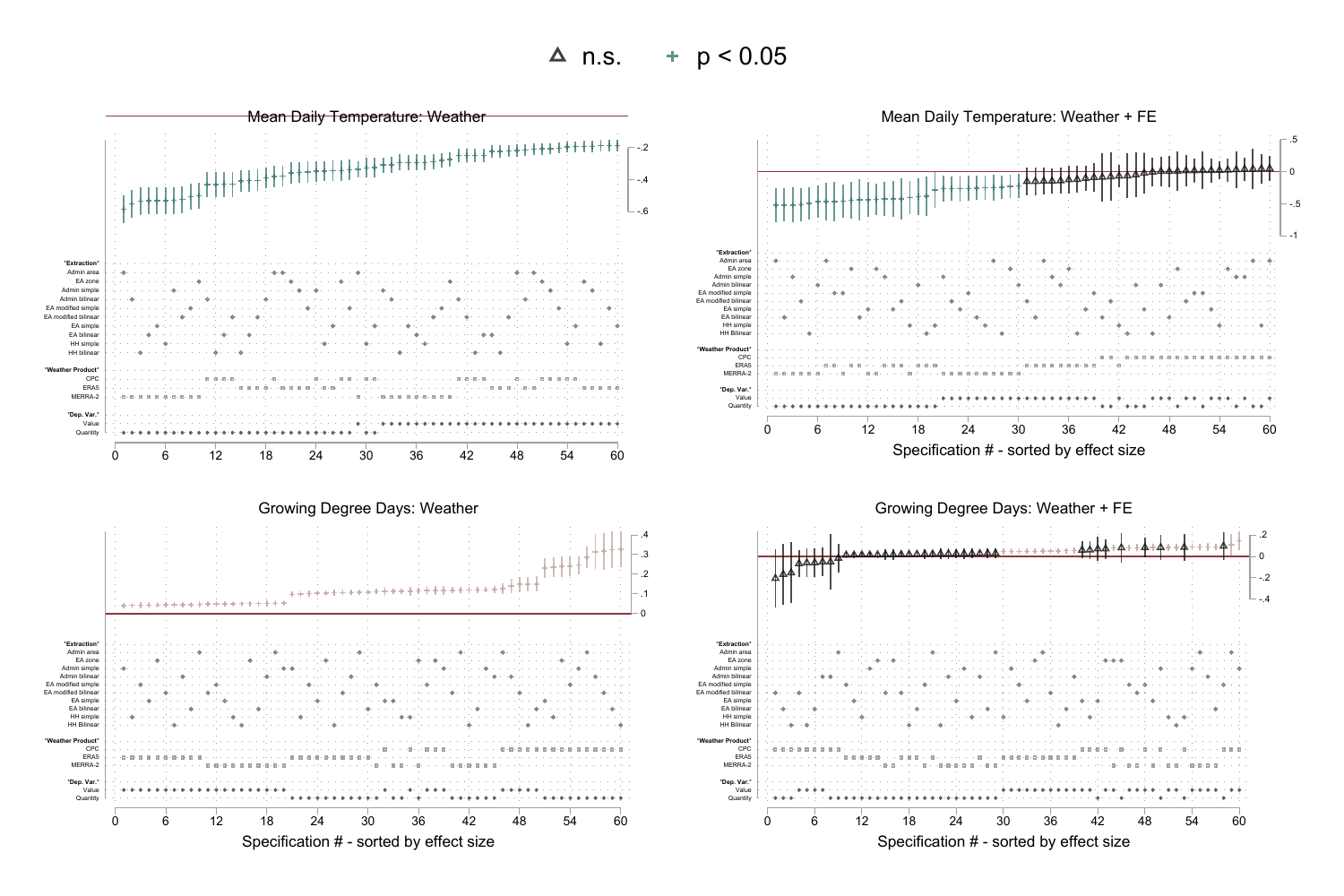}
		\end{center}
		\footnotesize  \textit{Note}: The figure presents specification curves where each panel presents results from a different model. Each panel includes 60 regressions, where each column represents a single regression. Significant and non-significant coefficients are designated at the top of the figure.
	\end{minipage}	
\end{figure}
\end{landscape}

\begin{landscape}
\begin{figure}[!htbp]
	\begin{minipage}{\linewidth}		
		\caption{Specification Curve for Temperature Variables in Niger}
		\label{fig:line_cty4_tp}
		\begin{center}
			\includegraphics[width=.9\linewidth,keepaspectratio]{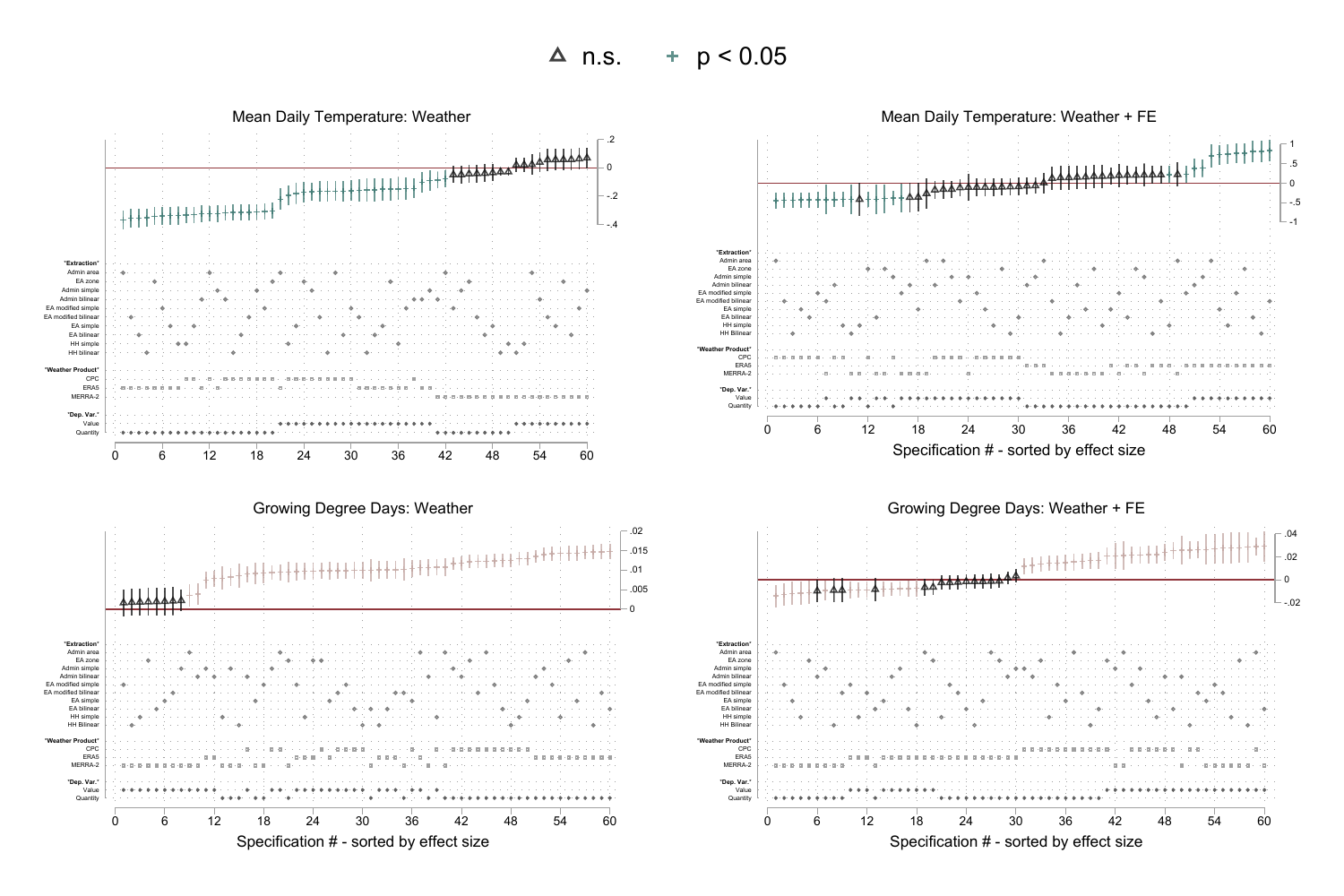}
		\end{center}
		\footnotesize  \textit{Note}: The figure presents specification curves where each panel presents results from a different model. Each panel includes 60 regressions, where each column represents a single regression. Significant and non-significant coefficients are designated at the top of the figure.
	\end{minipage}	
\end{figure}
\end{landscape}

\begin{landscape}
\begin{figure}[!htbp]
	\begin{minipage}{\linewidth}		
		\caption{Specification Curve for Temperature Variables in Nigeria}
		\label{fig:line_cty5_tp}
		\begin{center}
			\includegraphics[width=.9\linewidth,keepaspectratio]{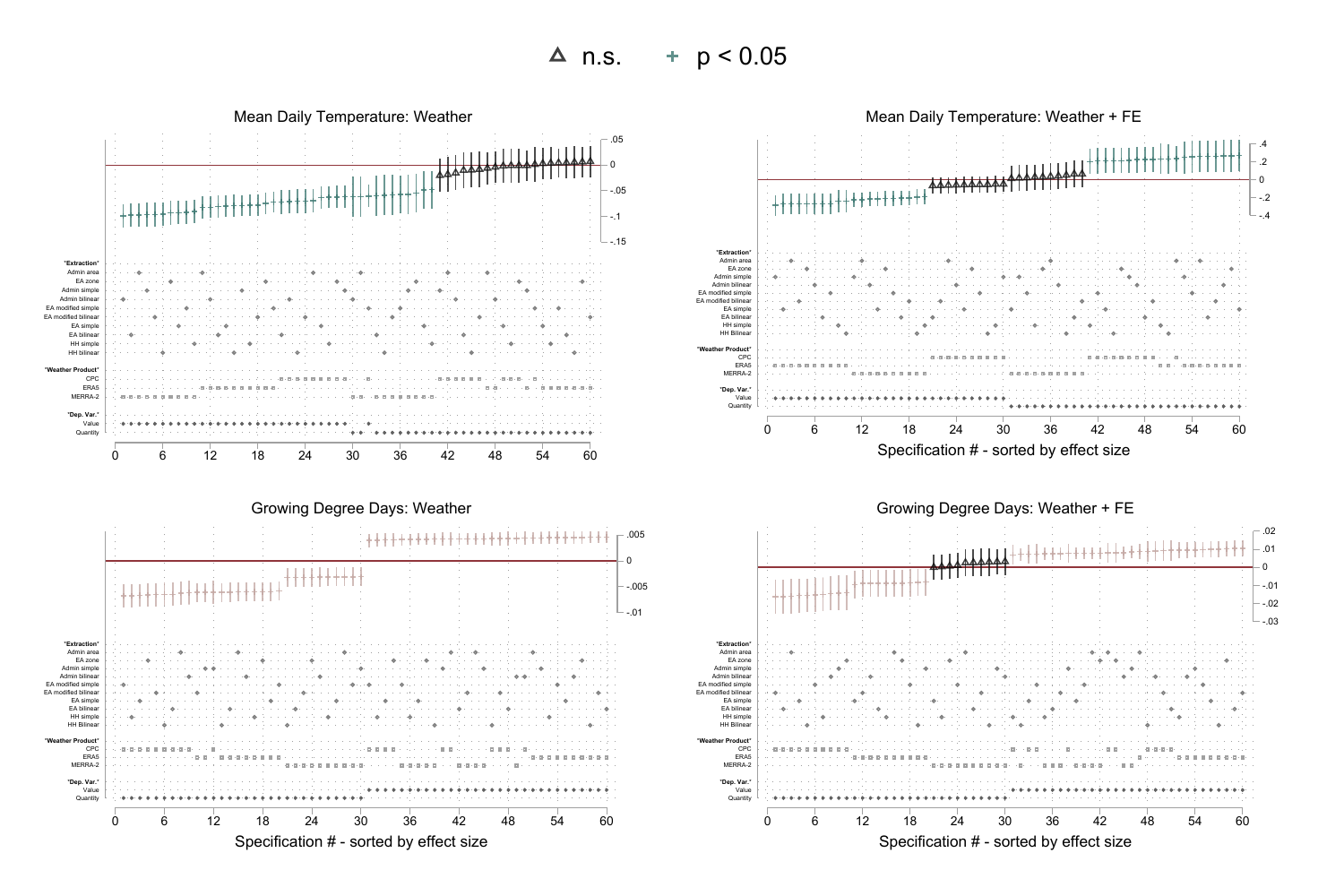}
		\end{center}
		\footnotesize  \textit{Note}: The figure presents specification curves where each panel presents results from a different model. Each panel includes 60 regressions, where each column represents a single regression. Significant and non-significant coefficients are designated at the top of the figure.
	\end{minipage}	
\end{figure}
\end{landscape}

\begin{landscape}
\begin{figure}[!htbp]
	\begin{minipage}{\linewidth}		
		\caption{Specification Curve for Temperature Variables in Tanzania}
		\label{fig:line_cty6_tp}
		\begin{center}
			\includegraphics[width=.9\linewidth,keepaspectratio]{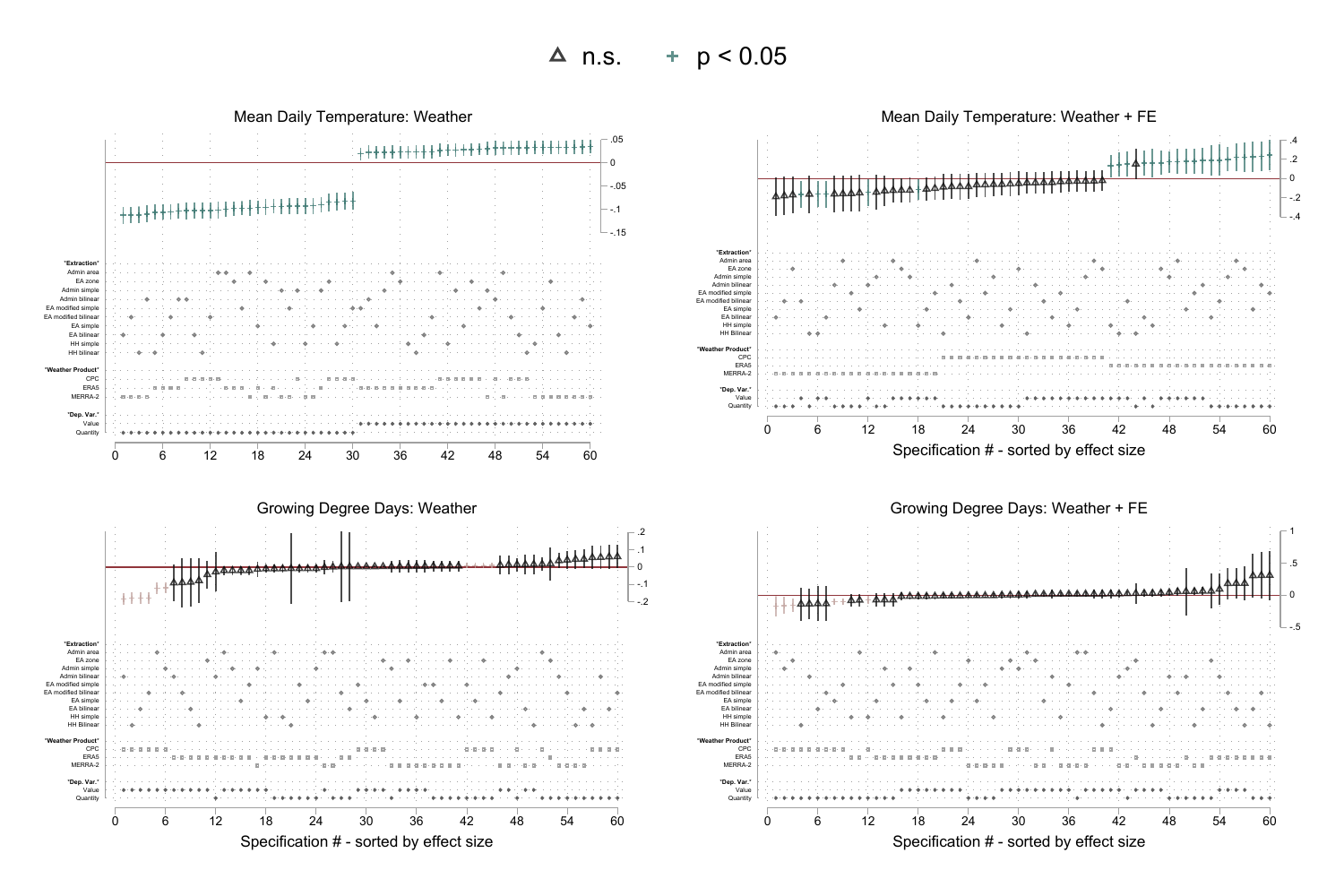}
		\end{center}
		\footnotesize  \textit{Note}: The figure presents specification curves where each panel presents results from a different model. Each panel includes 60 regressions, where each column represents a single regression. Significant and non-significant coefficients are designated at the top of the figure.
	\end{minipage}	
\end{figure}
\end{landscape}

\begin{landscape}
\begin{figure}[!htbp]
	\begin{minipage}{\linewidth}		
		\caption{Specification Curve for Temperature Variables in Uganda}
		\label{fig:line_cty7_tp}
		\begin{center}
			\includegraphics[width=.9\linewidth,keepaspectratio]{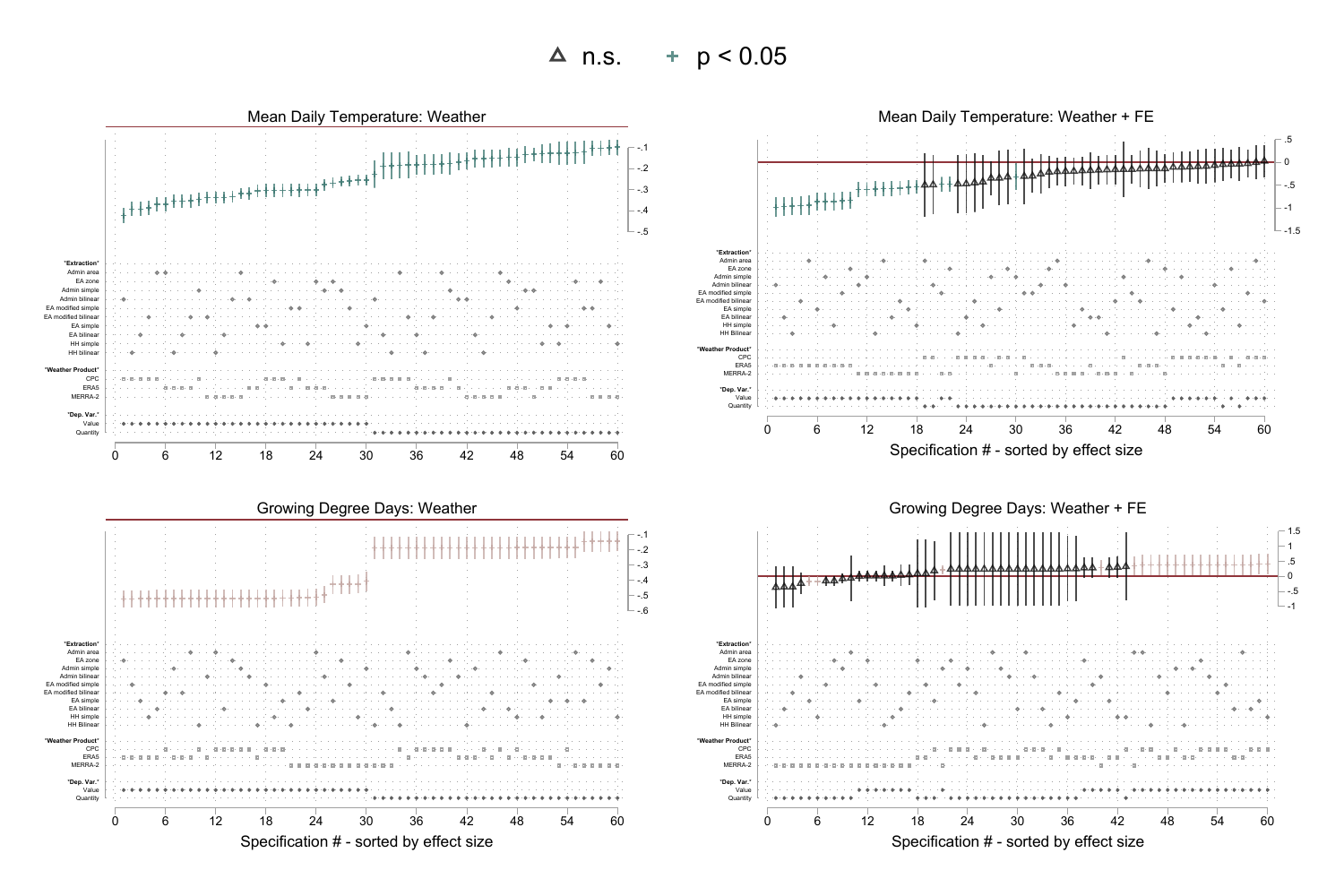}
		\end{center}
		\footnotesize  \textit{Note}: The figure presents specification curves where each panel presents results from a different model. Each panel includes 60 regressions, where each column represents a single regression. Significant and non-significant coefficients are designated at the top of the figure.
	\end{minipage}	
\end{figure}
\end{landscape}


\clearpage
\newpage
\appendix
\onehalfspacing

\begin{center}
	\section*{Online-Only Appendix to ``Privacy Protection, Measurement Error, and the Linking of Remote Sensing and Socioeconomic Survey Data''} \label{sec:app}
\end{center}


\section{Details on Remote Sensing Weather Data} \label{sec:appRS}

\setcounter{table}{0}
\renewcommand{\thetable}{A\arabic{table}}
\setcounter{figure}{0}
\renewcommand{\thefigure}{A\arabic{figure}}

\subsection{On Using Remote Sensing Products for Economics} \label{sec:appRS_sat}

Uncertainty is present in all model outputs, and weather datasets are no exception. Spatial datasets of weather variables, like precipitation and temperature, that are produced using remotely sensed data, are not direct measurements of the variable of interest. Satellite sensors provide spatially continuous observation of reflectance from the earth's surface in different parts of the magnetic spectrum. These values are used to estimate related phenomena, such as cloud presence, cloud top temperature or earth surface temperature. The continuous datasets are then used in combination with directly observed, but often sparsely distributed, gauge data to produce weather variables. Some inputs are common across products, but there are differences in other inputs as well as modeling techniques.

The type of analysis matters in assessing weather datasets for use in economic research. Is the goal to understand climate trends, capture characteristics of a particular agricultural season, or identify extreme weather events occurring in near real-time? This can help determine the relative importance of different dataset characteristics, such as spatial detail, temporal frequency and length of record, with respect to the intended analysis. The datasets used in this analysis were constrained by certain minimum criteria, leading to elimination of some commonly used datasets, such as the product from the Center for Climatic Research at the University of Delaware. We also did not consider proprietary datasets, preferring to use sources currently in the public domain. Despite the exclusions imposed by our minimum criteria, the datasets summarized in Table~\ref{tab:weather} represent a range of spatial resolutions and model types commonly used by economists. Further details on the specifics of each remote sensing product are provided below with the goal of providing economists with direction to a dataset that meets the requirements of their analysis.


\subsubsection{Africa Rainfall Climatology version 2 (ARC2)}

ARC2 is a merged gauge data and remote sensing product that provides daily rainfall outputs for the African continent. The dataset, produced by the National Oceanic and Atmospheric Administration (NOAA) Climate Prediction Center (CPC) provides improvements over ARC1 and a longer length of record compared to the rainfall estimate (RFE), the operational dataset of USAID's Famine Early Warning Systems Network (FEWSNET) program. Inputs are Global Telecommunications System (GTS) rain gauge data over Africa, geostationary Meteosat infrared (IR) imagery, and polar-orbiting microwave Special Sensor Microwave Image (SSM/I) and Advanced Microwave Sounding Unit (AMSU-B).

Validation efforts by \cite{ARC2} found that low reporting rates for some GTS stations degrades model performance in those regions. Other findings are a general tendency to underestimate rainfall, which is enhanced in areas of high relief or complex topography.

Data and technical documentation are available for download from \url{https://www.cpc.ncep.noaa.gov/products/international/data.shtml}.


\subsubsection{Climate Hazards group InfraRed Precipitation with Station data (CHIRPS)}

Like ARC2, the CHIRPS rainfall dataset builds on established techniques for merging gauge and remote sensing data. Produced by the Climate Hazards Group at University of California, Santa Barbara this dataset is designed for monitoring of drought and environmental change at a global level. To minimize latency, there are two products, a preliminary version with two day lag, and final output available at three weeks. Outputs are available at time-steps from six hours to three months. As inputs, CHIRPS makes use of a monthly climatology CHPclim, Tropical Rainfall Measuring Mission Multi-satellite Precipitation Analysis version 7 (TMPA 3B42 v7) and global Thermal Infrared Cold Cloud Duration (TIR CCD) from two NOAA archives. The remote sensing data are then merged with gauge data from five public archives, including the Global Historical Climatology Network (GHCN) and GTS, several private sources, and meteorological agencies. While targeted gauge data collection efforts resulted in a greater number of input stations for years prior to 2010, the number of stations going forward is more limited, particularly in Sub-Saharan Africa. Detailed metadata by country is available and may be a useful reference to determine if coverage for a region of interest is sufficient for the analysis.

Validation for select countries found that the climatology input CHPclim outperformed other climatology datasets in data sparse regions and complex terrain \citep{CHIRPS}. Furthermore, in an assessment of wet season statistics CHIRPS showed less bias than other rainfall sources and good correspondence with Global Precipitation Climatology Centre (GPCC) estimates. 

Data and technical documentation are available for download from \url{https://data.chc.ucsb.edu/products/CHIRPS-2.0/}.


\subsubsection{CPC Global Unified Gauge-Based Analysis of Daily Precipitation and Temperature}

NOAA's Climate Prediction Center (CPC) Unified Gauge-based (CPC-U) datasets for daily temperature and precipitation do not incorporate remote sensing data in the estimation of weather variables. Instead, an optimal interpolation (OI) technique is used on gauge data for precipitation, and Shepard's algorithm for temperature. CPC-U provides systematic global datasets for validation and climate monitoring. GTS is a primary input data source, with some national collections, but density is most sparse over Africa.

As to be expected, even though the OI interpolation performs better than other techniques, a cross-validation exercise shows performance to degrade significantly with increasing distance to nearest station \citep{CPC}. As a result, this dataset may not be suitable for analysis in some parts of Africa, with high spatial variation and low density of stations.

Data and technical documentation are available for download from \url{https://psl.noaa.gov/data/gridded/data.cpc.globalprecip.html} for precipitation and
\url{https://psl.noaa.gov/data/gridded/data.cpc.globaltemp.html} for temperature.


\subsubsection{European Centre for Medium-Range Weather Forecasts (ECMWF) ERA5}

ERA5, based on the European forecasting model ECMWF, is one of two assimilation model datasets used in this paper. The inputs are far too numerous to mention but include a range of satellite inputs as well as gauge datasets. There are a wide range of outputs as well, including 2-meter air temperature and rainfall, available at sub-daily intervals and differentiated vertically. ERA5 is coarser spatial resolution than the global and regional merged rainfall datasets, but more detailed than MERRA2. 

The sheer number and complexity of outputs can be a deterrent to the use of weather variables from assimilation models. Uncertainty or lack of understanding about inaccuracies associated with individual output variables of assimilation models, compared to other types of models, is another reason to carefully consider their suitability for particular research \citep{Parker16}. Nevertheless, reanalysis datasets are used in a broad range of applications and even outperform other gridded climate datasets in some settings \citep{ZandlerEtAl20}.

Data and technical documentation are available for download from \url{https://cds.climate.copernicus.eu}.


\subsubsection{Modern-Era Retrospective analysis for Research and Applications, version 2 (MERRA-2)}

The second reanalysis dataset used in this analysis is MERRA-2, a product of NASA's Goddard Earth Observing System, version 5 (GEOS-5) assimilation model. Specifically we make use of the variables T2MMEAN from the statD daily statistics collection, and PRECTOTLAND from the Land Surface Diagnostics collection. 
  
Data and technical documentation are available for download from \url{https://disc.gsfc.nasa.gov/}.


\subsubsection{Tropical Applications of Meteorology using SATellite data (TAMSAT)}

The TAMSAT rainfall dataset is the highest spatial resolution gridded dataset used in this analysis. Inputs are similar to other merged gauge and remote sensing products: Meteosat TIR imagery, purposefully collected archival (1983-2010) rain gauge data from meteorological agencies and other sources and GTS gauge data. Rainfall estimation is based on cold cloud duration (CCD) inferred from TIR and calibrated using gauge data within discrete calibration zones.

Validation of TAMSAT found a mean underestimation of rainfall of approximately four mm per dekad, though the bias was not always negative \citep{TAMSAT}. Due to differences in methodology from CHIRPS and ARC2 precipitation products, TAMSAT is not affected by inconsistency in gauge data inputs. This makes it suitable for placing rainfall variability in the context of a long-term climatology and thus detecting unusually wet or dry conditions.

Data and technical documentation are available for download from
\url{http://www.tamsat.org.uk/data/}.


\subsection{Defining Growing Season} \label{sec:appRS_gs}

We define growing season following the FAO \href{http://www.fao.org/agriculture/seed/cropcalendar/welcome.do}{crop calendar} for each country. Table~\ref{tab:growseason} presents details for each country on the growing season used, as well as whether that season spans years and whether it is unimodal or bimodal. Remote sensing data used in our analysis follows the defined growing season in each respective country.

Of the six countries, two (Malawi and Tanzania) span calendar years, which means that the growing season begins in one year and stretches into the year that follows. Take, for example, Malawi. The growing season in that country begins on 1 October and ends on 30 April. This means that it would begin 1 October 2021 and would end 30 April 2022. 

Similarly, of the six countries, two (Nigeria and Uganda) are bimodal. The season modality designates whether different regions within the countries have different growing seasons. In both Nigeria and Uganda, the northern part of the country has a different growing season from the southern part of the country. In these cases we designate the modality of the season, and also provide the growing season dates for both regions.


\begin{table}[htbp]	\centering
	\caption{Weather Variables \& Transformations} \label{tab:Wvar}
	\scalebox{0.9}
	{ \setlength{\linewidth}{.1cm}\newcommand{\contents}
		{\begin{tabular}{ll}
			\\[-1.8ex]\hline 
			\hline \\[-1.8ex]
			\multicolumn{2}{l}{\emph{\textbf{Panel A}: Rainfall}} \\
			\multicolumn{1}{l}{Daily rainfall} & \multicolumn{1}{l}{In mm} \\
			\multicolumn{1}{l}{Mean} & \multicolumn{1}{p{11cm}}{The first moment of the daily rainfall distribution for the growing season$^\dagger$} \\
			\multicolumn{1}{l}{Median} & \multicolumn{1}{p{11cm}}{The median daily rainfall for the growing season$^\dagger$} \\
			\multicolumn{1}{l}{Variance} & \multicolumn{1}{p{11cm}}{The second moment of the daily rainfall distribution for the growing season$^\dagger$} \\
			\multicolumn{1}{l}{Skew} & \multicolumn{1}{p{11cm}}{The third moment of the daily rainfall distribution for the growing season$^\dagger$} \\
			\multicolumn{1}{l}{Total} & \multicolumn{1}{p{11cm}}{Cumulative daily rainfall for the growing season$^\dagger$} \\
			\multicolumn{1}{l}{Deviations in total rainfall} & \multicolumn{1}{p{11cm}}{The z-score for cumulative daily rainfall for the growing season$^\dagger$} \\
			\multicolumn{1}{l}{Scaled deviations in total rainfall} & \multicolumn{1}{p{11cm}}{The z-score for cumulative daily rainfall for the growing season$^\dagger$} \\
			\multicolumn{1}{l}{Rainfall days} & \multicolumn{1}{p{11cm}}{The number of days with at least 1 mm of rain for the growing season$^\dagger$} \\
			\multicolumn{1}{l}{Deviation in rainfall days} & \multicolumn{1}{p{11cm}}{The number of days with rain for the growing season minus the long run average$^*$} \\
			\multicolumn{1}{l}{No rain days} & \multicolumn{1}{p{11cm}}{The number of days with less than 1 mm of rain for the growing season$^\dagger$} \\
			\multicolumn{1}{l}{Deviation in no rain days} & \multicolumn{1}{p{11cm}}{The number of days without rain for the growing season minus the long run average$^*$} \\
			\multicolumn{1}{l}{Share of rainy days} & \multicolumn{1}{p{11cm}}{The percent of growing season days with rain$^\dagger$} \\
			\multicolumn{1}{l}{Deviation in share of rainy days} & \multicolumn{1}{p{11cm}}{The percent of growing season days with rain minus the long run average$^\dagger$ $^*$} \\
			\multicolumn{1}{l}{Intra-season dry spells} & \multicolumn{1}{p{11cm}}{The maximum length of time (measured in days) without rain during the growing season$^\dagger$} \\
			\midrule
			& \\
			\multicolumn{2}{l}{\emph{\textbf{Panel B}: Temperature}} \\
			\multicolumn{1}{l}{Daily average temperature} & \multicolumn{1}{l}{In $^{\circ}$Celsius} \\
			\multicolumn{1}{l}{Daily maximum temperature} & \multicolumn{1}{l}{In $^{\circ}$Celsius} \\	
			\multicolumn{1}{l}{Mean} & \multicolumn{1}{p{11cm}}{The first moment of the daily temperature distribution for the growing season$^\dagger$} \\
			\multicolumn{1}{l}{Median} & \multicolumn{1}{p{11cm}}{The median daily temperature for the growing season$^\dagger$} \\
			\multicolumn{1}{l}{Variance} & \multicolumn{1}{p{11cm}}{The second moment of the daily temperature distribution for the growing season$^\dagger$} \\
			\multicolumn{1}{l}{Skew} & \multicolumn{1}{p{11cm}}{The third moment of the daily temperature distribution for the growing season$^\dagger$} \\
			\multicolumn{1}{l}{Growing degree days (GDD)} & \multicolumn{1}{p{11cm}}{The number of days within bound temperature for the growing season, following \cite{RS1991}$^\dagger$} \\
			\multicolumn{1}{l}{Deviation in GDD} & \multicolumn{1}{p{11cm}}{GDD for the growing season minus the long run average$^\dagger$ $^*$} \\
			\multicolumn{1}{l}{Scaled deviation in GDD} & \multicolumn{1}{p{11cm}}{The z-score for GDD} \\
			\multicolumn{1}{l}{Maximum temperature} & \multicolumn{1}{p{11cm}}{The average maximum daily temperature} \\
			\\[-1.8ex]\hline 
			\hline \\[-1.8ex]
			\multicolumn{2}{p{\linewidth}}{\footnotesize  \textit{Note}: The table presents definitions for included weather variables and transformations from weather sources defined in Table~\ref{tab:weather}. $^\dagger$Growing season determined for each country following \href{http://www.fao.org/agriculture/seed/cropcalendar/welcome.do}{FAO crop calendar} (see Table~\ref{tab:growseason}). $^*$For variables when ``long run'' is referenced, long run is defined as the entire length of the weather dataset. While each weather source has a different start date, to ensure blinding all datasets were shortened to 1983, which is the latest start date of the data sources.} \\
		\end{tabular}}
	\setbox0=\hbox{\contents}
    \setlength{\linewidth}{\wd0-2\tabcolsep-.25em}
    \contents}
\end{table}


\newpage
\begin{table}[htbp]	\centering
	\caption{Growing Seasons} \label{tab:growseason}
	\scalebox{0.9}
	{ \setlength{\linewidth}{.1cm}\newcommand{\contents}
		{\begin{tabular}{llll}
			\\[-1.8ex]\hline 
			\hline \\[-1.8ex]
			& \multicolumn{1}{c}{Growing Season} & \multicolumn{1}{c}{Span Calendar Years} & \multicolumn{1}{c}{Season Modality}  \\
			\multicolumn{1}{l}{\href{http://www.fao.org/giews/countrybrief/country.jsp?code=ETH}{Ethiopia}} & \multicolumn{1}{l}{1 March - 30 November} & \multicolumn{1}{c}{no} & \multicolumn{1}{c}{unimodal}  \\
			\multicolumn{1}{l}{\href{http://www.fao.org/giews/countrybrief/country.jsp?code=MWI}{Malawi}} & \multicolumn{1}{l}{1 October - 30 April} & \multicolumn{1}{c}{yes} & \multicolumn{1}{c}{unimodal}  \\
			\multicolumn{1}{l}{\href{http://www.fao.org/giews/countrybrief/country.jsp?code=NER}{Niger}} & \multicolumn{1}{l}{1 June - 30 November} & \multicolumn{1}{c}{no} & \multicolumn{1}{c}{unimodal}  \\
			\multicolumn{1}{l}{\href{http://www.fao.org/giews/countrybrief/country.jsp?code=NGA}{Nigeria}} & \multicolumn{1}{l}{\emph{North}: 1 May - 30 September} & \multicolumn{1}{c}{no} & \multicolumn{1}{c}{bimodal}  \\
			 & \multicolumn{1}{l}{\emph{South}: 1 March - 31 August} &  &  \\
			\multicolumn{1}{l}{\href{http://www.fao.org/giews/countrybrief/country.jsp?code=TZA}{Tanzania}} & \multicolumn{1}{l}{1 November - 30 April} & \multicolumn{1}{c}{yes} & \multicolumn{1}{c}{unimodal}  \\
			\multicolumn{1}{l}{\href{http://www.fao.org/giews/countrybrief/country.jsp?code=UGA}{Uganda}} & \multicolumn{1}{l}{\emph{North}: 1 April - 30 September}  & \multicolumn{1}{c}{no} & \multicolumn{1}{c}{bimodal} \\
	    	& \multicolumn{1}{l}{\emph{South}: 1 February - 31 July} & &  \\
			\\[-1.8ex]\hline 
			\hline \\[-1.8ex]
			\multicolumn{4}{p{\linewidth}}{\footnotesize  \textit{Note}: \footnotesize The table presents the growing season ranges, as defined by following FAO \href{http://www.fao.org/agriculture/seed/cropcalendar/welcome.do}{crop calendar} for each country, respectively.} \\
		\end{tabular}}
	\setbox0=\hbox{\contents}
    \setlength{\linewidth}{\wd0-2\tabcolsep-.25em}
    \contents}
\end{table}


\clearpage
\newpage
\section{Details on Household Data from the LSMS-ISA} \label{sec:appHH}

The World Bank Living Standards Measurement Study - Integrated Surveys on Agriculture (LSMS-ISA) is a household survey program that provides financial and technical assistance to national statistical offices in Sub-Saharan Africa for the design and implementation of national, multi-topic longitudinal household surveys with a focus on agriculture. The LSMS-ISA-supported countries include Burkina Faso, Ethiopia, Malawi, Mali, Niger, Nigeria, Uganda and Tanzania. We use the datasets from Ethiopia, Malawi, Niger, Nigeria, Uganda, and Tanzania in this work.\footnote{We intend to extend our analysis to include Mali. We do not intend to include Burkina Faso, due to issues with geo-reference locations which make its use incompatible with the project methodology.} More details on each country are included in the following sub-sections and details on samples are provided in Table~\ref{tab:lsms}. 

A common feature of the LSMS-ISA-supported surveys is that each sample household receives a multi-topic Household Questionnaire that elicit comprehensive socioeconomic information that also allows for the construction of consumption and income aggregates. Households engaged in agricultural activities additionally receive an Agriculture Questionnaire that elicits comprehensive information on smallholder crop, livestock and fishery activities and that allows for the construction of plot-level indicators of land and labor productivity and input use, among others. Finally, while the key variables that drive each survey's sampling design is household consumption and income, each survey provides a large sample of agricultural households in each round.

In our analysis, we only include households which did not move. Although the LSMS-ISA surveys follow individuals who ``split off'' and create new households, we do not include these movers in our analysis. 

\subsection{Ethiopia}

The LSMS-ISA data from Ethiopia includes three waves. Wave 1 (2011/12) includes 4,000 households in rural and small towns across the country \citepalias{ETH1}. This initial sample was followed in 2013/14 and 2015/16 \citepalias{ETH2, ETH3}. Beginning in Wave 2 (2013/14) the survey was also expanded to include 1,500 households in urban areas.

The Wave 1 data is representative at the regional level for the most populous regions (Amhara, Oromiya, Southern Nations, Nationalities, and People's Region, and Tigray). In Wave 2, in order to align with the existing Wave 1 design while ensuring that all urban areas were included, the population frame was stratified to provide population inferences for the same five domains as in Wave 1 as well as an additional domain for the city state of Addis Ababa. However, the sample size in both waves, is not sufficient to support region-specific estimates for each of the small regions (Afar, Benshangul Gumuz, Dire Dawa, Gambella, Harari, and Somalie). 

\subsection{Malawi}

The LSMS-ISA data from Malawi includes two separate surveys: (1) Integrated Household Survey, from which we include the first wave and (2) Integrated Household Panel Survey which includes three waves \citepalias{MWI1, MWI2, MWI3}. The two surveys are different in their representation of various households within the country. In this analysis, we rely only on the Integrated Household Panel Survey. 

The Integrated Household Panel Survey begins with Wave 1 in 2010 and includes 3,247 households from 204 enumeration areas that were visited as part of the Third Integrated Household Survey 2010/11 and that were designated as ``panel'' for follow-up, starting again in 2013. The sample was designed to be representative at the national-, urban/rural-, and regional-level at baseline. Wave 2 from 2013 aimed to track all panel households from Wave 1, including all individuals that changed locations between the waves. The Wave 2 household sample size was 4,000, including new households that were formed by split off individuals that were tracked. Finally, Wave 3 from 2016 aimed to track all households and split off individuals that were ever associated with a random half of 204 original enumeration areas that had been visited in 2010. The Wave 3 household sample was 2,500 households, including again new households that were formed by split off individuals that were tracked from previous rounds.

\subsection{Niger}

The LSMS-ISA data from Niger includes two rounds. In Wave 1. approximately 4,000 households in 270 Zones de Dénombrement \citepalias{NGR1}. The sample is nationally representative, as well as representative of Niamey, other urban, and rural areas. Households visited in Wave 1 were re-visited in Wave 2, including households and individuals who moved after the 2011 survey \citepalias{NGR2}. When the entire household moved within Niger, the household was found and re-interviewed in the second wave. When individuals from the household moved, one individual per household was selected to follow. This forms a sample of approximately 3,600 households in Wave 2. 

\subsection{Nigeria}

The LSMS-ISA data from Nigeria includes three waves \citepalias{NGA1, NGA2, NGA3}. The total sample consists of 5,000 panel households and is representative at the national level. Households are visited twice per wave of the Panel, both post-planting and post-harvest. The post-harvest visit is implemented jointly with a larger General Household Survey of 22,000 households (5,000 panel and 17,000 non-panel households). The sample is representative at the national level and provides reliable estimates of key socio-economic variables for the six zones in the country.

\subsection{Tanzania}

Three waves of the LSMS-ISA data from Tanzania are included in our analysis. The first wave includes 3,265 households and the sample is representative for the nation, and provides reliable estimates of key socioeconomic variables for mainland rural areas, Dar es Salaam, other mainland urban areas, and Zanzibar \citepalias{TZA1}. In Wave 2, all original households were targeted for revisit \citepalias{TZA2}. For those household members still residing in their original location, they were simply re-interviewed. For adults who had relocated, these individuals were tracked and re-interviewed in their new location with their new households. As a result of this, the sample size for the second round expanded to 3,924 households. Wave 3 adhered to the same tracking protocol as Wave 2, resulting in a final sample size of 5,015 households \citepalias{TZA3}.

\subsection{Uganda}

The LSMS-ISA from Uganda includes five waves, of which we use three in this analysis. Wave 1 (2009/10) includes approximately 3,200 households that were previously interviewed by the Uganda National Household Survey (UNHS) in 2005/06  \citepalias{UGA1}. The sample was designed to be representative at the national-, urban/rural- and regional-level. For subsequent waves, the Wave 1 sample was followed, including tracking of shifted and split-off households, for two additional rounds: 2010/11 and 2011/12 \citepalias{UGA2, UGA3}. Each round includes nearly 3,000 households. 


\clearpage
\newpage
\section{Anonymization Methods}\label{sec:appextraction}

\setcounter{table}{0}
\renewcommand{\thetable}{C\arabic{table}}
\setcounter{figure}{0}
\renewcommand{\thefigure}{C\arabic{figure}}

Extending section~\ref{sec:ext}, in this section, we present further evidence on Hypothesis $1$ ($H_0^1$ - different obfuscation procedures implemented to preserve privacy of farms or households have no impact on estimates of agricultural productivity). The following figures (Figures~\ref{fig:quad_cty1_rf} through \ref{fig:quad_cty1_rf}) pool the results from 77,760 regressions and then divide the pool into ten bins, one for each anonymization method for 7,776 regression results. Extending the results presented in the main text, these figures examine the differences in coefficients $(\beta_{1})$ and their relative significance by anonymization method.



\begin{landscape}
\begin{figure}[!htbp]
	\begin{minipage}{\linewidth}		
		\caption{Specification Curve for Rainfall Variables in Ethiopia}
		\label{fig:quad_cty1_rf}
		\begin{center}
			\includegraphics[width=.9\linewidth,keepaspectratio]{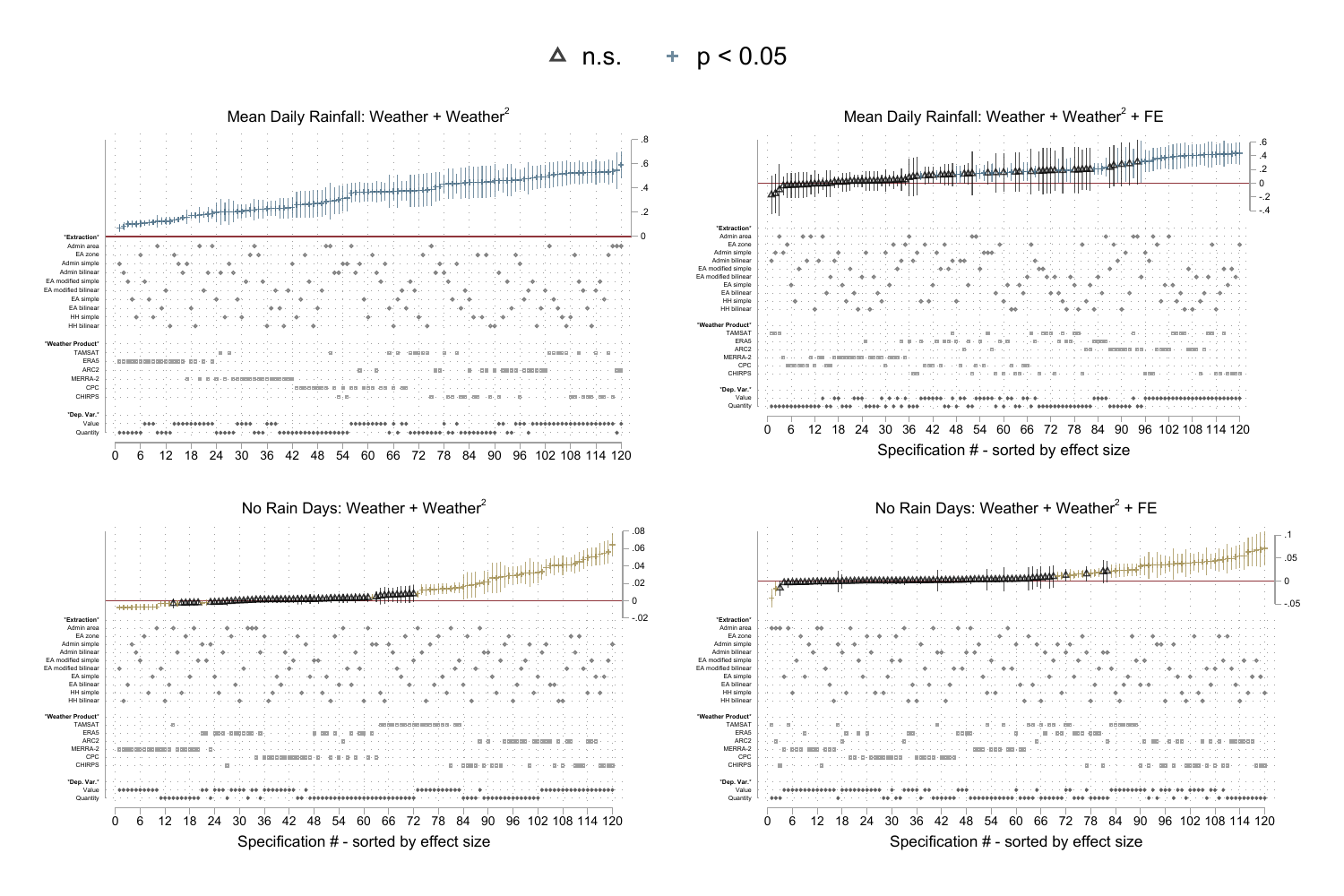}
		\end{center}
		\footnotesize  \textit{Note}: The figure presents specification curves where each panel presents results from a different model. Each panel includes 120 regressions, where each column represents a single regression. Significant and non-significant coefficients are designated at the top of the figure.
	\end{minipage}	
\end{figure}
\end{landscape}

\begin{landscape}
\begin{figure}[!htbp]
	\begin{minipage}{\linewidth}		
		\caption{Specification Curve for Rainfall Variables in Malawi}
		\label{fig:vquad_cty2_rf}
		\begin{center}
			\includegraphics[width=.9\linewidth,keepaspectratio]{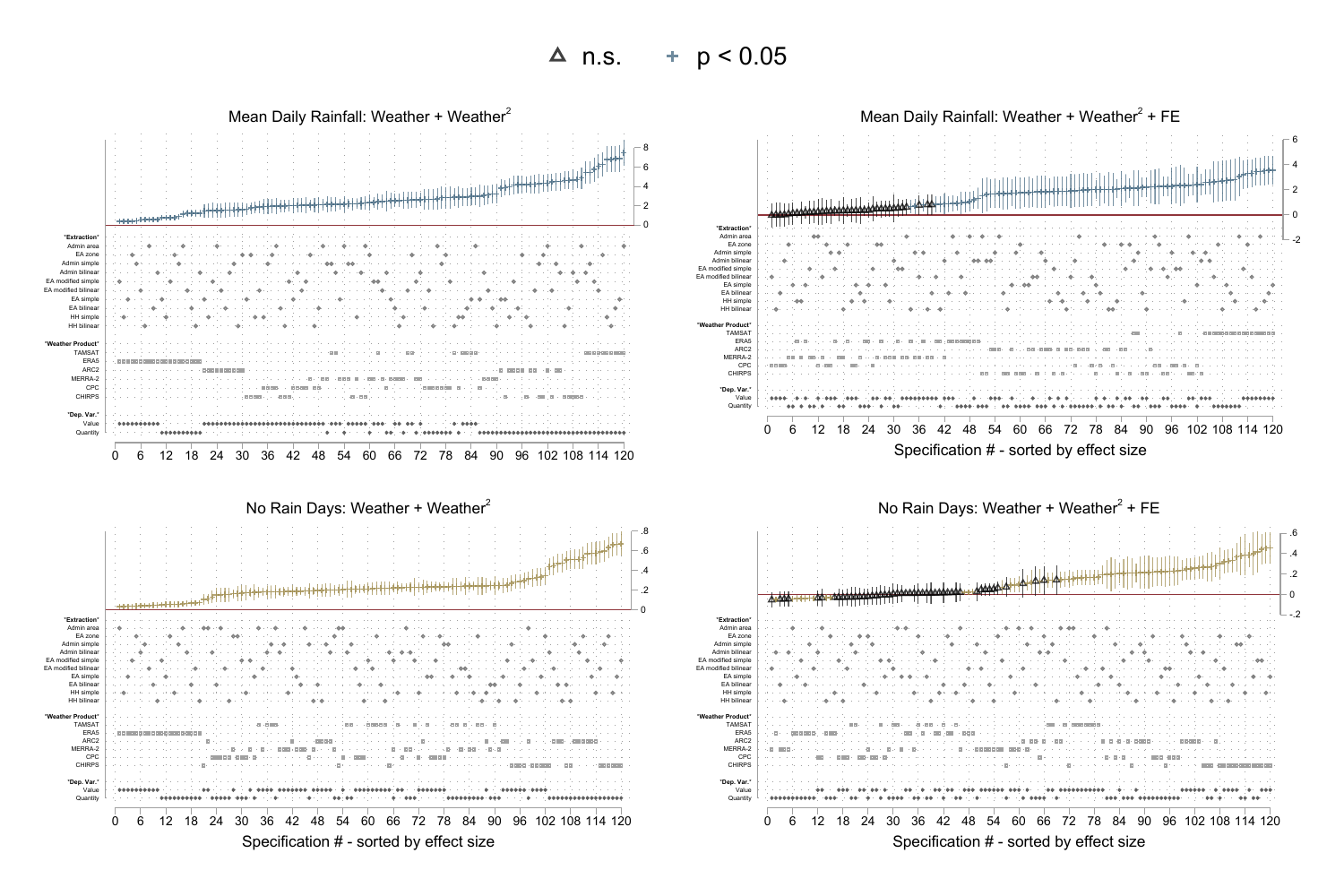}
		\end{center}
		\footnotesize  \textit{Note}: The figure presents specification curves where each panel presents results from a different model. Each panel includes 120 regressions, where each column represents a single regression. Significant and non-significant coefficients are designated at the top of the figure.
	\end{minipage}	
\end{figure}
\end{landscape}

\begin{landscape}
\begin{figure}[!htbp]
	\begin{minipage}{\linewidth}		
		\caption{Specification Curve for Rainfall Variables in Niger}
		\label{fig:quad_cty4_rf}
		\begin{center}
			\includegraphics[width=.9\linewidth,keepaspectratio]{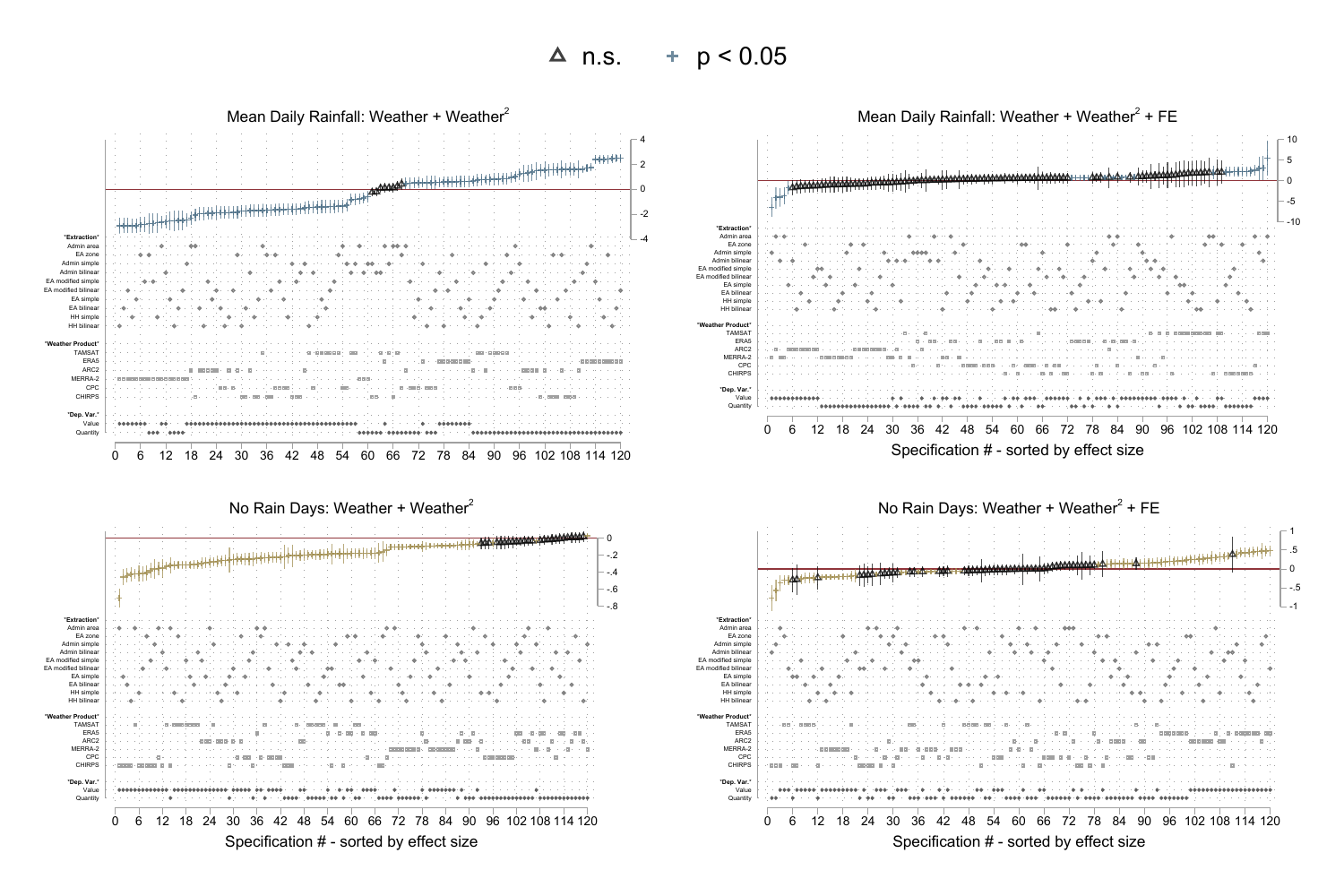}
		\end{center}
		\footnotesize  \textit{Note}: The figure presents specification curves where each panel presents results from a different model. Each panel includes 120 regressions, where each column represents a single regression. Significant and non-significant coefficients are designated at the top of the figure.
	\end{minipage}	
\end{figure}
\end{landscape}

\begin{landscape}
\begin{figure}[!htbp]
	\begin{minipage}{\linewidth}		
		\caption{Specification Curve for Rainfall Variables in Nigeria}
		\label{fig:quad_cty5_rf}
		\begin{center}
			\includegraphics[width=.9\linewidth,keepaspectratio]{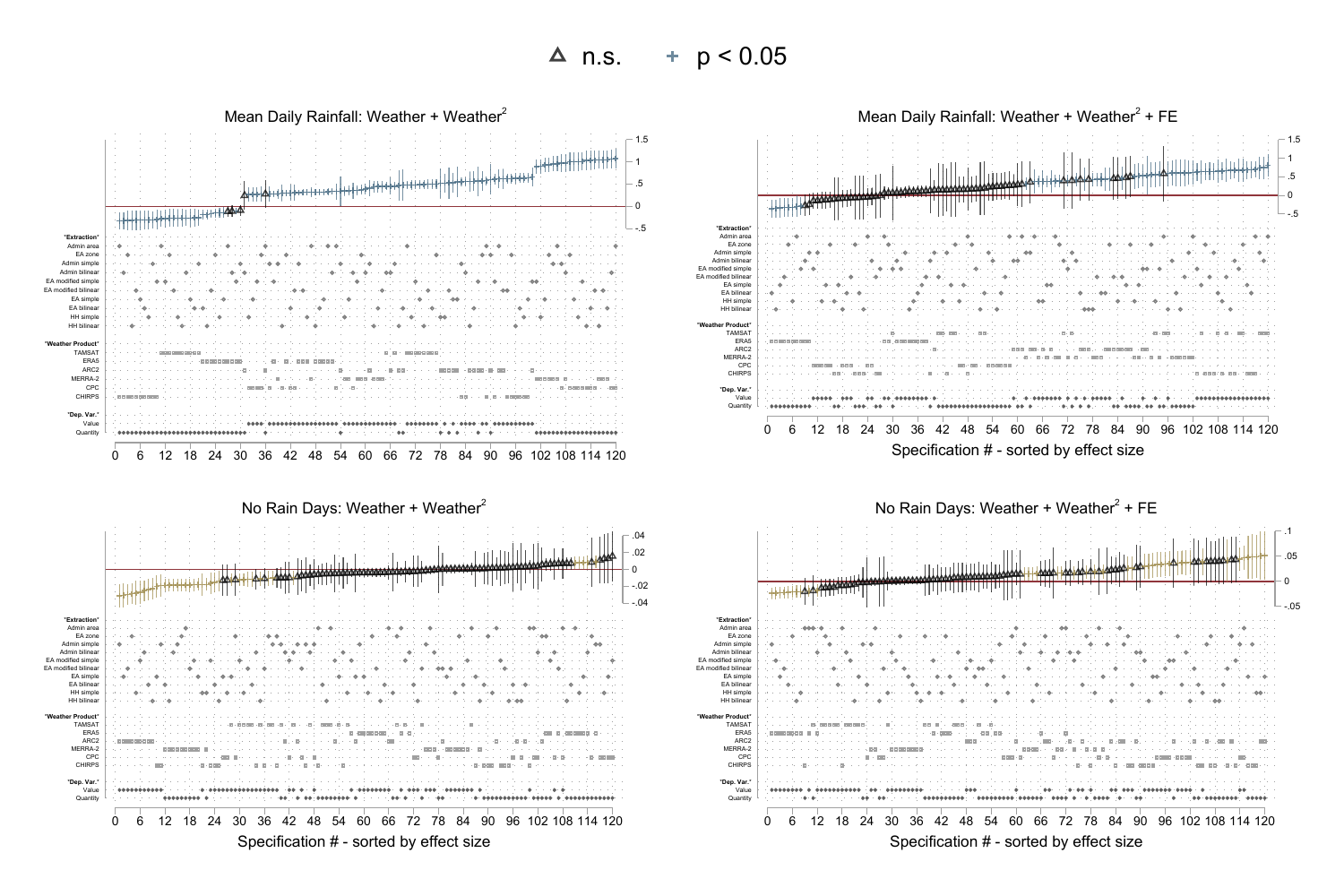}
		\end{center}
		\footnotesize  \textit{Note}: The figure presents specification curves where each panel presents results from a different model. Each panel includes 120 regressions, where each column represents a single regression. Significant and non-significant coefficients are designated at the top of the figure.
	\end{minipage}	
\end{figure}
\end{landscape}

\begin{landscape}
\begin{figure}[!htbp]
	\begin{minipage}{\linewidth}		
		\caption{Specification Curve for Rainfall Variables in Tanzania}
		\label{fig:quad_cty6_rf}
		\begin{center}
			\includegraphics[width=.9\linewidth,keepaspectratio]{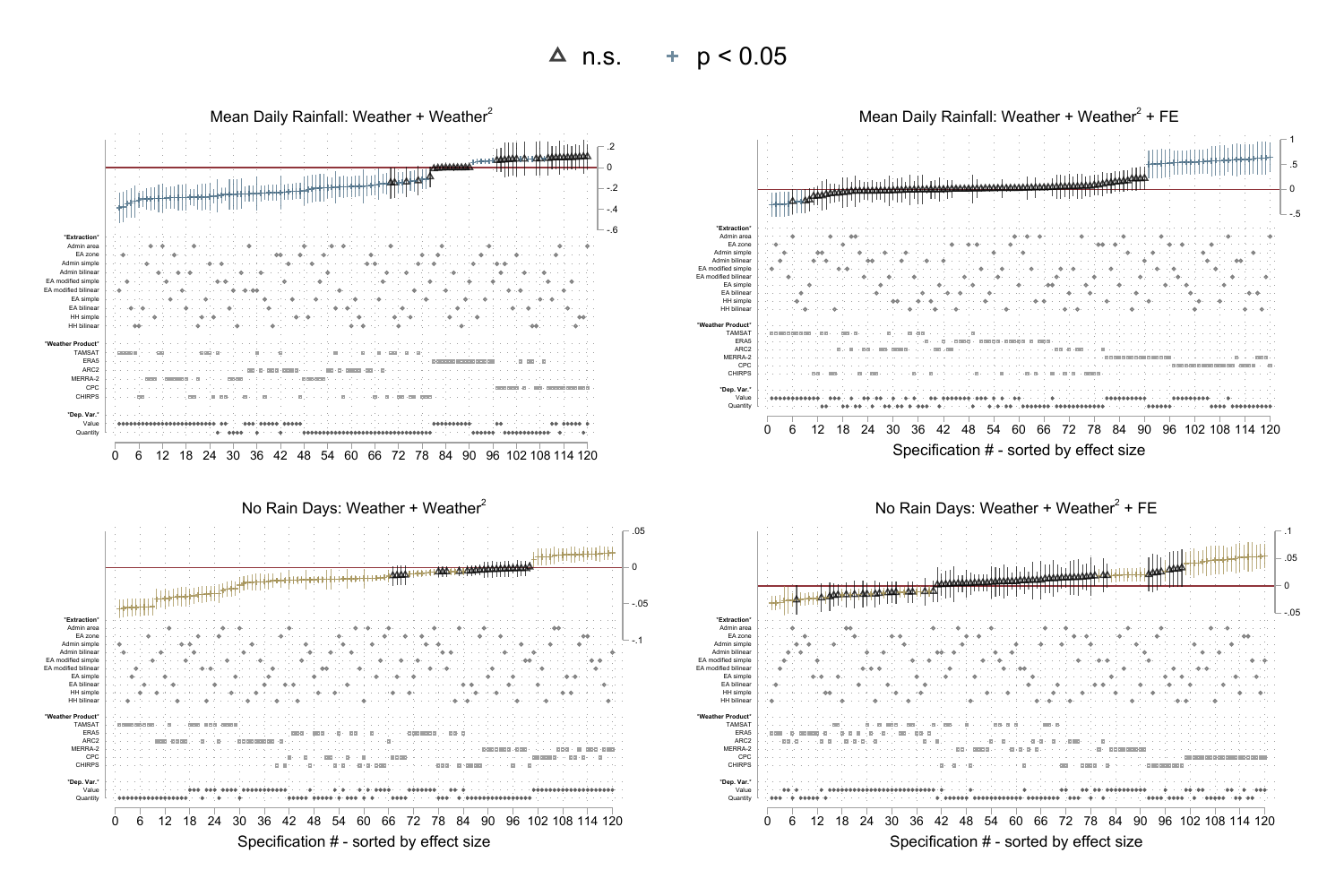}
		\end{center}
		\footnotesize  \textit{Note}: The figure presents specification curves where each panel presents results from a different model. Each panel includes 120 regressions, where each column represents a single regression. Significant and non-significant coefficients are designated at the top of the figure.
	\end{minipage}	
\end{figure}
\end{landscape}

\begin{landscape}
\begin{figure}[!htbp]
	\begin{minipage}{\linewidth}		
		\caption{Specification Curve for Rainfall Variables in Uganda}
		\label{fig:quad_cty7_rf}
		\begin{center}
			\includegraphics[width=.9\linewidth,keepaspectratio]{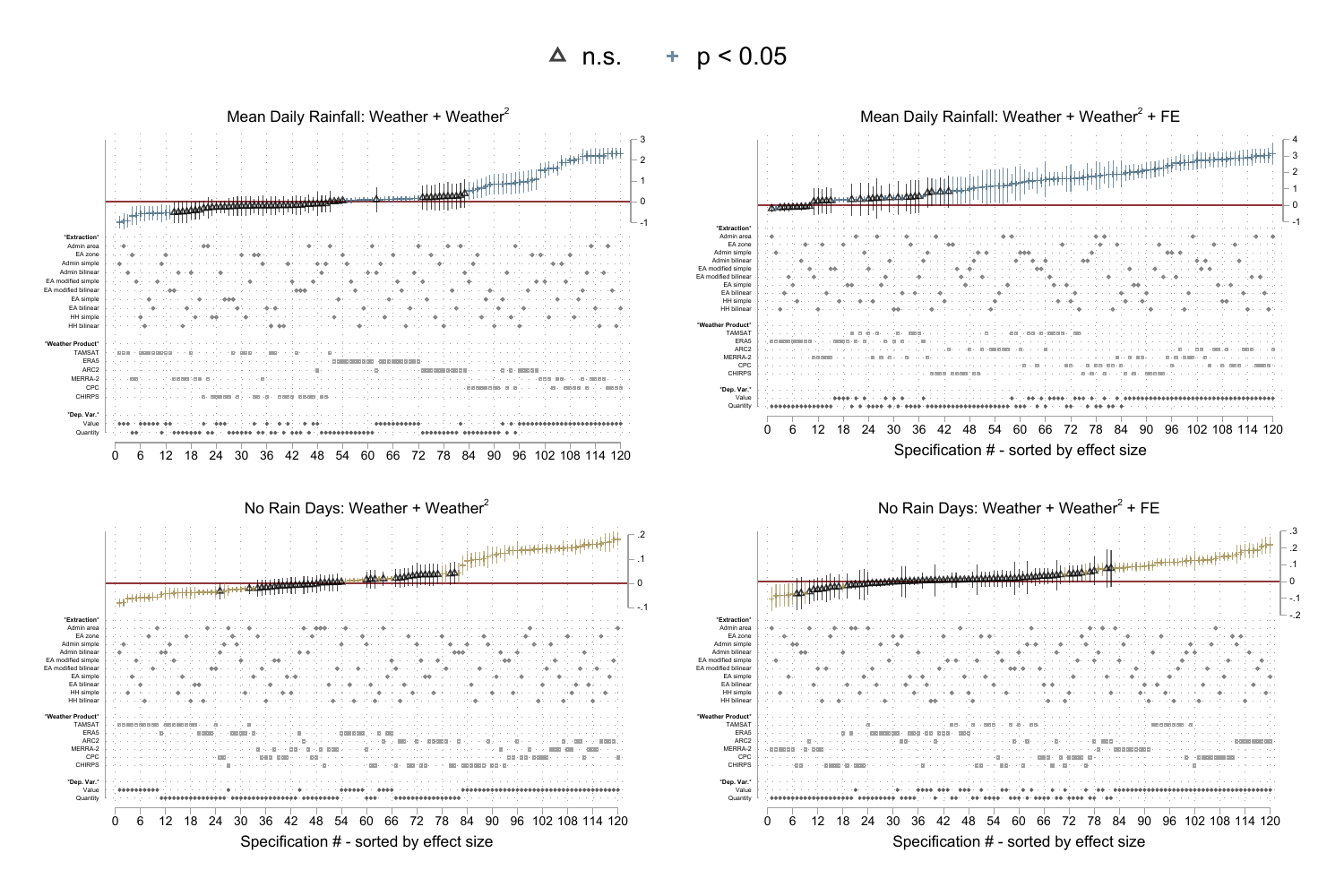}
		\end{center}
		\footnotesize  \textit{Note}: The figure presents specification curves where each panel presents results from a different model. Each panel includes 120 regressions, where each column represents a single regression. Significant and non-significant coefficients are designated at the top of the figure.
	\end{minipage}	
\end{figure}
\end{landscape}


\begin{landscape}
\begin{figure}[!htbp]
	\begin{minipage}{\linewidth}		
		\caption{Specification Curve for Temperature Variables in Ethiopia}
		\label{fig:quad_cty1_tp}
		\begin{center}
			\includegraphics[width=.9\linewidth,keepaspectratio]{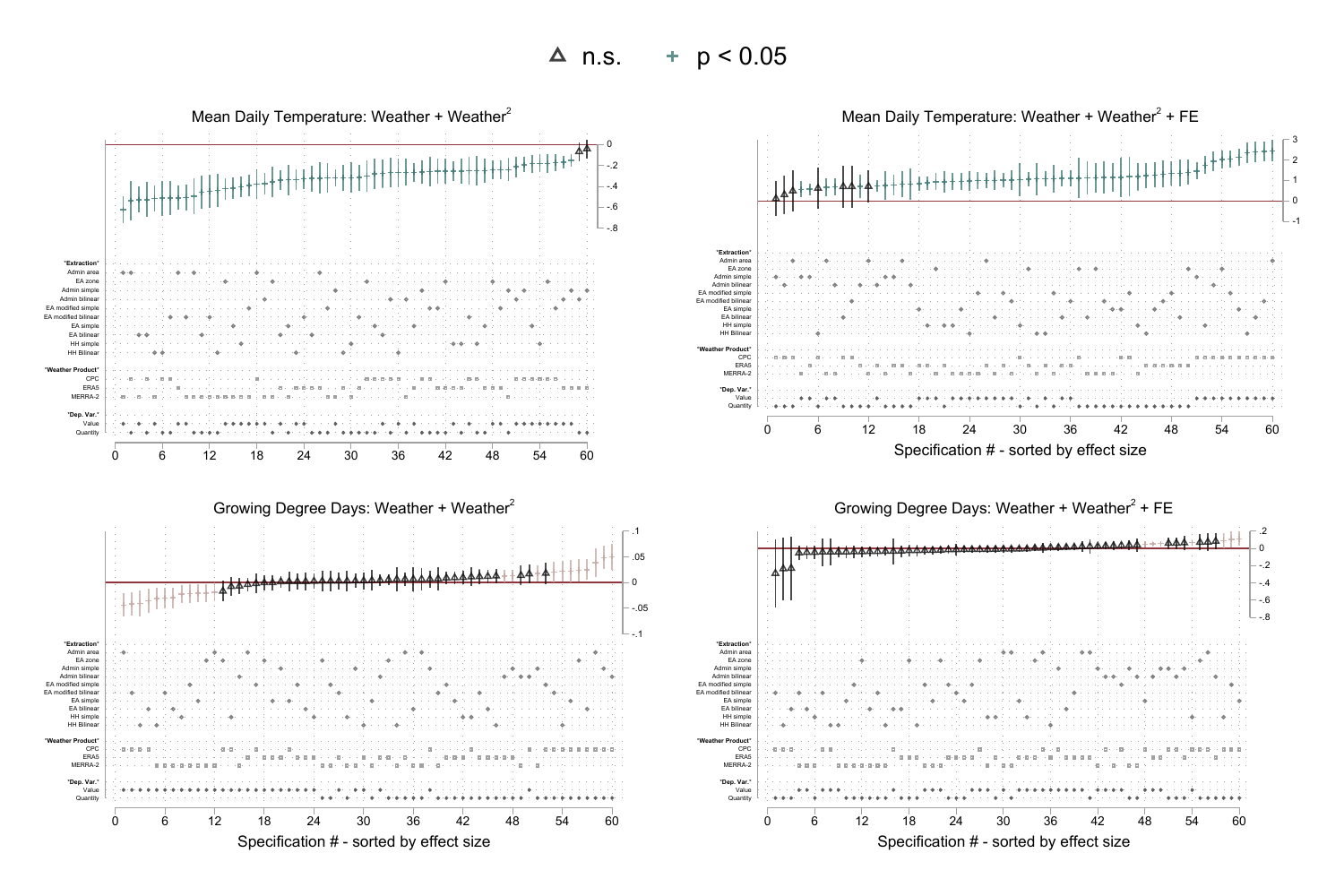}
		\end{center}
		\footnotesize  \textit{Note}: The figure presents specification curves where each panel presents results from a different model. Each panel includes 60 regressions, where each column represents a single regression. Significant and non-significant coefficients are designated at the top of the figure.
	\end{minipage}	
\end{figure}
\end{landscape}

\begin{landscape}
\begin{figure}[!htbp]
	\begin{minipage}{\linewidth}		
		\caption{Specification Curve for Temperature Variables in Malawi}
		\label{fig:quad_cty2_tp}
		\begin{center}
			\includegraphics[width=.9\linewidth,keepaspectratio]{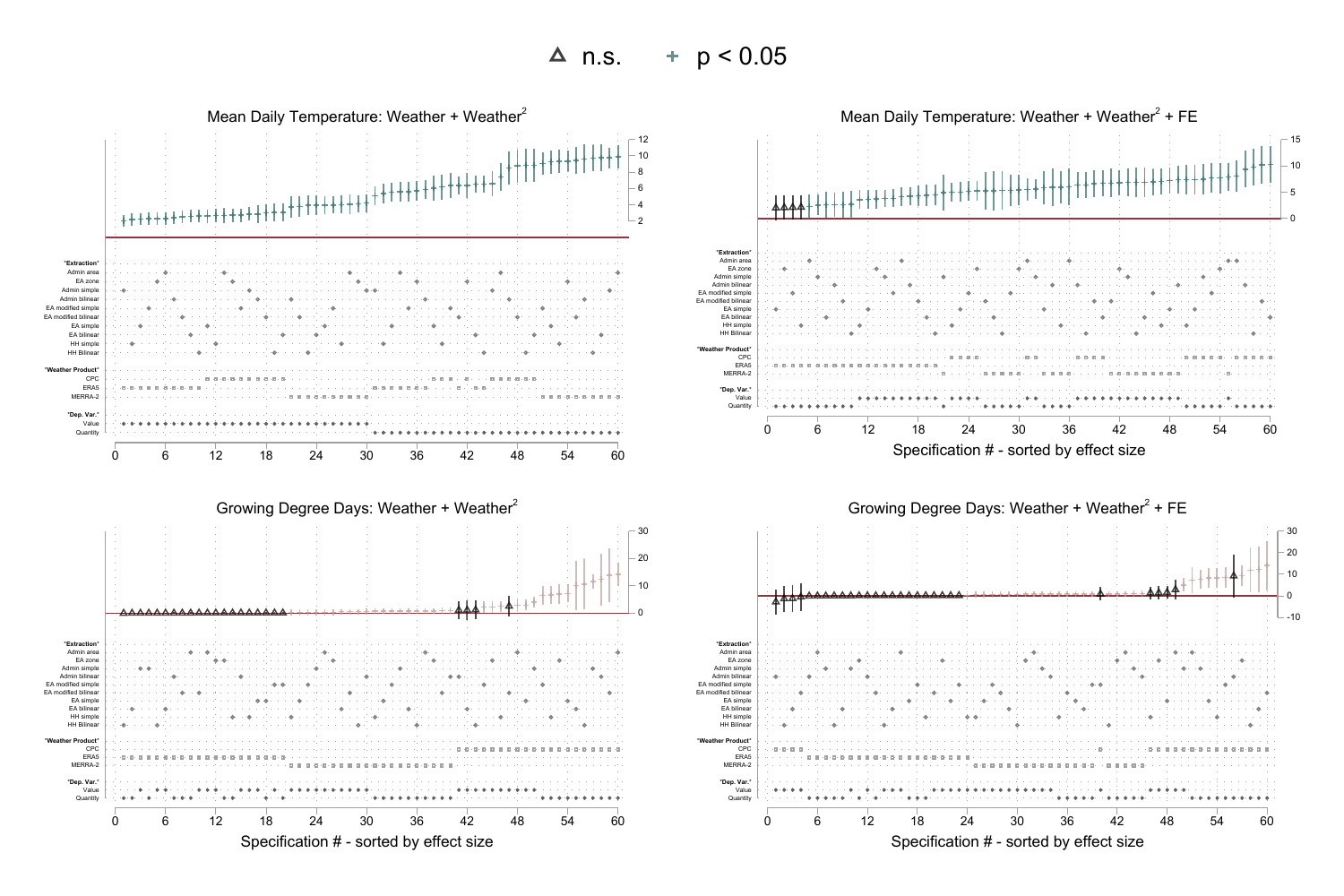}
		\end{center}
		\footnotesize  \textit{Note}: The figure presents specification curves where each panel presents results from a different model. Each panel includes 60 regressions, where each column represents a single regression. Significant and non-significant coefficients are designated at the top of the figure.
	\end{minipage}	
\end{figure}
\end{landscape}

\begin{landscape}
\begin{figure}[!htbp]
	\begin{minipage}{\linewidth}		
		\caption{Specification Curve for Temperature Variables in Niger}
		\label{fig:quad_cty4_tp}
		\begin{center}
			\includegraphics[width=.9\linewidth,keepaspectratio]{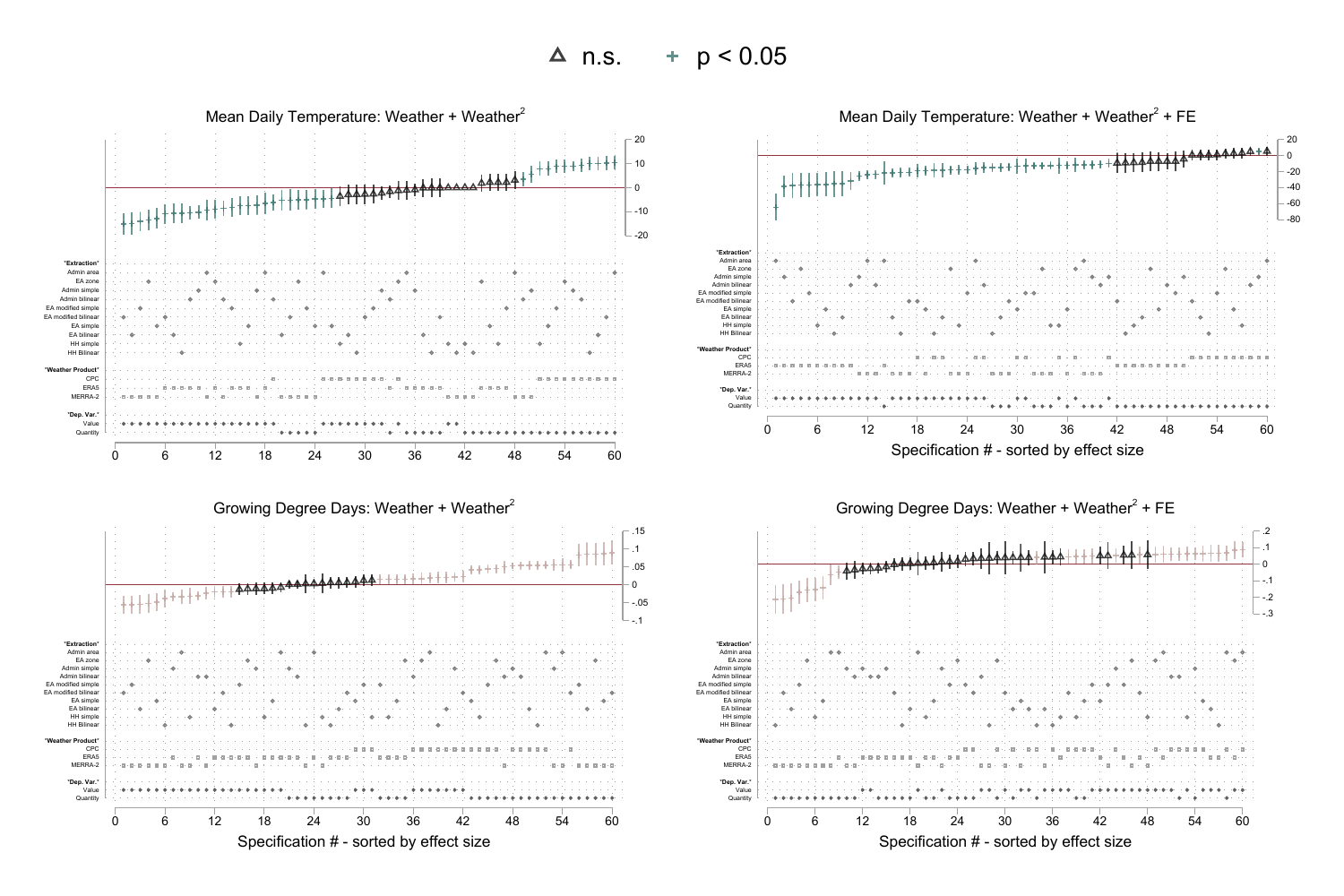}
		\end{center}
		\footnotesize  \textit{Note}: The figure presents specification curves where each panel presents results from a different model. Each panel includes 60 regressions, where each column represents a single regression. Significant and non-significant coefficients are designated at the top of the figure.
	\end{minipage}	
\end{figure}
\end{landscape}

\begin{landscape}
\begin{figure}[!htbp]
	\begin{minipage}{\linewidth}		
		\caption{Specification Curve for Temperature Variables in Nigeria}
		\label{fig:quad_cty5_tp}
		\begin{center}
			\includegraphics[width=.9\linewidth,keepaspectratio]{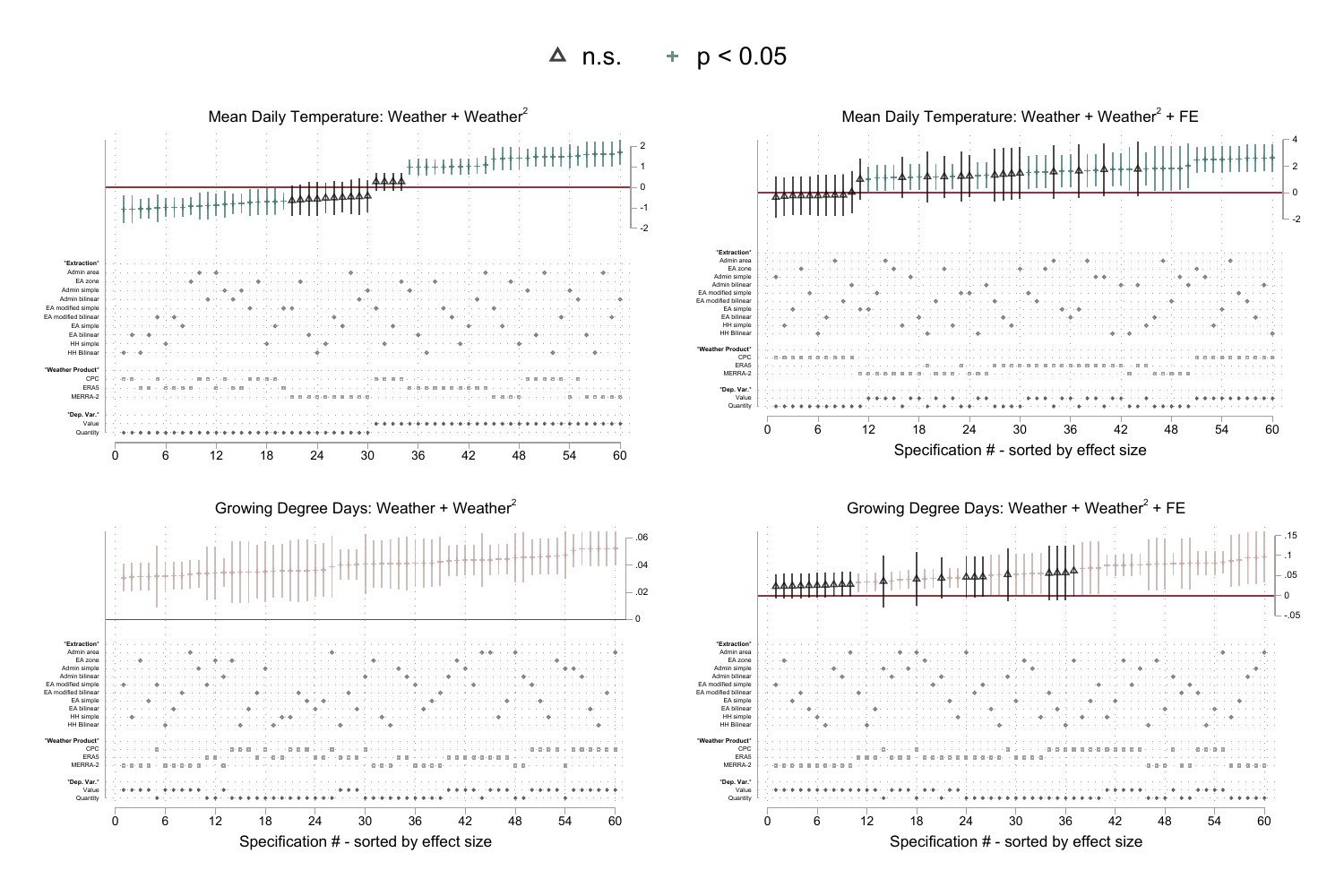}
		\end{center}
		\footnotesize  \textit{Note}: The figure presents specification curves where each panel presents results from a different model. Each panel includes 60 regressions, where each column represents a single regression. Significant and non-significant coefficients are designated at the top of the figure.
	\end{minipage}	
\end{figure}
\end{landscape}

\begin{landscape}
\begin{figure}[!htbp]
	\begin{minipage}{\linewidth}		
		\caption{Specification Curve for Temperature Variables in Tanzania}
		\label{fig:quad_cty6_tp}
		\begin{center}
			\includegraphics[width=.9\linewidth,keepaspectratio]{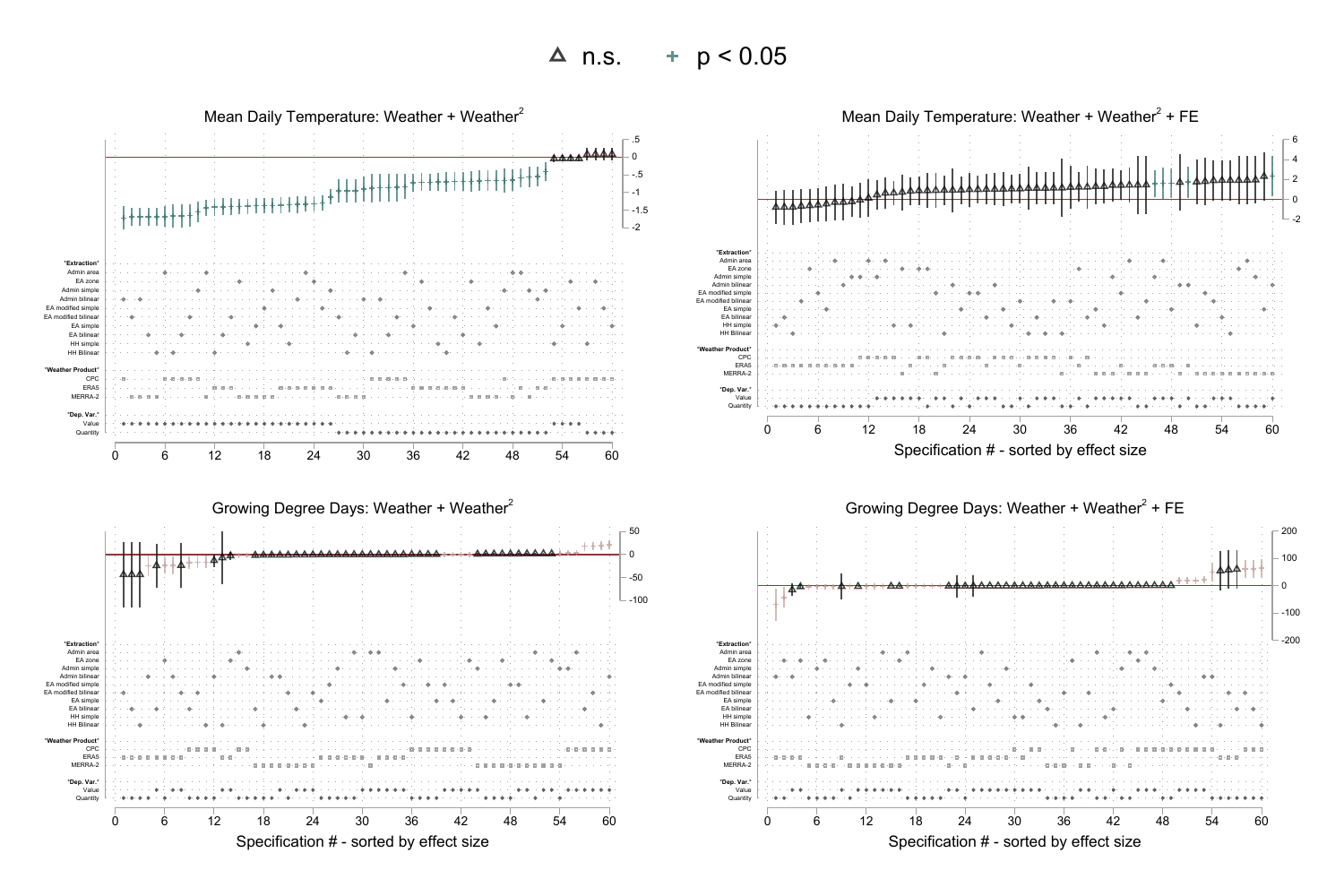}
		\end{center}
		\footnotesize  \textit{Note}: The figure presents specification curves where each panel presents results from a different model. Each panel includes 60 regressions, where each column represents a single regression. Significant and non-significant coefficients are designated at the top of the figure.
	\end{minipage}	
\end{figure}
\end{landscape}

\begin{landscape}
\begin{figure}[!htbp]
	\begin{minipage}{\linewidth}		
		\caption{Specification Curve for Temperature Variables in Uganda}
		\label{fig:quad_cty7_tp}
		\begin{center}
			\includegraphics[width=.9\linewidth,keepaspectratio]{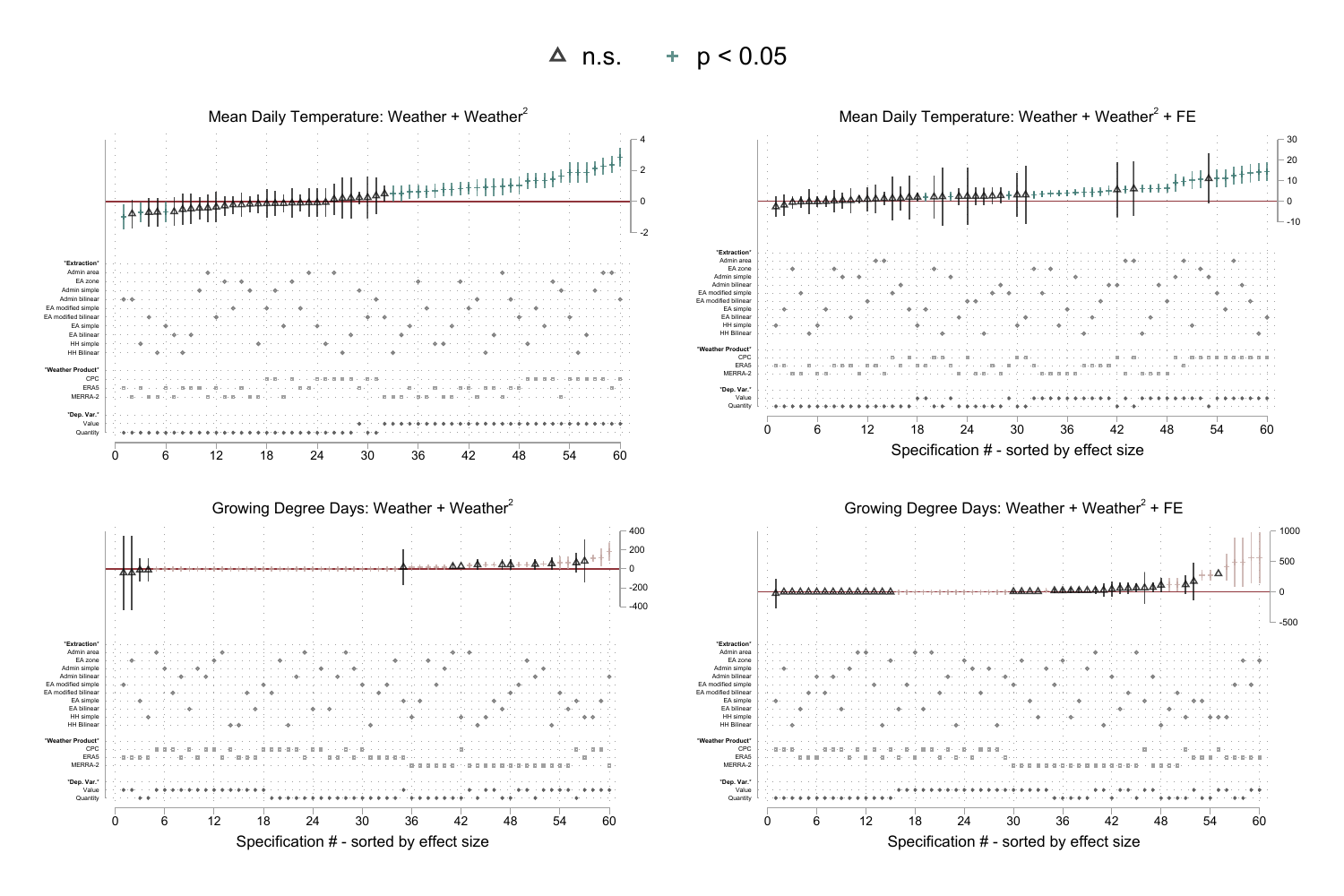}
		\end{center}
		\footnotesize  \textit{Note}: The figure presents specification curves where each panel presents results from a different model. Each panel includes 60 regressions, where each column represents a single regression. Significant and non-significant coefficients are designated at the top of the figure.
	\end{minipage}	
\end{figure}
\end{landscape}

\end{document}